\ifx\allfiles\undefined
\documentclass[a4paper,fleqn]{cas-sc}
\usepackage{rotating}
\usepackage{appendix}
\usepackage{graphicx}
\usepackage{caption}
\usepackage{tabularray}
\usepackage[authoryear]{natbib}
\usepackage{float}
\usepackage{tabularx}
\usepackage{makecell}
\usepackage{amsmath}
\usepackage{xcolor}
\usepackage{mathtools}
\usepackage{color}
\usepackage{booktabs}
\usepackage{array}
\usepackage{pdflscape}
\usepackage{longtable}
\usepackage{titlesec}
\titleformat{\section}{\normalfont\Large\bfseries}{\thesection}{1em}{}
\titleformat{\paragraph}
  {\normalfont\normalsize\bfseries}{\theparagraph}{2em}{}
\titlespacing*{\paragraph}{0pt}{0.5em}{0.5em}

\definecolor{rv}{rgb}{0, 0.501, 0.675}
\hypersetup{
  linkcolor = rv,  % 使用自定义颜色 rv
  citecolor = rv,  % 可以为引用也设置同样的颜色或另外选择
  urlcolor = rv    % 设置外部链接颜色
}
\def\tsc#1{\csdef{#1}{\textsc{\lowercase{#1}}\xspace}}
\tsc{WGM}
\tsc{QE}
\begin{document}
\let\WriteBookmarks\relax
\def\floatpagepagefraction{1}
\def\textpagefraction{. 001}

\shorttitle{} 

\shortauthors{ {Zhou et al. }}

\title [mode = title]{Twenty-Five Years of the Intelligent Driver Model: Foundations, Extensions, Applications, and Future Directions}
% Author Information
\author[label1,label2]{Shirui Zhou}
\ead{shiruizhou@tju.edu.cn}

\author[label3]{Shiteng Zheng}
\ead{stzheng1@bjtu.edu.cn}

% \author[label4]{Martin Treiber}
% \ead{martin@mtreiber.de}

\author[label1,label2]{Junfang Tian}
\cormark[1]

\ead{jftian@tju.edu.cn}

\author[label3]{Rui Jiang}

\ead{jiangrui@bjtu.edu.cn}

\author[label4]{H. M. Zhang}
\cormark[1]
\ead{hmzhang@ucdavis.edu}

\affiliation[label1]{organization={Institute of Systems Engineering, College of Management and Economics, Tianjin University},
addressline={No. 92 Weijin Road},
city={Nankai District},
postcode={300072},
state={Tianjin},
country={China}}
\affiliation[label2]{organization={Laboratory of Computation and Analytics of Complex Management Systems (CACMS),Tianjin University},
addressline={No. 92 Weijin Road},
city={Nankai District},
postcode={300072},
state={Tianjin},
country={China}}
\affiliation[label3]{organization={School of Systems Science, Beijing Jiaotong University},
addressline={No. 3 Shangyuancun},
city={Haidian District},
postcode={100044},
state={Beijing},
country={China}}
\affiliation[label4]{organization={Department of Civil and Environmental Engineering, University of California Davis},
            addressline={2001 Ghausi Hall Davis},
            city={Davis},
            postcode={95616},
            state={California},
            country={United States}}

% \affiliation[label4]{organization={Institute for Transport and Economics, Dresden University of Technology},
% addressline={Würzburger Str. 35},
% city={Dresden},
% postcode={D-01062},
% state={Saxony},
% country={Germany}}
\cortext[1]{Corresponding author. }

% \nonumnote{}

\begin{abstract}
The Intelligent Driver Model (IDM), proposed in 2000, has become a foundational tool in traffic flow modeling, renowned for its simplicity, computational efficiency, and ability to capture diverse traffic dynamics. Over the past 25 years, IDM has significantly advanced car-following theory and found extensive application in intelligent transportation systems, including driver assistance systems and autonomous vehicle control. However, IDM's deterministic framework and simplified assumptions face limitations in addressing real-world complexities such as stochastic variability, driver heterogeneity, and mixed traffic conditions. This paper provides a systematic review and critical reflection on IDM's theoretical foundations, academic influence, practical applications, and model extensions. While highlighting IDM's contributions, we emphasize the need to extend the model into a modular and extensible framework. Future directions include integrating stochastic elements, human behavioral insights, and hybrid modeling approaches that combine physics-based structures with data-driven methodologies. By reimagining IDM as a flexible modeling basis, this paper aims to inspire its continued development to meet the demands of intelligent, connected, and increasingly complex traffic systems.
\end{abstract}

% Keywords
\begin{keywords}
Intelligent driver model\sep Driving behavior modeling\sep Car following \sep Traffic dynamics
\end{keywords}

\maketitle
% Main text

{\section{Introduction}

Understanding and predicting traffic flow dynamics require accurate modeling of individual vehicle behavior, which serves as a foundation for both theoretical explorations and practical engineering applications in transportation systems. Among the various modeling approaches, car following (CF) models play a pivotal role by describing how drivers adjust their vehicle speed based on interactions with surrounding vehicles. These models provide essential tools for analyzing traffic stability, simulating diverse traffic scenarios, and optimizing intelligent transportation systems (ITS). Over the decades, several classical CF models have been proposed, including the GM model \citep{chandlerTrafficDynamicsStudiesCar1958}, the GHR model \citep{gazisNonlinearFollowLeaderModelsTraffic1961}, Gipps’ model \citep{gippsBehaviouralCarfollowingModelComputer1981}, the Wiedemann model \citep{wiedemann1992microscopic}, the Optimal Velocity Model (OVM, \citep{bandoDynamicalModelTrafficCongestion1995}), the Full Velocity Difference Model (FVDM, \citep{jiangFullVelocityDifferenceModel2001}), and Newell’s model \citep{newellSimplifiedCarfollowingTheoryLower2002}. Each of these models has contributed to advancing our understanding of driver behavior and traffic flow, offering varying levels of realism, computational efficiency, and theoretical rigor.

\begin{figure}[htbp]
    \centering
    \includegraphics[width=0.6\linewidth]{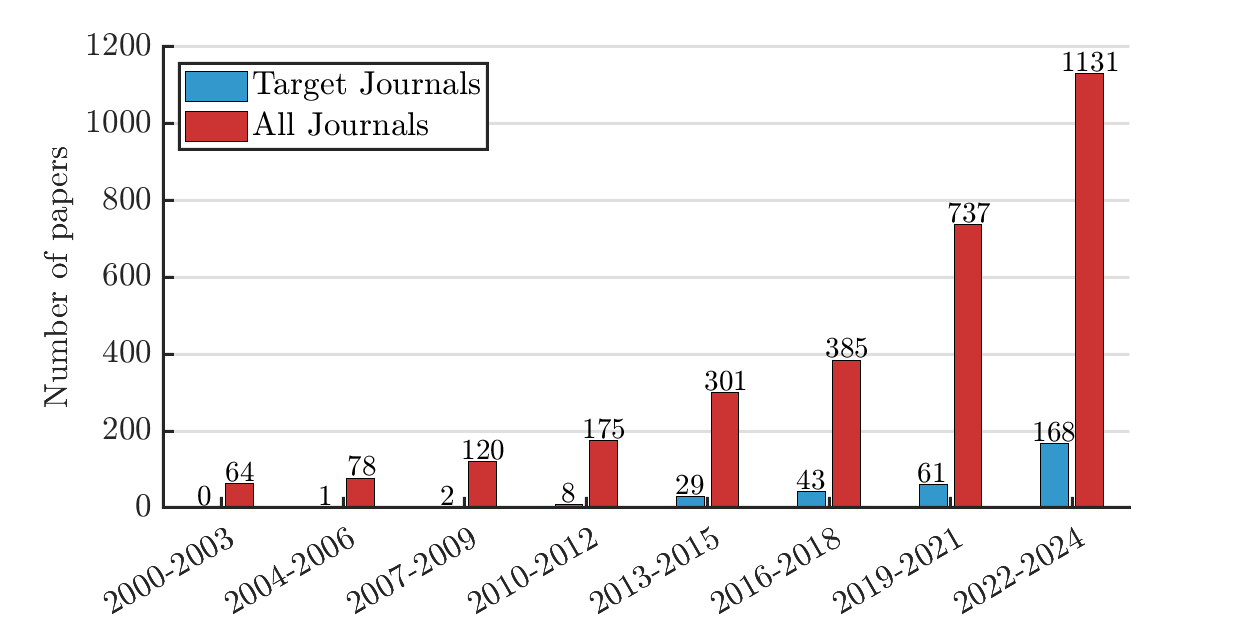}
    \caption{The red bars indicate the number of papers citing IDM in the Web of Science between 2000 and 2024. The blue bars indicate relevant papers published in the target list of journals (see Appendix) that at least use IDM for simulation or theoretical analysis rather than just citing it.}
    \label{fig:NumPaper}
\end{figure}

Among these, the Intelligent Driver Model (IDM), introduced by \cite{treiberCongestedTrafficStatesEmpirical2000}, has emerged as one of the most widely adopted CF models in traffic flow research and applications (see Fig. \ref{fig:NumPaper}). IDM is recognized for its simplicity, computational efficiency, and strong physical interpretability. Unlike earlier models, IDM conceptualizes vehicles as intelligent agents, dynamically adjusting their acceleration based on nonlinear interactions with the leading vehicle, while incorporating factors such as desired speed, safe distance, and relative velocity. This design allows IDM to capture a wide range of traffic states, including free flow, congestion, and stop-and-go oscillations, and to model complex phenomena like phase transitions, hysteresis, and multi-stability. Its versatility has made IDM a benchmark in both theoretical traffic flow studies and practical ITS applications.

Since its introduction in 2000, IDM has been extensively utilized in academic research and real-world applications, see Fig. \ref{fig:NumPaper}. Table \ref{tab:CFModel} indicates that the IDM has garnered  3458 citations, averaging 138.32 citations per year (CPY). It ranks first in both total citations and CPY among not only classic CF models but also more general models of transportation science. It has facilitated advancements in understanding CF behavior, deriving fundamental diagrams, and analyzing traffic flow stability. Moreover, IDM has been integrated into widely used platforms such as SUMO, enabling its application in evaluating vehicle-to-vehicle (V2V) and vehicle-to-infrastructure (V2I) communication technologies, as well as in testing advanced driver-assistance systems and autonomous vehicle control algorithms. However, despite its widespread adoption, IDM’s deterministic framework and simplified assumptions face challenges in capturing real-world traffic complexities, such as stochastic variability, driver heterogeneity, and the interactions between human-driven vehicles (HVs) and connected and autonomous vehicles (CAVs).

Given IDM’s foundational role in traffic flow modeling and the growing demands of modern transportation systems, this paper aims to systematically evaluate the model’s development, impact, and future potential over the past 25 years. Specifically, this study focuses on:

\begin{itemize}
    \item Summarizing IDM’s theoretical foundations and its key contributions to advancing CF theory, providing the basis for understanding traffic flow dynamics.
    \item Critically assessing IDM’s limitations and related extensions to address stochasticity, driver heterogeneity, and hybrid modeling approaches, thereby overcoming its shortcomings.
    \item Reviewing IDM’s diverse applications, including its contributions to traffic stability analysis, parameter calibration, traffic management, and interdisciplinary domains such as vehicular communications and infrastructure modeling.
    \item Proposing future research directions to enhance IDM’s adaptability and relevance in modern traffic systems, focusing on the integration of big data, hybrid modeling methodologies, digital twin technologies, and human-machine interaction.
\end{itemize}

This paper is organized as follows: Section~\ref{sec:fundamentals} introduces IDM’s theoretical foundations and key properties. Section~\ref{sec:limitations} discusses IDM’s limitations and extensions. Section~\ref{sec:applications} reviews IDM’s practical applications, and Section~\ref{sec:future} outlines future research directions. Finally, Section~\ref{sec:conclusion} concludes with reflections on IDM’s legacy and future potential.

\begin{table}[htbp]
\centering

\caption{Citations of famous classic models included in the Web of Science Core as of May 2025. \cite{wiedemann1992microscopic} is not included in the Web of Science, so the citation is collected in google scholar database.  }
\scriptsize

\label{tab:CFModel}
\resizebox{\linewidth}{!}{%
\begin{tblr}{
  width = \linewidth,
  colspec = {Q[113]Q[48]Q[225]Q[319]Q[140]Q[90]},
  cells = {c},
  cell{2}{1} = {r=10}{},
  cell{12}{1} = {r=4}{},
  hline{1,16} = {-}{0.08em},
  hline{2,12} = {-}{},
}
Classic Models & Citations & Classic Models & Title & Publication Source & Citations  Per Year(CPY)\\
Car Following Models & 3458 & {IDM \\\citep{treiberCongestedTrafficStatesEmpirical2000}} & Congested Traffic States in Empirical Observations and Microscopic Simulations & Physical Review E & 138.32\\
 & 2628 & {Optimal Velocity Model \\\citep{bandoDynamicalModelTrafficCongestion1995}} & Dynamical model of traffic congestion and numerical simulation & Physical Review E & 87.60\\
 & 1773 & {Gipps Model \\\citep{gippsBehaviouralCarfollowingModelComputer1981}} & A behavioural CF model for computer simulation & Transportation Research Part B & 40.30\\
 & 1318 & {Full Velocity Difference Model \\\citep{bandoDynamicalModelTrafficCongestion1995}} & Full velocity difference model for a car following theory & Physical Review E & 54.92\\
 & 985 & {GHR Model \\\citep{gazisNonlinearFollowLeaderModelsTraffic1961}} & Nonlinear follow-the-leader models of traffic flow & Operations Research & 15.39\\
 & 933 & {General Motor Model \\\citep{chandlerTrafficDynamicsStudiesCar1958}} & Traffic dynamics: studies in car following & Operations Research & 13.93\\
 & 881 & Pipes Model\citep{pipesOperationalAnalysisTrafficDynamics1953} & An operational analysis of traffic dynamics & Journal of applied physics & 12.24\\
 & 827 & Newell Nonlinear Model\citep{newellNonlinearEffectsDynamicsCar1961a} & Nonlinear effects in the dynamics of car following & Operations Research & 12.92\\
 & 641 & {Newell Simplified Model \\\citep{newellSimplifiedCarfollowingTheoryLower2002}} & A simplified car following theory: a lower order model & Transportation Research Part B & 27.87\\
 & 378 & {Wiedemann Model\\\citep{wiedemann1992microscopic}} & Microscopic traffic simulation: the simulation system MISSION & Project ICARUS (V1052) Final Report & 11.45\\
More General Models of Transportation Science & 3423 & {NaSch\\\citep{nagelCellularAutomatonModelFreeway1992a}} & A cellular automaton model for freeway traffic & Journal De Physique & 103.76\\
 & 3319 & {LWR \\\citep{lighthillKinematicWavesIITheory1955}} & On kinematic waves  II. a theory of traffic flow on long crowded roads & Proceedings of the Royal Society of London & 47.41\\
 & 2276 & {Cell Transmission Model\\\citep{daganzoCellTransmissionModelDynamic1994}} & The cell  transmission model-a dynamic representation of highway traffic consistent with the hydrodynamic theory & Transportation Research Part B & 73.42\\
 & 1591 & {Vickrey's bottleneck model\\\citep{vickreyCongestionTheoryTransportInvestment1969}} & Congestion theory and transport investment & American Economic Review & 28.41
\end{tblr}
}
\end{table}

\begin{table}[htbp]
\centering
\caption{Abbreviations and full names of concepts in IDM-related research.}
\scriptsize
\label{tab:abb}
\begin{tblr}{
  hline{1,15} = {-}{0.08em},
  hline{2} = {-}{},
    row{odd} = {gray9}   % 偶数行背景灰色，便于区分
}
Abbreviations & Full names\\
ACC & Adaptive Cruise Control\\
CACC & Cooperative Adaptive Cruise Control\\
HV & Human-driven Vehicle\\
AV & Autonomous Vehicle\\
CV & Connected Vehicle\\
CAV & Connected and Autonomous Vehicle\\
RL  & Reinforcement Learning\\
V2X & Vehicle to Everything\\
FF & Free Flow\\
CF & Car Following\\
MCMC & Markov Chain Monte Carlo\\
VANET & Vehicular Ad-hoc Network\\
VSL & Variable Speed Limits
\end{tblr}
\end{table}

\section{Fundamental Introduction and Analysis of the IDM}
\label{sec:fundamentals}

This section aims to provide a comprehensive overview of the IDM’s foundational features, focusing on its mathematical structure, behavioral assumptions, and its role as a key model for understanding individual vehicle dynamics and emergent traffic phenomena. We begin by conducting a formal analysis of IDM’s mathematical framework to clarify its core principles and assumptions. Following this, we investigate how IDM captures traffic fluctuations and vehicle interactions, leading to instabilities and emergent flow patterns. Additionally, we analyze spatiotemporal traffic phases, emphasizing IDM’s ability to model complex traffic phenomena across different temporal and spatial scales. For clarity, Table \ref{tab:abb} summarizes the abbreviations frequently used in IDM-related research, along with their full names.

\subsection{Model Structure and Philosophies}

\subsubsection{Model Structure}

The formulation of basic IDM is as follows:

\begin{equation}
a_n(t) = a \left[ 1 - \left( \frac{v_n(t)}{v_0} \right)^{\delta }- \left( \frac{s^{*}_n(t)}{s_n(t)} \right)^2 \right]
\label{equ:IDM1}
\end{equation}

where $ a_n(t) $ represents the vehicle's acceleration, $ a $ is the maximum acceleration, $ v_n(t) $ is the current velocity, $ v_0 $ is the desired velocity, $ s_n(t) $ is the actual spacing, and $ s^{*}_n(t) $ is the desired dynamic spacing. The exponent $\delta$ determines how quickly acceleration decreases as a vehicle approaches its desired speed. The term \( \left(\frac{s^*}{s}\right)^2 \) is used to describe the nonlinear sensitivity of drivers to changes in vehicle spacing, particularly as the actual spacing \( s \) approaches the desired spacing \( s^* \). The quadratic form of this term is closely aligned with Stevens' power law, which states that the perceived intensity of a stimulus follows a power-law relationship with the actual stimulus. Specifically, the exponent \( n = 2 \) in IDM reflects a balance between capturing drivers’ heightened sensitivity to small spacings and maintaining numerical stability in the model. While a linear form (\( n = 1 \)) would underestimate drivers' rapid responses in critical situations, and a cubic form (\( n = 3 \)) would lead to excessive sensitivity, the quadratic term provides a reasonable approximation of how drivers respond nonlinearly to perceived distance changes, as suggested by Stevens' law for psychophysical scaling \citep{parkDriversCharacteristicsPerceptionLead2001}.

The desired spacing $ s^*_n(t) $ is defined as:

\begin{equation}
s^*_n(t) = s_0 + \max \left( T v_n(t) + \frac{v_n(t) \Delta v_n(t)}{2\sqrt{ab}}, 0 \right)
\label{equ:IDM2}
\end{equation}

where $ s_0 $ is the minimum spacing, $ T $ is the desired time headway, $ \Delta v_n(t) = v_n(t) - v_{n-1}(t) $ is the velocity difference to the preceding vehicle $n-1$, and $ b $ is the comfortable deceleration. The model captures critical driving behaviors by decreasing acceleration as the vehicle approaches its desired speed and ensuring safety through dynamic spacing.

The third term in the desired gap formula, $\frac{v \Delta v}{2\sqrt{ab}}$, reflects the principle of \textit{kinematic deceleration}. This term is derived from the classical braking distance formula:

\begin{equation}
d = \frac{v^2}{2a},
\end{equation}

which quantifies the distance required to stop under constant deceleration. In the IDM, this component ensures that the vehicle maintains a sufficient safety margin when approaching a slower lead vehicle. By incorporating the relative speed $\Delta v$ between vehicles, the model enables smooth and anticipative deceleration, preventing abrupt braking and reducing the risk of collisions. This "intelligent braking strategy" is a core feature of the IDM and one of the key reasons for its ease of use and popularity.

The safety property of the IDM arises directly from the structure of its desired gap function. The model ensures that a following vehicle maintains the desired time headway of $T$, while dynamically reacting to closing speeds through the interaction term. When the actual gap $s$ falls below the desired gap $s^*$, the braking term in the acceleration equation becomes dominant, triggering rapid deceleration to prevent potential collisions. This mechanism guarantees collision-free operation as long as the physical braking limits are not exceeded (e.g., under extreme driver behavior or external disturbances). Moreover, $s_0 +T v_n(t)$ is the classical linear model that gives rise to the triangular fundamental diagram.

The collision-free nature of the IDM has been analytically validated and confirmed through simulation studies (e.g., \cite{treiberTrafficFlowDynamics2013}). This robust safety property makes IDM particularly well-suited for applications such as Advanced Driver Assistance Systems, where proactive collision avoidance is critical. On the other hand, it makes IDM (or any safe CF model) not suitable for evaluating driving safety.

Beyond its safety-centric design, another key advantage of the IDM lies in its explicit representation of dynamic time headway variations. In the desired spacing formula (\ref{equ:IDM2}), the term $T v_n(t)$ represents the nominal space, while the additional term $\frac{v_n(t) \Delta v_n(t)}{2\sqrt{ab}}$ dynamically adjusts the effective spacing based on the closing speed $\Delta v_n(t)$. The desired time headway, defined as $s^*_n(t)/v_n(t)$, therefore adapts to both velocity and relative speed variations. This explicit coupling between spacing and velocity not only provides a clear physical interpretation but also enhances the model's ability to replicate real-world driving behavior under diverse traffic conditions. Such a feature is particularly important for capturing the dynamics of congested or mixed traffic flows.}

Moreover, as outlined in Equations (\ref{equ:IDM1}) and (\ref{equ:IDM2}), the IDM is a nonlinear CF model characterized by six parameters: $a$, $b$, $v_0$, $\delta$, $s_0$, and $T$. According to \cite{treiberMicroscopicCalibrationValidationCarFollowing2013}, these parameters are intuitive, each is related to a specific driving regime: the desired speed $v_0$ is relevant for cruising in free-flow scenarios, the desired time gap $T$ pertains to steady-state CF, the minimum gap $s_0$ is associated with creeping and standing traffic, and the maximum acceleration $a$ and desired deceleration $b$ relate to non-steady traffic flow. The exponent $\delta$ governs this reduction: a larger $\delta$ delays the reduction in acceleration as the vehicle approaches the desired velocity. In the limit $\delta \to \infty$, the acceleration curve aligns with that of the Gipps model, while $\delta=1$ replicates the overly smooth acceleration behavior characteristic of the optimal velocity model. This design not only gives IDM strong physical meaning but also provides analytical value.

\subsubsection{Model Philosophies}

In addition to the parameters with obvious physical meanings, IDM also effectively captures the acceleration behavior of drivers by balancing two key components: a non-interaction term and an interaction term. The model defines the desired speed as the maximum attainable speed in free-flow conditions (FF), where drivers can accelerate smoothly to maintain this speed without hindrance. However, in CF scenarios, the driver's ability to reach the desired speed is constrained by the leading vehicle, although the goal to achieve this speed remains unchanged. The IDM acceleration formula incorporates a non-interaction term, reflecting the driver's acceleration in the absence of external influences, and an interaction term, which accounts for the deceleration required due to the presence of a leading vehicle. In FF conditions, acceleration is primarily determined by the non-interaction term, while in CF conditions, both terms significantly influence the dynamics. The relative weight of these terms in IDM is dependent on the driver's current speed, the speed of the leading vehicle, and the distance between them. These factors collectively ensure that the IDM provides a nuanced and realistic depiction of vehicle dynamics across diverse traffic scenarios.

Hence, IDM is widely used not only because it has a clear physical meaning, but also because it was developed with some basic assumptions built on the modeling philosophy \citep{treiberTrafficFlowDynamics2013}:

\begin{enumerate}
    \item \textbf{Acceleration Conditions}: The acceleration must satisfy specific general conditions to ensure a complete model.

    \item \textbf{Equilibrium Distance}: The minimum safe following distance to the leading vehicle should be no less than a designated "safe distance," calculated as $s_0 + vT$, where $s_0$ is the minimum gap and $T$ is the time gap to the leading vehicle.

    \item \textbf{Braking Strategy}: An intelligent braking approach governs interactions with slower vehicles or obstacles:
    \begin{itemize}
        \item Under normal conditions, braking is gradual, smoothly reducing deceleration to zero as the vehicle approaches a steady state or stops.
        \item In critical situations, the deceleration exceeds comfortable levels until safety is restored, after which it resumes normal deceleration.
    \end{itemize}

    \item \textbf{Smooth Mode Transitions}: Transitions between driving modes, such as from acceleration to CF, are smooth. The model ensures that the jerk, or the derivative of acceleration, remains finite, indicating continuous differentiability across variables.

    {\item \textbf{Model Parsimony}: The model should be as simple as possible, aiding in effective calibration.}
\end{enumerate}

{\subsubsection{Connections with Classic Models}

The modeling philosophy of IDM is not an isolated concept but builds upon a rich history of CF theories. Many of its principles, including safety constraints, equilibrium distances, and smooth transitions between driving states, are inspired by or extend ideas from earlier CF models. To fully understand its motivations and distinctive characteristics, it is essential to examine its relationship to foundational models such as the GM model \citep{chandlerTrafficDynamicsStudiesCar1958}, the OVM \citep{bandoDynamicalModelTrafficCongestion1995} and the Gipps model \citep{gippsBehaviouralCarfollowingModelComputer1981}. IDM synthesizes and extends their strengths while addressing their inherent limitations, resulting in a model that is both comprehensive and robust.

The GM model, one of the earliest CF frameworks, highlights the importance of relative speed (velocity difference) as a key factor influencing driver behavior. Specifically, it assumes that the acceleration of a following vehicle is proportional to the velocity difference and inversely proportional to the headway between vehicles. This foundational approach captures drivers' tendency to react to both distance and speed differences but oversimplifies real-world behavior by assuming a linear relationship between these variables. Moreover, the GM model does not account for explicit safety constraints, such as minimum headways or emergency braking mechanisms. IDM addresses these shortcomings by introducing a nonlinear interaction term that captures the effects of both relative speed and headway more realistically. Additionally, IDM incorporates explicit safety measures, ensuring collision-free behavior and smooth transitions between free-flow and CF conditions.

The OVM emphasizes the importance of "optimal spacing" in CF behavior. It assumes that drivers adjust their acceleration to minimize the difference between their current speed and a desired speed, which depends solely on the headway to the leading vehicle. This simple and intuitive framework provides insights into equilibrium spacing and CF dynamics. However, the OVM assumes instantaneous driver reactions, neglecting the finite reaction times observed in real-world driving. It also lacks explicit safety constraints, such as minimum headways or braking strategies, making it insufficient for modeling close-following scenarios or sudden deceleration events. IDM improves upon OVM by incorporating safety constraints and modeling the dynamic interactions between relative speed and headway, thereby addressing these deficiencies.

The Gipps model introduces the concept of "safe speed," explicitly ensuring collision-free behavior by predicting the position of the leading vehicle after a reaction time interval and limiting acceleration and deceleration accordingly. This focus on safety is a significant advancement compared to earlier models. However, the Gipps model relies on piecewise-defined functions, which are not smooth or continuous. This lack of smoothness makes it unsuitable for analyzing higher-order dynamics, such as jerk (the rate of change of acceleration), and hinders its ability to simulate smooth transitions between driving states. IDM inherits the focus on safety constraints from the Gipps model but improves upon it with a continuous and differentiable formulation. This smoothness not only ensures finite jerk but also enables realistic transitions between driving modes, such as accelerating to a desired speed or braking to avoid collisions.

The FVDM, which extends the OVM, incorporates the influence of both the velocity difference and the optimal spacing, making it a hybrid model that combines features of the GM and OVM frameworks. While FVDM provides a more realistic description of CF behavior by including velocity differences, its structural design does not inherently ensure accident-free dynamics. This limitation arises because FVDM lacks explicit safety constraints like those in IDM. On the other hand, the simpler structure of FVDM, inherited from the OVM, makes it more amenable to theoretical analysis. IDM, while more complex, ensures accident-free behavior by construction, as it incorporates both safety constraints and a nonlinear interaction term. However, the added complexity of IDM's structure makes it more challenging to analyze theoretically compared to FVDM.

In summary, IDM bridges the conceptual gaps between these classical models while addressing their specific limitations. From the GM model, IDM inherits the emphasis on velocity differences but refines it with nonlinear dynamics. From the OVM and FVDM, IDM adopts the concept of desired spacing while ensuring safety through explicit constraints. From the Gipps model, it incorporates the idea of safe speeds but improves smoothness and differentiability. This unification of ideas, combined with its ability to model diverse traffic scenarios, explains why IDM has become a cornerstone of CF modeling in the past decades.

\subsection{Basic Properties of IDM}

Before applying the IDM in traffic simulations and related research, it is critical to understand its fundamental properties to ensure reliable and realistic results. This section focuses on two key aspects: numerical properties (including numerical update schemes and parameter calibration) and traffic stability, emphasizing their relevance for practical applications such as Advanced Driver Assistance Systems and large-scale traffic simulations.

\subsubsection{Numerical Schemes of the IDM}

For numerical solution schemes of the IDM, \cite{treiberComparingNumericalIntegrationSchemes2015} compared four explicit integration methods: Euler method, ballistic update, Heun method (trapezoidal rule), and standard fourth-order Runge-Kutta (RK4). The study found that the ballistic scheme is the best for all practical purposes. Particularly, it is better than Runge-Kutta 4 (RK4) and Euler. The higher consistency order 4 of RK4 is of no use when lane changes or other discrete events are included reducing the overall consistency order to 1. While Euler has the same consistency order 1 as the ballistic scheme, the prefactor of the numerical error is much smaller. Moreover, the ballistic scheme with larger update time gaps could be considered as a special case of the human factor "limited attention span/action points" with an action (i.e., a changed acceleration) only occurring at each timestep. Most researchers and simulators, including SUMO, use this scheme.}

From the parameter calibration perspective, \cite{treiberMicroscopicCalibrationValidationCarFollowing2013} explored methodologies for calibrating CF models based on IDM. They emphasized criteria for assessing model quality in addition to RMSE, including completeness, robustness, parameter orthogonality, and intuitive parameters aligned with plausible values. The study highlighted that global calibration based on gaps is more reliable than local calibration or those based on speeds or accelerations. Moreover, they found that a sampling rate of 1 Hz suffices, with data completeness and a minimum interval of 300 seconds being crucial for accurate calibration, while data smoothing had minimal impact. \cite{sharmaMoreAlwaysBetterImpact2019b} found that there is no one-to-one correspondence between IDM parameters and driving regimes. Specifically, the acceleration behavior of the IDM driver in a specific state is controlled by the interaction of multiple IDM parameters.This highlights the importance of considering parameter interactions and dataset characteristics when interpreting IDM calibration results. \cite{punzo_we_2015} conducted a comprehensive analysis of IDM parameters, identifying the \textit{minimum time headway} $T$ as the most influential parameter in determining model performance and stability. Sensitivity analysis revealed that $T$ explains the majority of performance variability, followed by the \textit{acceleration exponent} $\alpha$ and \textit{maximum acceleration} $a$. To balance model simplicity and performance, they proposed a simplified IDM configuration by fixing less influential parameters, achieving faster convergence while maintaining comparable accuracy. This aligns with other findings emphasizing the importance of prioritizing key parameters during calibration for robust model performance.

{\subsubsection{Traffic Stability}

From the perspective of traffic flow mechanisms, the IDM has been extensively investigated for its ability to capture traffic instabilities, such as stop-and-go waves, while ensuring safety by avoiding vehicle collisions through its design. These instabilities arise from delays in drivers adjusting their speed to actual conditions, leading to oscillations where vehicles repeatedly decelerate and accelerate rather than maintaining a steady pace. Such oscillations propagate upstream, amplifying congestion, increasing travel times, fuel consumption, and emissions, with significant implications for traffic efficiency and safety.

The ability of the IDM to reproduce these instability mechanisms is rooted in its \textit{string stability} properties, which depend on parameters such as the sensitivity of the equilibrium speed to gap changes $v'_e(s)$, the driver's maximum acceleration $a$, the comfortable deceleration $b$, and the desired time headway $T$. Here $v_e(s)$ represents the equilibrium speed that a driver aims to maintain based on the current vehicle gap $s$ under steady-state traffic conditions.  Higher sensitivity $v'_e(s)$ increases instability, while larger values of $a$ and $b$ improve stability by enabling smoother reactions. The desired time headway $T$ is particularly critical, as shorter $T$ values lead to higher instability due to reduced reaction time.

For low speeds $v_e \to 0$, the stability condition simplifies to:
\begin{equation}
    a \geq \frac{s_0}{T^2},
\end{equation}
where $s_0$ is the minimum gap. This condition indicates that at low speeds, the driver's maximum acceleration $a$ must exceed a threshold determined by the minimum gap $s_0$ and the desired time headway $T$ to maintain traffic flow stability.

Furthermore, the factors influencing string stability can be described as follows:
\begin{itemize}
    \item \textbf{Sensitivity to gap changes $v'_e(s)$}: Higher sensitivity increases the tendency for string instability, as vehicles overreact to changes in the gap.
    \item \textbf{Maximum acceleration $a$ and comfortable deceleration $b$}: Larger values enhance stability by allowing drivers to react more smoothly to disturbances.
    \item \textbf{Desired time headway $T$}: Shorter time headways lead to increased instability because of reduced reaction times.
\end{itemize}

The interplay of these parameters determines the overall stability of traffic flow modeled by IDM. For more insights and mathematical formulations, readers are referred to \cite{treiberTrafficFlowDynamics2013}.

Over the past 25 years, IDM has found wide application in traffic simulations and research precisely because of its ability to model traffic instabilities and stop-and-go waves. Contrary to what is often stated, the stability of the IDM is not a problem for practical applications. In fact, IDM is often parameterized to intentionally reproduce traffic flow instabilities for studying phenomena like congestion and wave propagation. However, if a more stable IDM is required, the acceleration parameter $a$ is the main lever to control stability. The higher the value of $a$, the more stable the IDM (and its derivatives) becomes. For example, realistic values such as $a > 1.5  \text{m/s}^2$ yield a stable IDM under nearly all other plausible parameter combinations. While such configurations will not produce traffic waves, they demonstrate the flexibility of the IDM to adapt to different stability requirements in practice. This adaptability has solidified the IDM's position as one of the most widely used CF models in traffic simulations.

Despite its practical utility, the IDM does exhibit limitations when compared to empirical observations. For instance, while IDM can reproduce certain instability phenomena, such as upstream-propagating convective instabilities and transitions between convective and absolute instabilities, its predictive accuracy for traffic fluctuations is limited due to inherent oversimplifications. In \cite{jiangTrafficExperimentRevealsNature2014}, experimental evidence shows that traffic oscillations exhibit a concave growth pattern, where the standard deviation of vehicle acceleration increases concavely as the disturbance propagates upstream in a platoon (see Fig. \ref{fig:IDMGrowth}). This finding has been further validated using various datasets, including NGSIM data and controlled experiments, such as the 25- and 51-vehicle platoon experiments in Hefei (\cite{jiangExperimentalFeaturesCarfollowingBehavior2015}; \cite{tianEmpiricalAnalysisSimulationConcave2016}; \cite{tianCellularAutomatonModelSimulating2016}). Together, these studies confirm that the concave growth of traffic oscillations is a universal phenomenon observed under different traffic conditions and experimental settings. In contrast, traditional CF models, such as the IDM, inherently predict a convex growth pattern of oscillations due to the linear instability of their steady-state solutions \citep{wangStabilityAnalysisStochasticLinear2020}. This convex growth highlights a critical limitation of IDM, as it fails to capture the empirically observed concave growth, which is essential for accurately modeling traffic flow dynamics.

Moreover, the IDM assumes a deterministic framework that overlooks the stochastic and heterogeneous nature of driver behavior, which plays a significant role in traffic stability. Further insights from \cite{tianRoleSpeedAdaptationSpacing2019} suggest that the fundamental mechanism driving traffic fluctuations lies in the competition between speed adaptation and stochastic effects, rather than being solely attributed to reaction delay as assumed in the IDM. Their analysis of CF behavior in a 25-car platoon experiment shows that reaction delay has a negligible effect on fluctuation growth, while the interplay of speed adaptation and stochastic disturbance dominates the dynamics. This evidence challenges IDM’s core assumptions and indicates that its deterministic nature limits its ability to accurately model real-world instability.

\begin{figure}
        \centering
        \includegraphics[width=0.5\linewidth]{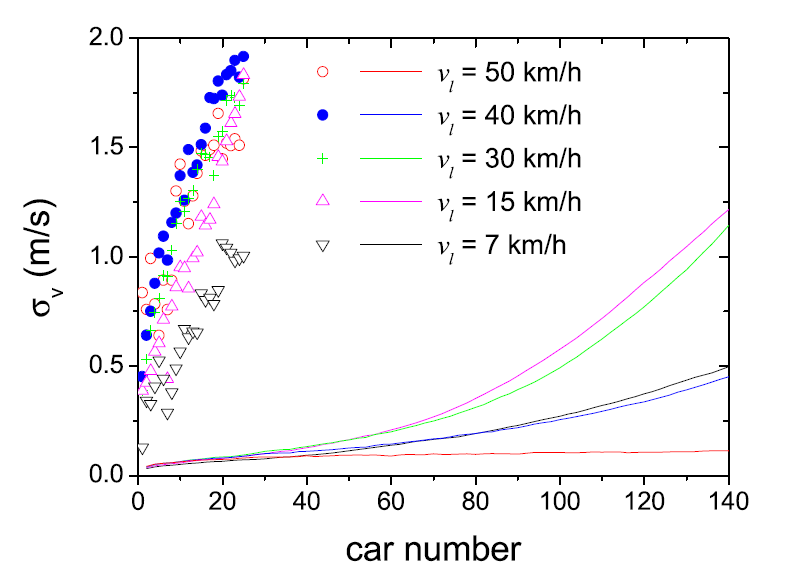}
        \caption{Simulation results (solid lines) of the standard deviation $\sigma_v$ of the time series of the velocity of each car of IDM from \cite{jiangTrafficExperimentRevealsNature2014}. The curves represent the simulation results while the markers such as circles represent the experimental results. The different colors represent different experiment runs in which the velocity of leading vehicle $v_l$ is 7,15,30,40 and 50 km/h.}
        \label{fig:IDMGrowth}
    \end{figure}

\subsubsection{Spatiotemporal Patterns Division}

Classic traffic theory forms the foundation by classifying traffic into free flow and congested flow, where congested flow encompasses a variety of spatiotemporal traffic patterns. Free flow is characterized by high vehicle speeds and low densities, with minimal interaction between vehicles. Congested flow, on the other hand, includes a range of traffic states such as stop-and-go waves, wide-moving jams, and localized clusters. IDM, through its nonlinear acceleration framework, can reproduce several distinct traffic patterns, including pinned localized clusters (PLC), moving localized clusters (MLC), triggered stop-and-go waves (TSG), oscillatory congested traffic (OCT), and homogeneous congested traffic (HCT) \citep{treiberThreephaseTrafficTheoryTwophase2010}. These patterns arise continuously, with smooth transitions between states, as shown in Fig.~\ref{fig:Twophase}. This flexibility and ability to model traffic instabilities have established IDM as a benchmark model for simulating traffic dynamics within the two-phase framework.

\begin{figure}
    \centering
    \includegraphics[width=0.75\linewidth]{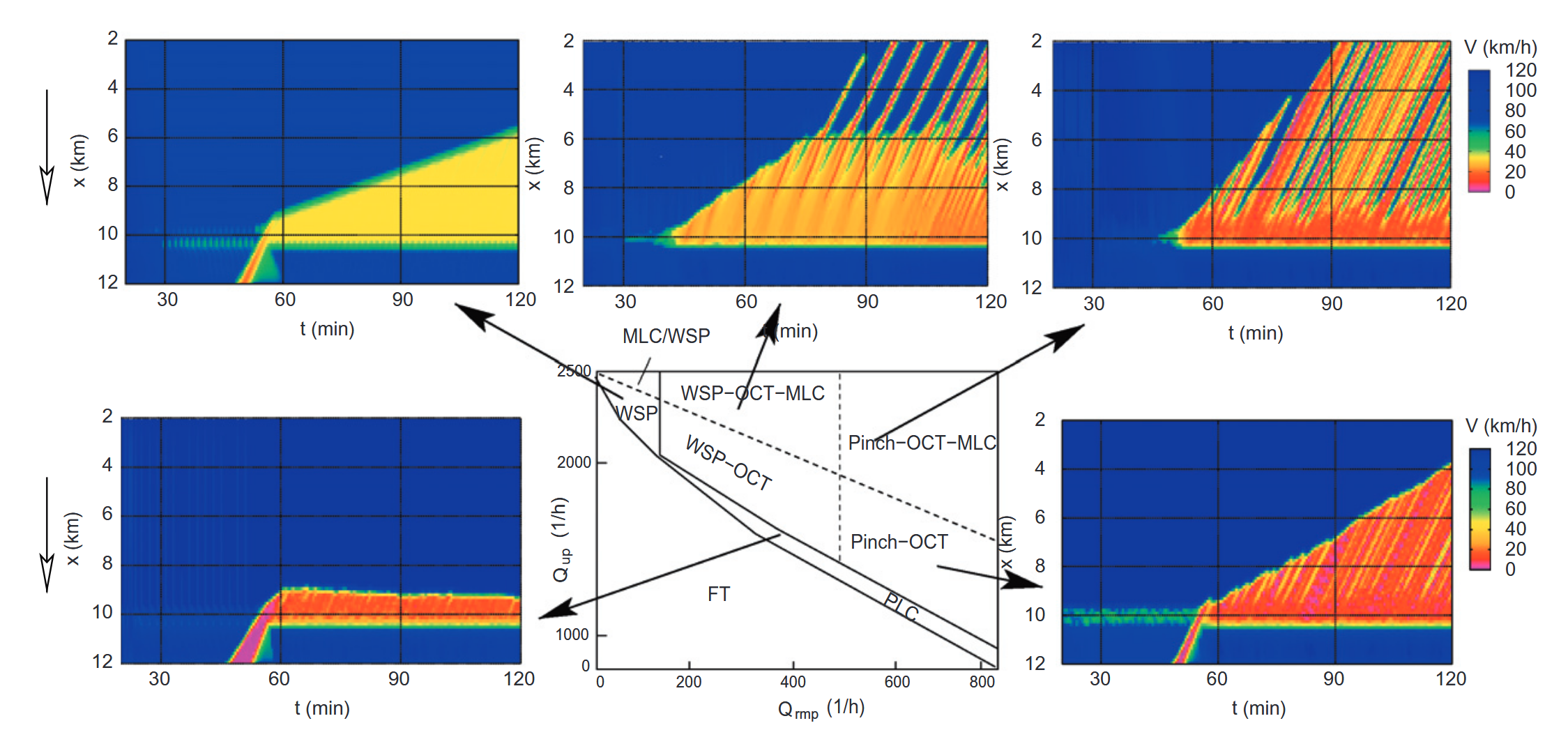}
    \caption{Dynamic phase diagram of on-ramp-induced congested traffic patterns simulated by IDM from \cite{treiberThreephaseTrafficTheoryTwophase2010}. The abbreviations denote free traffic (FT), pinned localized cluster (PLC), moving localized cluster (MLC), homogeneous congested traffic (HCT), oscillatory congested traffic (OCT), and triggered stop-and-go (TSG) pattern.}
    \label{fig:Twophase}
\end{figure}

}

{ \section{Critical Discussion of the IDM and Its Extensions}
\label{sec:limitations}
The bibliometric analysis in the previous section provides a comprehensive overview of IDM’s academic impact and identifies key studies that have shaped its development. By synthesizing insights from these core contributions, several critical limitations of IDM have been highlighted. These include its inability to fully capture stochastic variability, driver heterogeneity, and real-time adaptability, which are essential for modeling complex traffic behaviors in real-world scenarios. Addressing these limitations has become a focal point of research, driving the development of diverse extensions to enhance IDM’s functionality and applicability.

Broadly, these extensions can be categorized into three main directions. First, \textbf{physics-based extensions} focus on refining IDM by incorporating additional behavioral or physical principles to better capture human-vehicle interactions. These efforts include both \textit{deterministic approaches}, which improve the model’s core mechanisms, and \textit{stochastic approaches}, which introduce randomness to represent driver variability. Second, \textbf{physics-data-driven hybrid extensions} integrate IDM’s interpretability with the flexibility and adaptability of machine learning and statistical techniques, enabling the model to perform effectively in data-rich and complex traffic environments. Lastly, \textbf{comparative studies with other CF models} evaluate IDM’s strengths and weaknesses, offering valuable insights into its positioning within the broader landscape of CF research.

The following sections systematically review these extensions. We begin with physics-based models, discussing both deterministic and stochastic approaches, followed by hybrid methodologies that combine physical and data-driven techniques. Finally, we examine comparative studies, highlighting their role in contextualizing IDM and its extensions in the broader CF research domain.

\begin{table}
\centering
\caption{Summary of IDM Extensions, Features, and Use Cases}
\scriptsize

\label{tab:IDM_extensions}
\begin{tblr}{
  width = \linewidth,
  colspec = {Q[150]Q[160]Q[350]},
    cells = {m},
  hline{1,10} = {-}{0.08em},
hline{2} = {-}{},
  row{odd} = {gray9},   % 偶数行背景灰色，便于区分
}
\textbf{Model} & \textbf{Improvements / Features} & \textbf{Use Cases} \\ 
IDM & - & Basis for realistic models; calibration reference; testing V2V/V2I communication; benchmark for AI models; physics-informed ML; traffic simulation \\
IDM+ / IIDM & Steady-state time-gap parameter $T$ & Same use cases as IDM with a triangular fundamental diagram \\
IDM-ACC & Reduces sensitivity to cut-ins/lane changes; includes jerk control & Adaptive cruise control; longitudinal control for autonomous vehicles (validated on test tracks/public roads) \\
IDMM (IDM with Memory) & Considers adaptation to traffic state & Models observed traffic flow dynamics \\
Stochastic IDM Variants (e.g., HDM, 2D-IDM, Action-Point IDM, etc.) & Stochasticity, action points, noise, random parameters & Reproduce platoon experiments (e.g., concave growth) \\
Dynamic Time-Gap IDM Variants (e.g., 2D-IDM, Variance-Driven IDM) & Risk allostasis, task difficulty homeostasis, indifference region & Describe naturalistic data and platooning experiments \\
Reaction Time IDM Variants (e.g., HDM, Multi-Anticipative IDM) & Reaction time, estimation errors, multi-anticipation & Realistic modeling of human drivers \\
IDM-Based Lane-Changing Model (MOBIL) & Lane-changing behavior while preserving driver traits & Scenarios with lane changes \\
\end{tblr}
\end{table}

\subsection{Limitations of IDM}

Despite its simplicity and computational efficiency, IDM exhibits several critical limitations that hinder its ability to fully capture the complexity of real-world traffic systems. These limitations primarily stem from its deterministic nature, oversimplified behavioral assumptions, and lack of flexibility in addressing diverse traffic scenarios. Key issues include:

\begin{itemize}
    \item \textbf{Omission of Human Factors:} IDM does not explicitly incorporate essential human driving characteristics, such as reaction time, estimation errors, and multi-vehicle anticipation. This omission weakens its ability to reproduce realistic driver behavior and traffic stability, particularly under stop-and-go wave conditions.

    \item \textbf{Deterministic Nature:} IDM's deterministic framework fails to reflect the inherent stochastic variability and heterogeneity in driver behavior, limiting its applicability in scenarios characterized by high uncertainty and variability.

    \item \textbf{Inaccurate Oscillation Dynamics:} The model predicts convex growth patterns in traffic oscillations, which contradict empirical evidence showing concave growth. This discrepancy reduces its effectiveness in modeling traffic flow instabilities.

    \item \textbf{Steady-State Gap and Deceleration Issues:} IDM often produces unrealistic equilibrium gaps and exaggerated deceleration behaviors, particularly at high speeds or during lane changes. This leads to inefficiencies in traffic flow and unrealistic braking patterns.

    \item \textbf{Over-Simplified Assumptions in Complex Scenarios:} IDM's simplified structure limits its ability to model complex scenarios such as merging, lane-changing, and obstacle avoidance, resulting in behaviors that deviate from real-world observations.

    \item \textbf{Limited Adaptability to Modern Traffic Systems:} With its lack of mechanisms for cooperative driving or information-sharing, IDM is less effective in modeling connected and autonomous vehicle (CAV) interactions, as well as mixed traffic environments.
\end{itemize}

These shortcomings underscore the need for advancing IDM through methodical refinements and extensions. Various extended models have been proposed to address these issues by incorporating features such as stochastic variability, human behavioral factors, and enhanced adaptability for modern intelligent transportation systems. Table~\ref{tab:IDM_extensions} provides an overview of these IDM extensions, summarizing their addressed limitations and application scenarios.

To further illustrate the advancements in IDM, the following sections will delve into its extensions from two key perspectives: deterministic and stochastic models. Deterministic extensions primarily focus on enhancing the baseline IDM by addressing specific weaknesses, such as reaction time and equilibrium gap issues, while maintaining its computational efficiency. Stochastic extensions, on the other hand, incorporate randomness and human behavioral variability, enabling more realistic modeling in complex and uncertain traffic environments. Together, these extensions represent significant strides in improving IDM's robustness and applicability across a wide range of traffic scenarios.

\subsection{Physics-Based Extensions of IDM}

The development of IDM extensions for modeling driver behavior has seen significant advancements, which can be broadly categorized into the \textbf{deterministic IDM family} and \textbf{stochastic IDM family}. Given the rising focus on automated driving, a dedicated subsection is introduced to specifically address \textbf{IDM extensions tailored for automated driving} . Moreover, we also highlight their role in overcoming IDM's limitations and enhancing applicability in this rapidly growing field.

\subsubsection{Deterministic IDM Extensions}

Deterministic extensions of IDM focus on addressing specific limitations, such as reaction time, equilibrium gaps, and oscillation dynamics, through modifications grounded in physical principles. As summarized in Table~\ref{tab:IDM_extensions}, these extensions include models like IDM+, which resolves the misalignment of the time-gap parameter with steady-state conditions, and IDMM (IDM with memory), which incorporates driver adaptation to traffic states.

For example, \cite{treiber_delays_2006} emphasized the importance of incorporating reaction time and human factors. While IDM does not directly account for reaction time, it can be extended to include factors such as reaction delays, estimation errors, and multi-vehicle anticipation. Although these factors can undermine traffic stability, they may be compensated by introducing spatial and temporal expectation mechanisms. This explains why IDM and its deterministic extensions, such as the \textbf{Human Driver Model (HDM)}, effectively reproduce empirical traffic dynamics despite their simplicity. It further stabilizes traffic flow by increasing oscillation wavelength and reducing the gradient of stop-and-go waves, aligning better with empirical observations. Similarly, \cite{kestingAdaptiveCruiseControlDesign2008} extended IDM for simulating adaptive cruise control systems, creating the \textbf{IDM-ACC}, which captures smoother and more stable driving patterns characteristic of ACC-equipped vehicles. Studies such as \cite{li_evaluation_2017,li_evaluating_2017} demonstrated IDM-ACC's ability to evaluate the impact of CACC systems on highway safety and efficiency, showing its robust applicability in automated driving. Despite differences in model components, HDM and IDM-ACC often produce similar patterns due to their shared foundation in IDM's "intelligent braking strategy" (see subsections 2.1.2 and 2.2.1).

Other deterministic extensions further refine IDM to address specific weaknesses. For instance, \textbf{Improved IDM (I-IDM)} proposed by \cite{treiberTrafficFlowDynamics2013} eliminates exaggerated deceleration at speeds exceeding the desired speed $v_0$, ensuring equilibrium gaps align with empirical observations. \textbf{Sigmoid-IDM} \citep{chen_sigmoid-based_2024} incorporates a sigmoid function to better handle traffic oscillations, while \textbf{Task-Difficulty IDM (TD-IDM)} \citep{saifuzzamanIncorporatingHumanFactorsCarFollowingModels2014} accounts for task-difficulty considerations to model human factors in heterogeneous environments.

\subsubsection{Stochastic IDM Extensions}

Stochastic IDM extensions aim to address the deterministic nature of the original IDM by introducing variability to better represent random driver behavior, heterogeneity, and traffic uncertainties. These models enhance IDM's realism and applicability in scenarios involving traffic instabilities, mixed traffic environments, and connected driving. As summarized in Table \ref{tab:IDM_extensions}, stochastic extensions such as 2D-IDM, Action-Point IDM, and other variants effectively capture phenomena like concave oscillation growth and driver heterogeneity.

\cite{jiangTrafficExperimentRevealsNature2014} introduced the two-dimensional IDM (2D-IDM), which revealed the concave growth pattern of traffic oscillations through time-varying headway analysis. Building on this, \cite{xiong_speed_2022} utilized 2D-IDM to establish a dynamic programming model for recommended speed at isolated signalized intersections in connected environments, characterizing random human driving behavior. Further advancements were made by \cite{tianImproved2DIntelligentDriver2016}, who proposed the Improved 2D-IDM (I2D-IDM). This model successfully simulated synchronized flow and traffic oscillations by incorporating defensive driving behavior at high speeds, thus improving realism in high-speed scenarios.

To address traffic oscillations, \cite{treiberIntelligentDriverModelStochasticity2018} introduced IDM with white noise and IDM with action points, showing that even basic stochastic elements can effectively reproduce observed fluctuation patterns. Similarly, \cite{tianRoleSpeedAdaptationSpacing2019} proposed the Stochastic Speed Adaptation Model (SSAM), which utilized the function formulation of desired spacing of IDM (see Equation (2)) and captures critical traffic instability characteristics such as speed adaptation and spacing indifference. By doing so, SSAM successfully reproduces the concave growth patterns observed in empirical data.

Other stochastic extensions have focused on modeling inter-driver and intra-driver variability. For instance, \cite{lindorfer_modeling_2018} presented the Enhanced HDM, which incorporates stochastic Wiener processes to model driving errors, variable reaction times, and driver distractions. \cite{zhangGenerativeCarFollowingModelConditioned2022c} proposed a time-varying IDM that emphasizes both inter-driver and intra-driver heterogeneity, enhancing the model's ability to simulate real-world variability in human driving behavior.

\subsubsection{IDM Extensions for Automated Driving}

In addition to the characterization of driving behavior, IDM has also been extended to various aspects of autonomous driving modeling, such as obstacle avoidance, vehicle merging, ecological driving, network attacks, and information error modeling. \cite{sharath_enhanced_2020} proposed AV-IDM for two-dimensional motion planning, specifically in mixed traffic scenarios involving obstacle avoidance in which the surrounding obstacles like vehicles and road boundaries are treated as stimuli. \cite{chen_safety_2024} developed a model focusing on freeway merging areas, redefining the desired safety distance in IDM to enhance safety during merging under AV environments. \cite{ward_probabilistic_2017} proposed a probabilistic extension of IDM for AVs, incorporating normal distribution noise to improve interaction-aware planning in merge scenarios. \cite{zhou_impact_2017} introduced the Multi-AV cooperative (MA-C) IDM, integrating cooperative components to assess freeway merging under various AV ratios. \cite{zhou_cooperative_2023} introduced the EIDM for cooperative control in mixed traffic, combining multiple vehicle state information to mitigate the negative impacts of HVs. As to eco-driving of AV, \cite{dongFlexibleEcoCruisingStrategyConnected2024a} proposed a Flexible Eco-Cruising Strategy  for CAVs, employing a hierarchical control framework to optimize driving lane planning and speed, thus reducing driving costs in various traffic conditions. When addressing cybersecurity and information errors,\cite{wang_modeling_2020} developed the CIDM to model CAV behavior under cyberattacks, utilizing dynamic communication topologies to analyze the impact of manipulated information.  \cite{berkhahn_traffic_2022} introduced IDMrm, incorporating random misperception (rm) into IDM to model traffic dynamics at intersections with potential perception errors. \cite{li_unraveling_2024}presented the Dynamic Cooperative IDM (DC-IDM), addressing information errors and their impact on CAVs. \cite{liu_connected_2022} proposed a microscopic model for mixed traffic flow unsignalized intersections for autonomous driving, which incorporates the Ornstein–Uhlenbeck process into IDM to model the random perception errors of vehicles at intersections. \cite{luo_modeling_2024} proposed the VC-IDM, integrating a generalized force model to analyze platoon behavior and defense mechanisms under cyberattack scenarios.

\textbf{Summary}: The IDM and its extensions owe much of their success to the foundational "intelligent braking strategy," which ensures collision-free behavior when physically possible \cite{treiberTrafficFlowDynamics2013}. This feature not only enhances IDM's safety and robustness but also serves as the unifying principle across all deterministic and stochastic variants, making IDM a reliable and flexible framework for both research and practical applications. In deterministic extensions, models like IDM-ACC and HDM refine IDM's ability to reproduce empirical traffic dynamics, improve stability, and adapt to automated driving scenarios. Meanwhile, stochastic extensions, such as 2D-IDM and SSAM, introduce variability to better capture real-world phenomena like driver heterogeneity, traffic oscillations, and random behavior, thereby enhancing IDM's realism in diverse traffic conditions.

Despite its strengths, IDM's deterministic nature limits its ability to fully replicate the complexity of human driving behavior, particularly the inherent randomness and adaptability of drivers in dynamic traffic environments. This limitation becomes especially pronounced in scenarios involving traffic instabilities or stop-and-go waves, where stochastic elements play a critical role in capturing variability and unpredictability.

In conclusion, IDM remains a robust and widely applicable CF model, particularly for automated driving systems where safety and deterministic behavior are paramount. However, to further expand its applicability to human-driven vehicles (HVs), future research should prioritize the integration of stochastic elements and context-specific behavioral adaptations, enabling IDM to better address the variability and complexity of real-world traffic dynamics.}

\subsection{Hybrid Physics and Data-Driven Extensions}

Hybrid physical and data-driven extensions aim to overcome the limitations of both classical and data-driven models. Classical models, such as IDM, accurately represent traffic mechanisms but rely on strong assumptions and struggle with data uncertainties. On the other hand, data-driven methods, while flexible and adaptive, depend heavily on high-quality data and often lack interpretability. By combining these approaches, hybrid models offer a robust framework that balances adaptability with clarity. Typically, the main approach to extend IDM in data-driven vehicle-following modeling is to combine IDM with reinforcement learning (RL) methods and deep learning or machine learning techniques.

\subsubsection{Reinforcement Learning-Based Models}

Reinforcement learning (RL)-based extensions of IDM aim to improve decision-making and vehicle control by integrating data-driven adaptability with IDM’s physically interpretable structure. For instance, \cite{brito_learning_2022} extended IDM using deep reinforcement learning to predict and identify leading vehicles, estimate motion plans, compute control commands (e.g., acceleration), and optimize trajectories in dense traffic, thereby enhancing safety and feasibility. Similarly, \cite{caoTrustworthySafetyImprovementAutonomous2022a} developed a decision framework for RL, incorporating IDM as a lower performance bound, where autonomous vehicles (AVs) only adopt RL policies if they outperform IDM. 

Additionally, \cite{baiLongitudinalControlAutomatedVehicles2024a} combined Twin Delayed Deep Deterministic Policy Gradient (TD3)-based RL with IDM to achieve improved longitudinal control, safety, and efficiency. \cite{kerbel_shared_2024} incorporated IDM’s expected acceleration as a state variable in shared learning to enhance fleet-wide fuel economy and adaptability. Furthermore, \cite{hartRobustCarfollowingBasedDeep2024} proposed an RL model with reward functions grounded in IDM’s assumptions, using parameters such as the desired time gap, maximum and comfortable accelerations/decelerations, and desired speed to guide learning.

\subsubsection{Machine and Deep Learning-Based Models}

The integration of IDM with machine learning (ML) and deep learning (DL) methods has enabled advanced modeling of vehicle behavior by blending data-driven insights with the established physical frameworks of IDM. For example, \cite{chen_investigating_2020} developed an LSTD-IDM model that combines long-term and short-term driving characteristics by merging IDM with longitudinal driving data. \cite{moPhysicsInformedDeepLearningParadigm2021c} proposed a physics-informed deep learning framework (PIDL-CF), blending IDM with deep learning to predict driver acceleration under various conditions.

Hybrid models have also extended IDM with probabilistic and transformer-based approaches. \cite{bhattacharyya_hybrid_2022} introduced a hybrid particle-filtering method combined with collaborative IDM extensions. \cite{gengPhysicsInformedTransformerModelVehicle2023b} presented the Transformer-IDM (PIT-IDM) for enhanced vehicle trajectory prediction by integrating IDM with deep learning techniques. Similarly, \cite{kangTrajectoryBasedEmbeddingRandomCoefficients2023b} proposed PGM-IDM, combining trajectory-based embeddings with probabilistic graphical models to improve vehicle-following responses. 

Further advancements include \cite{liuQuantileRegressionPhysicsInformedDeepLearning2023c}, which introduced QRPIDL-CF, fusing quantile regression and deep learning with IDM’s physical insights. \cite{tang_grounded_2024} proposed a grounded relational reasoning approach that incorporates IDM’s expected distance into the reward function of a Markov decision process, while \cite{zhang_bayesian_2024} developed a memory-augmented Bayesian calibration (MA-IDM) to address parameter uncertainties and behavioral differences in IDM, significantly enhancing its predictive accuracy.

{\textbf{Summary}: The development of hybrid IDM extensions represents a promising direction in vehicle-following modeling, successfully combining the interpretability of physics-based models with the adaptability of data-driven methods. These approaches improve IDM’s capability to handle complex traffic scenarios, such as dense traffic, heterogeneous driver behavior, and uncertain data inputs. By enhancing IDM with reinforcement learning, hybrid models can dynamically adapt to changing traffic conditions, while deep learning integration enables the modeling of nuanced driver behaviors.

However, hybrid approaches also face several challenges. First, the reliance on high-quality and diverse data for training remains a significant limitation, as the absence of representative data may lead to poor generalization. Second, while combining physical and data-driven models enhances interpretability compared to purely data-driven methods, hybrid models still risk losing physical consistency due to the black-box nature of machine learning components. Finally, the computational cost of training and deploying hybrid models can be prohibitively high, particularly for real-time applications in autonomous driving. 

Overall, while hybrid IDM extensions provide a powerful toolkit for addressing the limitations of traditional and data-driven models, future work must focus on balancing interpretability, computational efficiency, and data requirements to ensure their practical applicability.}

{\subsection{Comparisons with Other Models}

\subsubsection{Deterministic Model Comparisons}

Many studies have compared their proposed models with IDM, but relatively few have conducted comprehensive and unified evaluations that fairly compare all models, including IDM. One such study is \cite{punzoCalibrationCarFollowingDynamicsAutomated2021}, where deterministic models, including IDM, Gipps' model, and the full velocity difference model (FVDM) with constant-time headway (CTH) and SIGMOID spacing policies, were systematically evaluated across multiple datasets and calibration settings. IDM demonstrated strong and consistent calibration performance, achieving low errors in speed and acceleration while performing well in spacing accuracy. Although Gipps' model slightly outperformed IDM in reproducing spacing for both human-driven and autonomous vehicle datasets, IDM maintained a more balanced performance across all three measures: spacing, speed, and acceleration. The robustness of IDM across diverse datasets and calibration settings, especially when using Pareto-efficient objective functions such as NRMSE(s,v,a), underscores its versatility as a reliable model for CF dynamics.

\subsubsection{Challenges in Comparing with Stochastic Models}

While comparisons between deterministic models are well-established, IDM and stochastic models are often compared qualitatively rather than quantitatively. This is because stochastic models inherently produce non-unique outputs, and their calibration processes are complex, involving additional randomness and parameter variability. These factors make direct comparisons with deterministic models like IDM challenging and often unfair. Additionally, the absence of a unified framework for calibrating stochastic models and comparing them with deterministic models hinders systematic evaluation. Such a framework is crucial for enabling fair and consistent comparisons, helping to elucidate the advantages and limitations of stochastic models relative to deterministic counterparts like IDM.

\textbf{Summary}: While IDM has proven to be a versatile and robust model across various datasets and scenarios, its limitations—such as the lack of stochastic variability, over-simplification of human factors, and challenges in adapting to cooperative driving—necessitate further refinement. Comparisons with deterministic models highlight IDM’s balanced performance across key metrics, but its deterministic nature limits its applicability in more variable scenarios. Comparisons with stochastic models remain a significant challenge due to the absence of a standardized evaluation framework. Future research should prioritize developing such frameworks to enable fair and systematic comparisons, thereby advancing the understanding of both deterministic and stochastic CF models.

\section{Applications of IDM}
\label{sec:applications}
This section provides a structured review of the applications of IDM in traffic research and related domains. The review is organized into five thematic areas: traffic stability analysis, calibration frameworks, traffic bottleneck management and control, section and network-level optimization, and interdisciplinary applications. While certain studies may overlap across these themes, each is assigned to the category that best represents its primary focus to ensure clarity and coherence.

The review begins by exploring IDM's role in traffic stability analysis, focusing on its use in understanding reaction-based instabilities, string stability, and mixed traffic dynamics. Next, the discussion shifts to calibration frameworks, where IDM has been employed to address challenges in parameter estimation, error modeling, and behavioral feature extraction. Applications in bottleneck management, such as intersections and ramp merging, are then highlighted, with IDM widely used to model vehicle interactions under cooperative and signal optimization scenarios. The section further extends to road segment and network-level studies, where IDM is leveraged for evaluating traffic capacity, stability, and emissions under connected and autonomous vehicle (CAV) environments. Finally, the review touches on IDM's interdisciplinary applications, such as its use in vehicular communication systems and civil engineering for infrastructure load modeling, showcasing its versatility beyond traditional traffic modeling domains.}

\subsection{Applications in Traffic Stability Analysis}

A significant focus has been on understanding traffic stability through IDM. \cite{kestingHowReactionTimeUpdate2008} initiated this exploration by analyzing the influence of reaction time and speed adaptation on traffic stability, using IDM to reveal various instability mechanisms. Building on this, \cite{jinEquivalenceContinuumCarFollowingModels2016a} demonstrated the equivalence between continuum and CF models, illustrating that higher-order IDM can be transformed into continuous models with consistent stability conditions. \cite{monteil_linear_2014} advanced the discussion by examining string stability in mixed traffic flows and the impact of multiple anticipations in cooperative CF models. \cite{montaninoHomogeneousHeterogeneousTrafficFlows2021b} further delved into string stability in mixed and heterogeneous traffic flows, incorporating uncertainties in model parameters. They later proposed a unified modeling framework \citep{montaninoStringStabilityMixedHeterogeneous2021b} to assess stability, considering driver and vehicle heterogeneity and using IDM to verify this framework. \cite{bouadiStochasticFactorsStringStability2022b} extended these concepts into the realm of stochastic models, investigating string stability in stochastic CF models using IDM with stochastic variants and the generalized Lyapunov equation to establish stability conditions.

{\textbf{Summary}: IDM has served as a foundational tool for probing traffic stability, facilitating analytical insights from reaction-based instabilities to string stability in heterogeneous and stochastic contexts. Yet, much of this literature remains confined to linearized scenarios, externally imposed uncertainties, and idealized driver behaviors, limiting its explanatory power in real-world complexity. As traffic systems grow increasingly interconnected and adaptive, future research must re-situate IDM within hybrid, data-informed architectures that move beyond string stability toward capturing emergent, nonlinear, and multi-agent dynamics.}

\subsection{Applications in Calibration Framework}

Research on IDM-based calibration can be categorized into several key areas. Firstly, the development of calibration frameworks focuses on effectively handling Model Performance (MoP), Goodness of Fit (GoF), and trajectory completeness. Secondly, model evaluation involves calibrating various CF models to compare their performance across different metrics. Thirdly, algorithm design aims to create efficient optimization algorithms for calibration processes. Lastly, behavior feature extraction uses IDM-calibrated data to classify different driving behaviors and vehicle types. 

For the development of calibration frameworks, building on this focus on error, \cite{jin_reducing_2014} explored error accumulation in CF models calibrated with vehicle trajectory data, deriving a dynamic error model based on acceleration. \cite{longBiScaleCarFollowingModelCalibration2024a} introduced a bi-scale CF model calibration using corridor-level trajectories, highlighting macro measurement error accumulation. To address the validation aspect, \cite{treiberValidationTrafficFlowModels2012} developed a quantitative method for calibrating and validating traffic flow models based on the spatiotemporal evolution of congested traffic patterns. In terms of parameter importance and simplification, \cite{punzo_we_2015} conducted sensitivity analysis on driving behavior models, finding leader trajectory more important than parameters, and proposed a simplified model with faster convergence. For setting robust calibration objectives, \cite{punzoCalibrationCarFollowingDynamicsAutomated2021} proposed robust guidelines for selecting MoP and GoF for calibrating CF dynamics. They used Pareto efficiency and indifference curves for function comparison. Focusing on data integrity, \cite{sharmaMoreAlwaysBetterImpact2019b} investigated the impact of vehicular trajectory completeness on the calibration and validation of CF models. Finally, addressing behavioral pattern recognition, \cite{zhengMultiObjectiveCalibrationFrameworkCapturing2023b} proposed a multi-objective calibration framework to capture behavioral patterns of autonomously-driven vehicles, validating with empirical observations of ACC vehicle dynamics. \cite{zhouCalibrationStochasticCarFollowing2025a} established the calibration framework of stochastic CF models and revealed the essential differences between deterministic and stochastic CF model calibration using various stochastic IDM variants.

As for the development of the optimization algorithms for calibration, \cite{rahman_improving_2015} proposed a stochastic calibration method using Bayesian estimation and MCMC simulation. \cite{liGlobalOptimizationAlgorithmTrajectory2016b} developed a global optimization algorithm combining direct and gradient searches for CF models. \cite{zhongCrossEntropyMethodProbabilisticSensitivity2016b} introduced a calibration method using the Cross-Entropy Method and sensitivity analysis under noisy data. \cite{fardCopulaBasedEstimationDistributionAlgorithm2019b} suggested a copula-based algorithm to explore complex parameter interactions in traffic models. \cite{wangIntegratedSelfConsistentMacroMicroTraffic2024a} created an integrated macro-micro calibration framework using enhanced optimization and deep learning.

In addition to the above theme, calibration is crucial for model evaluation. \cite{zhuModelingCarFollowingBehaviorUrban2018b} investigated the applicability of CF models for Chinese drivers, highlighting IDM's effectiveness and portability on Shanghai highways. \cite{campi_roundabouts_2024} assessed various models in roundabout scenarios, with IDM excelling in capturing CAV behavior. \cite{zhouExperimentalFeaturesEmissionsFuel2023b} used four classical emission and fuel consumption models and found IDM effective at the vehicle pair level for energy consumption and emissions predictions. However, at the platoon level, predicted results were almost constant, contrasting with experimental observations.

Based on calibrated IDM, different behavior patterns and vehicle types can be extracted. \cite{ossenHeterogeneityCarFollowingBehaviorTheory2011b} used data collected by helicopter to calibrate the IDM to identify differences in following behaviors between different driver types, showing significant heterogeneity. \cite{makridis_response_2020} estimated the response time of vehicles equipped with ACC by calibrating IDM parameters and found that it was comparable to that of human drivers. \cite{hu_autonomous_2023} calibrated IDM parameters to distinguish the empirical following behaviors of AVs and HVs in Waymo data. \cite{li_automated_2023} proposed an AV identification algorithm based on IDM calibration results, that is, using error results to distinguish AVs from HVs.

{\textbf{Summary}: IDM has played a pivotal role in the development of calibration frameworks, supporting advances in model performance metrics, optimization algorithms, and behavioral feature extraction. From error modeling to cross-scenario validation, IDM has been widely applied due to its analytical tractability and interpretability. However, most calibration efforts still focus on deterministic model variants, while the calibration of stochastic IDM-family models remains notably underdeveloped due to challenges in parameter identifiability, noise disentanglement, and the lack of robust likelihood functions \citep{punzoCalibrationCarFollowingDynamicsAutomated2021}. This limits their empirical applicability despite their theoretical promise. Moving forward, calibration should be re-conceptualized not merely as parameter fitting but as a behavioral inference task, requiring methods that can accommodate uncertainty, heterogeneity, and model adaptivity across varying traffic regimes.}

\subsection{Applications in Traffic Bottleneck Management and Control}

Next, we will focus on two typical bottlenecks, intersections and ramps, and summarize the IDM-related research in the field of traffic bottlenecks from the aspects of intersection system evaluation, vehicle trajectory planning and signal optimization, and ramp merging management.

\subsubsection{Intersection Systems}

Intersection research mainly includes evaluation of the intersection system, vehicle trajectory planning and signal optimization research, as well as the coordinated optimization research of the two. \cite{joererVehicularNetworkingPerspectiveEstimating2014a} introduced a method to evaluate vehicle collision probability at intersections, using modified traffic simulators based on IDM to assess intelligent transportation system safety, especially when traffic rules are disregarded. \cite{schrader_global_2024} employed Sobol global sensitivity analysis to identify how simulation input parameters like IDM's parameter impact performance metrics of the intersection system. For vehicle trajectory planning related research, IDM is mainly used as a simulation model for background vehicles or manually driven vehicles \citep{chen_human-like_2023,dong_comparative_2022,ghosh_traffic_2022,guoCoordinationConnectedAutomatedVehicles2023a,jiangEcoApproachingIsolatedSignalized2017a,medina_cooperative_2018,stebbinsCharacterisingGreenLightOptimal2017a,wang_learning_2023,xiaoDecisionMakingAutonomousVehiclesRandom2024a,zhangPlatoonCenteredControlEcoDrivingSignalized2023b,zhang_risk-aware_2023}.In addition,\cite{zhouCooperativeSignalFreeIntersectionControl2022b} used IDM as the microscopic simulation model for CAV under the coorperative intersection control framework. For the signal optimization, IDM is also used as the simulation foundation \citep{cao_gain_2022,sohn_waterfilling_2023,yang_enhancing_2024}. \cite{duCoupledVehicleSignalControlMethod2021b} developed an efficient coupled vehicle signal control method for optimizing traffic signal timing and CAV trajectories in mixed traffic environments, using IDM as the artificially driven vehicle model. \cite{huang_reservation-based_2023} proposed a reservation-based cooperative ecological driving model for optimizing CAV trajectories and intersection signals at isolated intersections, using background vehicles simulated by IDM. \cite{liEcoDrivingSystemElectricVehicles2018a} integrated an electric vehicle eco-driving system and signal control in a V2X environment, using IDM to formulate non-connected electric vehicle following rules and a hybrid algorithm for optimization. \cite{liang_equitable_2020} combined connected vehicle technology for traffic signal control at isolated intersections, optimizing platoon dissipation based on IDM estimation of vehicle proximity.

\subsubsection{Ramp Merging}

For ramp scenarios, the main focus is on research related to collaborative merging. \cite{baselt_merging_2014,chen_deep_2023,hou_cooperative_2023,ito_coordination_2019,maProgressiveVirtualRiskBasedVehicle2024a,tangNovelHierarchicalCooperativeMerging2022b,xiongManagingMergingCAVLane2022b,zhou_state-constrained_2019} proposed various coorperative merging methods by using IDM to simulate a HV in a mixed traffic merging scenario. \cite{guNetworkTrafficInstabilityAutomated2022b} use macroscopic or network fundamental diagrams  to study the impact of CAVs on network flow destabilization caused by turning and merging maneuvers. They used IDM, HDM and cooperative IDM to simulate the AV, HV and CAV, respectively. \cite{silgu_combined_2022}  proposed a controller integrating ramp metering and variable speed limits for freeway traffic, tested via SUMO simulations with CACC and HVs.  For the system evaluation of ramp, \cite{xiaoUnravellingEffectsCooperativeAdaptive2018a} analyzed the impact of CACC deactivation on traffic flow at merging bottlenecks, using simulations to evaluate capacity and queue discharge rates.

{\textbf{Summary}: IDM has become a foundational simulation tool in bottleneck management, enabling evaluations of intersection safety, signal control strategies, and ramp merging coordination in mixed traffic. Its integration into intersection studies supports both trajectory planning and cooperative signal optimization, while in ramp scenarios, IDM facilitates the modeling of human-driven vehicles in increasingly complex CAV-HV interactions. However, most applications treat IDM as a passive background model rather than an active behavioral hypothesis, often sidestepping the calibration and validation of IDM under high-interaction, high-uncertainty conditions. Particularly in cooperative merging and signal-free intersection control, the use of IDM-family models lacks rigorous behavioral grounding or stochastic robustness. Future work should re-express IDM in these contexts not merely as a simulation agent, but as a control-informed, behaviorally adaptive module, with a stronger emphasis on empirical validation, uncertainty modeling, and multi-agent coordination under real-world constraints.}

\subsection{Applications in Section and Network Control}

Among the research related to the direct application of IDM, the vehicle control especially under the connected and autonomous environment is most popular. We divide these literature into four types: individual longitudinal control (car following),
Individual lateral control (lane change), integration of individual lateral and longitudinal coordinated control.

\subsubsection{Vehicle Longitudinal and Lateral control}

In individual longitudinal control, scholars tend to design the control algorithm of the automatic vehicle based on physical methods or data-driven methods. Therefore, IDM is still used as a vehicle control model for background vehicles, ACC vehicles, manually driven vehicles or some low-connectivity scenarios \citep{juPredictiveCruiseControllerElectric2023a,kamal_efficient_2016,kamalLookAheadDrivingSchemesEfficient2022a,li_deep_2022,li_effect_2015,liu_semantic_2023,liu_potential_2023,liuSpatiotemporalTrajectoryPlanningAutonomous2024a,luLearningDriverSpecificBehaviorOvertaking2018a,ma_energetic_2022,maPersonalizedDrivingBehaviorsFuel2022a,manolis_real_2020,shiConnectedAutomatedVehicleCooperative2021b,shiDeepReinforcementLearningBased2023b,suo_model-free_2024,tangHighwayDecisionMakingMotionPlanning2022a,tas_making_2018,toghi_social_2022,valiente_prediction-aware_2024,wegener_automated_2021,wuFlowModularLearningFramework2022a,wu_hybrid_2023,yan_unified_2023,yangIsolatedIntersectionControlVarious2016a,yuAssessmentDynamicPlatoonInformation2023a,yuanPredictiveEnergyManagementStrategy2019,zhang_stackelberg_2024,zhangNoMoreRoadBullying2023a,zhou_development_2020,milanesModelingCooperativeAutonomousAdaptive2014b,fernandes_platooning_2012,vivekCyberphysicalRisksHackedInternetconnected2019a}. Among these, \cite{zhangNoMoreRoadBullying2023a} and \cite{manolis_real_2020} integrated the IDM into the simulation software VISSIM and AIMSUN, respectively.

Compared with longitudinal control, there are relatively few studies on lateral control, and IDM is generally used as a longitudinal control model to study how to perceive, cooperate, make decisions, and adjust in lane change decisions based on the methodologies of game theory, model predictive control, reinforcement learning and etc. like \citep{chen_adversarial_2022,gongGameTheoryBasedDecisionMakingIterative2023a,guIntegratedEcoDrivingAutomationIntelligent2022a,hwang_autonomous_2022,li_bounded_2024,liuMultistepCooperativeLaneChange2024a,liu_interaction-aware_2023,scheel_recurrent_2022,sheng_cooperation-aware_2023,sunberg_improving_2022,wangModelingBoundedRationalityDiscretionary2022a,zhang_game_2020}.

For the individual lateral and longitudinal coordinated control, research usually integrate two types of models into one unified framework \citep{kanagaraj_self-driven_2018,li_developing_2023,mullakkal-babu_hybrid_2021}.

\subsubsection{Evaluation of CAV Systems on Road section}

In the road section scale, many research focus on evaluating the performance of the CAV systems. For example,several research evaluate the stability and the capacity of this system \citep{qin_analytical_2021,qinStabilityAnalysisConnectedVehicles2023a,sunStabilityExtensionCarFollowingModel2023b,sunRelationshipCarFollowingString2020b,shang_impacts_2021,talebpourEffectInformationAvailabilityStability2018a,talebpourInfluenceConnectedAutonomousVehicles2016a,wang_stability_2019,yao_linear_2021,zhou_analytical_2021}. \cite{alhariqiImpactVehicleArrangementMixed2023a,zhou_modeling_2020} evaluated the impact of different following arrangements in a platoon on the road emissions. \cite{mullakkal-babu_comparative_2022} evaluated the adverse impact of takeover behavior on mixed traffic flow at different takeover time levels for fully automated vehicles. \cite{rahman_longitudinal_2018} explored the impact of CV platooning on highways on longitudinal traffic safety at different market penetration rates, and studied the role of the allocation of CV (i.e., whether they are limited to specific lanes) on traffic safety. \cite{wangEgoEfficientLaneChangesConnected2022a} studied the impact of lane change on the sytem. \cite{wuTimeDependentPerformanceAnalysis80211pBased2020a} investigated the impact of disturbances caused by acceleration and deceleration of the lead vehicle, gusts, and uncertainties in the platoon control system, such as aerodynamic drag and rolling resistance torque, on platoon communication.
\cite{yao_fuel_2021} studied the impact of CAVs on fuel consumption and emissions of mixed highway traffic flows. \cite{yaoAnalysisImpactMaximumPlatoon2023a} explored the impact of the maximum size of an autonomous platoon on mixed traffic flows. \cite{zhou_impact_2021} studied
the impact of degradation of CAV platooning control modes on the fundamental diagram. Autonomous driving car testing is an important part of autonomous driving research and development, and IDM also plays an important role in this as the background vehicle \citep{sun_scenario-based_2022,li_scegene_2022,sun_adaptive_2022,cao_autonomous_2023,zhang_accelerated_2024,zhang_bayesian_2024}. 

\subsubsection{Traffic optimization of road section and network}

Furthermore, some system management research on the road section and network have been conducted. Some tried to improve the traffic capacity of road section based on platoon control.  Most of them use IDM for simulating the HV like \cite{ferreira_impact_2012,ge_optimal_2017,heEcoDrivingAdvisoryStrategiesPlatoon2018a,liu_joint_2022,montaninoTrajectoryDataReconstructionSimulationBased2015b,wangRollingHorizonControlFramework2014b,wang_optimal_2022,wang_trajectory_2022}, while other platoon control related research always use IDM as the degradation model of CAV. Some use the VSL to optimize the CAV system in which IDM is used to simulate HV or CAV \citep{khondakerVariableSpeedLimitMicroscopic2015a,li_developing_2022,li_integrated_2017,luTD3LVSLLaneLevelVariableSpeed2023a,othman_analysis_2022}. \cite{mahajan_improving_2024,schakel_improving_2014} proposed a speed-based lane guidance system and a new on-board advisory system, respectively, to provide recommendations on lanes, vehicle speeds, and headway, optimizing lane allocation under high-traffic conditions to reduce capacity degradation. \cite{wang_cooperative_2020} studied a cooperative ecological driving system for signal corridors, where the CAV switches between a leading vehicle and a following vehicle, and IDM is used to simulate a CAV with network communication failure.

On the road network scale, various scenarios are simulated with IDM as the underlying model. \cite{huang_characterizing_2023} considered different signal control schemes and demand loading patterns, and focused on the overall network level to consider the traffic flow patterns brought about by autonomous driving mixed traffic. \cite{claes_decentralized_2011} proposed a distributed vehicle path planning method based on predictive information.

{\textbf{Summary}: IDM serves as a foundational component in section and network-level traffic control, frequently used to model human-driven or fallback behaviors in studies of CAV longitudinal-lateral coordination, platoon dynamics, and system-wide optimization. Its integration into mixed traffic simulations underpins evaluations of capacity, stability, safety, and energy consumption across both microscopic and macroscopic scales. However, IDM is often treated as a static behavioral proxy, lacking dynamic adaptation or calibration across diverse control regimes. This simplification becomes especially problematic when IDM is used to simulate degraded CAV behavior or background vehicles in reinforcement learning and game-theoretic frameworks. Moreover, the model’s deterministic structure limits its ability to capture emergent behaviors in stochastic, high-density environments. Future research should reposition IDM not only as a benchmarking tool but as an adaptive module within learning-based, uncertainty-aware, and multi-agent control systems, capable of reflecting the evolving interplay between autonomy, connectivity, and human variability across networked infrastructures.}

\subsection{Interdisciplinary Applications}

 IDM is widely recognized for its versatility across numerous research domains beyond its core application in traffic modeling and management of road segments and networks. IDM finds extensive use in areas such as vehicle self-organizing networks, communication algorithms and their associated network attacks within vehicular communications, and even in civil engineering for studies related to bridge loads.

\subsubsection{Applications in vehicle communications}

In terms of in-vehicle networks, IDM is used as a fundamental model for vehicle operation. \cite{wangOptimalFeedbackControlLaw2023a} and \cite{wang_analytical_2024} have significantly advanced the understanding of cyberattacks on automated vehicles by developing an optimal feedback control law and a comprehensive framework for modeling attacks on adaptive cruise control vehicles. \cite{dongDynamicManagerSelectionAssisted2022a} introduced a resource allocation algorithm for 5G-V2X platoons, optimizing communication performance under finite block length constraints, crucial for ensuring ultra-reliable low-latency communications. \cite{jiaDisturbanceAdaptiveDesignVANETenabledVehicle2014a} and \cite{jiaNetworkConnectivityPlatoonBasedVehicular2014a} delved into VANET-enabled platoon architectures, emphasizing disturbance adaptation and network connectivity, which are vital for effective safety message transmission under varying system parameters. Moreover,  \cite{segataAutomaticEmergencyBrakingRealistic2013a} assessed the integration of automatic emergency braking systems within VANETs, while \cite{tianSelfOrganizedRelaySelectionCooperative2017a} proposed a decentralized relay selection algorithm based on non-cooperative game theory to enhance cooperative dynamics. 

\subsubsection{Applications in civil engineering}

\cite{ge_intelligent_2022} introduced an advanced traffic flow load monitoring system that integrates machine vision and weigh-in-motion  systems, leveraging deep learning for precise vehicle detection and extraction of key parameters for traffic flow load simulation.  \cite{zhou_hybrid_2024} proposed a two-step Hybrid Virtual-Real Traffic Simulation method to replicate the spatiotemporal distribution of bridge loads, demonstrating the broad applicability of IDM in various interdisciplinary fields.

{\textbf{Summary}: The adaptability of IDM has enabled its interdisciplinary expansion beyond classical traffic modeling into domains such as vehicular communications, network security, and civil infrastructure assessment. In vehicular networks, IDM supports the evaluation of cyber-physical threats, communication reliability, and cooperative control under V2X and VANET architectures, often serving as the behavioral backbone in simulations of control feedback and message propagation. In civil engineering, IDM-driven traffic simulations are increasingly used to estimate dynamic load effects on infrastructure, bridging traffic theory with structural analysis. Yet, these applications often inherit the behavioral simplifications of the base model, overlooking the contextual nuances of driver behavior under cyber disruptions or infrastructural constraints. To unlock its full interdisciplinary potential, future work should integrate IDM into co-simulation platforms with tighter behavioral-physical coupling, enabling high-fidelity, domain-specific extensions that reflect not only vehicle dynamics but also information flow, structural feedback, and adversarial interactions.}

\section{Future research directions}
\label{sec:future}
Based on the literature review and the analysis of limitations in previous research, several significant gaps and new opportunities for future studies can be identified, as outlined below.

\subsection{Combination of Big Data}

Current IDM research faces several limitations. Despite the widespread use of open-source datasets like the NGSIM, Naples data set \citep{punzoAnalysisComparisonMicroscopicTraffic2005}, Hefei data set \citep{jiangExperimentalFeaturesCarfollowingBehavior2015}, and the HighD data set \citep{krajewskiHighDDatasetDroneDataset2018}, data scarcity remains a significant challenge, affecting both empirical and experimental data acquisition. The limited availability and scope of these datasets make it difficult to extract universal behavioral or macro-level characteristics. Additionally, insufficient data volume constrains model validation and refinement. This situation not only hampers the development of mechanism-driven models like IDM but also limits the effective training of data-driven models, such as those based on machine learning and deep learning. Issues of data quality and diversity are also prominent, as existing datasets may lack representation of specific traffic conditions or regions. Furthermore, inadequate data collection frequency and precision hinder the ability to capture dynamic traffic changes.

Integrating big data into IDM research can significantly address current limitations through both empirical and experimental data approaches. Empirical data, gathered from traffic monitoring systems, floating car data, and intelligent infrastructure, offers a vast and dynamic dataset. This abundance of data provides a robust foundation for testing and validating models across diverse scenarios. By leveraging these large-scale datasets, IDM can be further refined and integrated with machine learning and deep learning techniques, enhancing traffic prediction capabilities. This integration enables more accurate and adaptive models that can efficiently respond to real-time traffic conditions. On the experimental data front, the advantages are equally compelling. Conducting controlled traffic experiments enables researchers to delve deeper into the mechanisms of driver behavior and traffic flow. Experimental data provides the opportunity to isolate and study specific variables, minimizing interference and uncovering fundamental characteristics and mechanisms \citep{tianCarFollowingDynamicsExperiments2023}. This detailed insight is crucial for guiding model development and improving theoretical frameworks. By bridging micro-level decision-making processes with macro-level traffic phenomena, experimental data facilitates a comprehensive understanding of traffic dynamics. This approach not only enriches the theoretical foundation of IDM but also supports the development of models that are both innovative and aligned with real-world complexities.

\subsection{Hybrid Modelling Methodology}

Hybrid modeling, which combines data-driven and mechanism-driven approaches, offers significant advantages over purely mechanistic or data-driven models. It leverages the strengths of both methodologies: the adaptability and pattern recognition capabilities of data-driven models, and the theoretical rigor and interpretability of mechanism-driven models like \citep{yuanModelingStochasticMicroscopicTraffic2020,yuanMacroscopicTrafficFlowModeling2021}. This synergy can lead to more robust and versatile traffic models, capable of addressing complex, real-world scenarios. However, the integration process faces several limitations.

Firstly, the integration often lacks seamless fusion, resulting in limited adaptability and flexibility in practical applications. Secondly, data-driven models and mechanism-driven models often have different underlying structures and assumptions, which makes combining them smoothly a challenge. This can lead to a lack of cohesion in the hybrid model, where components operate in isolation rather than in a unified manner. Additionally, the calibration and tuning processes for each model type can be complex and may not align well, resulting in inefficiencies or conflicts in parameter settings.  Additionally, hybrid models can exhibit inconsistent performance across different traffic scenarios, particularly in data-scarce or anomalous situations, affecting their robustness.Moreover, the lack of standardized metrics can lead to unfair comparisons and inconsistent assessments just as \cite{wangComparingHundredsMachineLearning2024} pointed out.  Furthermore, The challenge of integrating heterogeneous data sources persists, as effectively combining different data types to enhance model performance remains complex. Lastly, computational complexity and real-time applicability are significant constraints; integrated models may demand substantial computational resources, hindering their use in real-time traffic management. Addressing these challenges requires developing more flexible integration frameworks and efficient algorithms to support the advancement and application of IDM models.

\subsection{Emerging Technologies}

The integration of autonomous driving and V2X technology poses both opportunities and challenges for IDM development. While many studies have explored these technologies, the models developed are often scenario-specific and lack broad applicability. Additionally, there is inconsistency in how vehicle behaviors are defined; IDM is used to simulate HV in some cases, ACC vehicles in others, and even CAV, see Section 3.3. This inconsistency can lead to confusion and limit the effectiveness of simulations. Therefore, there is a pressing need for a universal microscopic behavior model in intelligent connected environments, akin to IDM, that can accurately simulate a variety of vehicle types and support diverse traffic scenarios. Such a model would facilitate more efficient traffic management and optimization in highly interconnected systems.

Digital twin technology offers significant benefits for IDM development by creating high-fidelity digital replicas of real-world traffic systems. This approach allows for the rapid and cost - effective collection of extensive data, which is crucial for model development and training and enhances the accuracy and robustness of IDM. Digital twins enable the simulation of diverse traffic scenarios, providing a platform for rigorous model testing and optimization. Additionally, they support system-wide optimization by identifying and addressing potential traffic bottlenecks and safety issues. Thus, digital twin technology provides essential support for IDM innovation and application, driving advancements in intelligent transportation systems.

\subsection{Human-Machine Interaction Integration}

The integration of IDM with Human-Machine Interaction (HMI) represents a promising direction for advancing both driver behavior modeling and intelligent transportation systems. As semi-autonomous and autonomous vehicles become more prevalent, the interaction between human drivers and automated systems introduces new challenges and opportunities for IDM development. Traditional IDM focuses on simulating CF behaviors based on deterministic or stochastic rules, yet it often overlooks the dynamic and context-sensitive nature of human decision-making in shared control scenarios. For instance, transitions between manual and automated driving modes, driver reactions to system suggestions, and trust in automation are critical factors that influence vehicle behavior.

Future research should aim to extend IDM by incorporating cognitive and psychological models of driver behavior, enabling more accurate simulation of these interactions. This may involve integrating real-time data from physiological sensors (e.g., eye-tracking, heart rate monitors) and external HMI systems (e.g., dashboards, alerts) to adjust IDM parameters dynamically based on the driver's state. Furthermore, HMI-integrated IDM can play a pivotal role in designing adaptive interfaces that balance automation and human control, ensuring smooth transitions and reducing driver workload. By capturing the interplay between human decisions and automated systems, such models would not only enhance the realism of traffic simulations but also contribute to safer and more efficient vehicle systems in complex driving environments.

{\section{Conclusions and Discussion}
\label{sec:conclusion}

After a quarter-century of development, IDM has become a cornerstone of microscopic traffic flow modeling. Its conceptual elegance, physical interpretability, and computational efficiency have made it a widely adopted framework in both academic inquiry and practical applications, including traffic simulations, driver assistance systems, and autonomous vehicle control. However, its very ubiquity has also contributed to a certain inertia in the field—dampening critical engagement with its underlying assumptions and limiting the exploration of alternative modeling paradigms. This paper does not aim to offer an uncritical celebration of IDM, but rather seeks to reframe the discourse: to acknowledge its foundational value, while emphasizing the need to re-express and re-engineer the model in light of new technological, behavioral, and methodological demands.

Our analysis reveals that while IDM remains powerful, it is fundamentally incomplete. The model is grounded in a mechanical and rationalist view of driving behavior, which, while effective in controlled scenarios, does not fully capture the complexity of modern traffic systems. As real-world traffic becomes increasingly heterogeneous, stochastic, and influenced by algorithmic decision-making, IDM’s limitations become more apparent. Its deterministic nature, limited treatment of human variability, and unrealistic responses under certain traffic conditions—particularly during congestion or at high speeds—suggest that IDM, in its current form, is misaligned with the realities of contemporary mobility ecosystems.

Thus, we contend that the future of IDM lies not in preserving it as a fixed standard, but in redefining it as a basis for innovation. IDM should no longer be seen as a monolithic solution, but as a modular and extensible framework—one that can be systematically expanded with stochastic elements, integrated with human behavioral insights, and enhanced through hybrid modeling that combines physics-based structure with data-driven learning. Promising directions include the development of context-specific extensions tailored to autonomous vehicles, cooperative driving systems, and eco-driving applications. In this light, IDM’s role should evolve from being a standalone model to becoming an interpretable backbone within a broader ecosystem of intelligent transportation modeling tools—serving not just as a simulation engine, but as a conceptual foundation for next-generation traffic systems.

Furthermore, our review identifies a pressing methodological gap: the absence of standardized, rigorous frameworks for evaluating and comparing CF models across deterministic and stochastic paradigms. Without such benchmarks, IDM’s continued dominance may reflect historical momentum rather than actual superiority. Advancing traffic flow theory requires not only more sophisticated models, but also a clearer understanding of what constitutes model robustness, generalizability, and practical utility under diverse real-world conditions.

Ultimately, this paper does not seek to close the chapter on IDM, but to open a more critical and generative conversation about its future. We hope to shift the perspective from viewing IDM as a fixed modeling artifact to engaging with it as a living theory-in-practice, one that must adapt to meet the demands of intelligent, connected, and increasingly unpredictable traffic environments. IDM’s legacy is not just as a model that captured the past, but as a lens through which we must now reimagine what CF modeling can and should become.

In this sense, IDM is not the end of a modeling journey—it is the beginning of a new phase of theoretical exploration and methodological innovation.}

\section{Acknowledgments}
% 72222021 田老师优青
% 72010107004 马老师国际合作
% 72288101 高老师基础科学中心
% W2411064 姜老师国际合作
This work is supported by the National Natural Science Foundation of China (Grant No. 72222021, W2411064, 72288101, 71931002), Young Scientific and Technological Talents (Level One) in Tianjin (Grant No. QN20230104) and Beijing-Tianjin-Hebei Basic Research Cooperation Project (Grant No. G2024210009).

\section*{Appendix}

Table \ref{tab:target_journals} is a list of target journals included in the literature review. We have combined the list of \cite{punzoCalibrationCarFollowingDynamicsAutomated2021} and related engineering research community are considered to represent the state of the art in the field.
\begin{table}[!h]
\centering
\scriptsize

\caption{Target Journals Included in the Literature Review.}
\label{tab:target_journals}
\begin{tblr}{
width = 0.7\linewidth,
hline{1,8} = {-}{0.08em}, % 表头和底部分隔线加粗
hline{2-7} = {-}{}, % 普通分隔线
row{odd} = {gray9}, % 奇数行背景设置为灰色
colspec = {Q[180]Q[350]}, % 调整列宽比例，使内容适应新的总宽度
cells = {m}, % 单元格垂直居中对齐
}
\textbf{Category} & \textbf{Journal Name} \\ 
\textbf{Transportation Science and Engineering} & 
Transportation Science \newline
Transportation Research Part A: Policy and Practice \newline
Transportation Research Part B: Methodological \newline
Transportation Research Part C: Emerging Technologies \newline
Transportation Research Part D: Transport and Environment \newline
Transportation Research Part E: Logistics and Transportation Review \newline
Transportation Research Part F: Traffic Psychology and Behaviour \newline
Accident Analysis and Prevention \newline
Transportmetrica A: Transport Science \newline
Transportmetrica B: Transport Dynamics \\ 
\textbf{Electrical and Computer Engineering} & 
IEEE Transactions on Intelligent Transportation Systems \newline
IEEE Transactions on Intelligent Vehicles \newline
IEEE Transactions on Vehicular Technology \newline
IEEE Transactions on Automation Science and Engineering \newline
IEEE Transactions on Robotics \newline
IEEE Transactions on Transportation Electrification \newline
Vehicular Communications \\ 
\textbf{Energy and Environment} & 
Applied Energy \newline
Energy \\ 
\textbf{Physics and Interdisciplinary Studies} & 
Physical Review E \newline
Physica A: Statistical Mechanics and Its Applications \newline
Plos One \\ 
\end{tblr}
\end{table}

% \clearpage 
\bibliographystyle{cas-model2-names}

\bibliography{Ref20241025}

@article{parkDriversCharacteristicsPerceptionLead2001,
  title = {Drivers' {{Characteristics}} in the {{Perception}} of a {{Lead Vehicle}}'s {{Deceleration Level}}},
  author = {Park, Kyung S. and , Ahn J., Lee and {and Koh}, Bong K.},
  year = {2001},
  month = jun,
  journal = {International Journal of Cognitive Ergonomics},
  volume = {5},
  number = {2},
  pages = {125--136},
  publisher = {Routledge},
  issn = {1088-6362},
  doi = {10.1207/S15327566IJCE0502_3},
  urldate = {2025-05-21}
}

@article{zhouCalibrationStochasticCarFollowing2025a,
  title = {On the Calibration of Stochastic Car Following Models},
  author = {Zhou, Shirui and Zheng, Shiteng and Xu, Tu and Treiber, Martin and Tian, Junfang and Jiang, Rui},
  year = {2025},
  month = jun,
  journal = {Transportation Research Part B: Methodological},
  volume = {196},
  pages = {103224},
  issn = {0191-2615},
  doi = {10.1016/j.trb.2025.103224},
  urldate = {2025-05-06}
}

@article{wangStabilityAnalysisStochasticLinear2020,
  title = {Stability {{Analysis}} of {{Stochastic Linear Car-Following Models}}},
  author = {Wang, Yu and Li, Xiaopeng and Tian, Junfang and Jiang, Rui},
  year = {2020},
  month = jan,
  journal = {Transportation Science},
  volume = {54},
  number = {1},
  pages = {274--297},
  issn = {0041-1655, 1526-5447},
  doi = {10.1287/trsc.2019.0932},
  urldate = {2022-08-26},
  langid = {english},
  lccn = {2}
}

@article{vivekCyberphysicalRisksHackedInternetconnected2019a,
  title = {Cyberphysical Risks of Hacked Internet-Connected Vehicles},
  author = {Vivek, Skanda and Yanni, David and Yunker, Peter J. and Silverberg, Jesse L.},
  year = {2019},
  month = jul,
  journal = {PHYSICAL REVIEW E},
  volume = {100},
  number = {1},
  pages = {012316},
  publisher = {Amer Physical Soc},
  address = {College Pk},
  issn = {2470-0045, 2470-0053},
  doi = {10.1103/PhysRevE.100.012316},
  urldate = {2024-11-28},
  langid = {english}
}

@article{wiedemann1992microscopic,
  title={Microscopic traffic simulation: the simulation system MISSION, background and actual state},
  author={Wiedemann, Reiter and Reiter, U},
  journal={Project ICARUS (V1052) Final Report},
  volume={2},
  pages={1--53},
  year={1992}
}

@article{daganzoCellTransmissionModelDynamic1994,
  title = {The Cell Transmission Model: {{A}} Dynamic Representation of Highway Traffic Consistent with the Hydrodynamic Theory},
  shorttitle = {The Cell Transmission Model},
  author = {Daganzo, Carlos F.},
  year = {1994},
  journal = {Transportation research part B: methodological},
  volume = {28},
  number = {4},
  pages = {269--287},
  publisher = {Elsevier},
  urldate = {2024-11-12}
}

@article{lighthillKinematicWavesIITheory1955,
  title = {On Kinematic Waves {{II}}. {{A}} Theory of Traffic Flow on Long Crowded Roads},
  author = {Lighthill, {\relax MJ} and Whitham, {\relax GB}},
  year = {1955},
  month = may,
  journal = {Proceedings of the Royal Society of London. Series A. Mathematical and Physical Sciences},
  volume = {229},
  number = {1178},
  pages = {317--345},
  issn = {0080-4630, 2053-9169},
  doi = {10.1098/rspa.1955.0089},
  urldate = {2024-11-12},
  copyright = {https://royalsociety.org/journals/ethics-policies/data-sharing-mining/},
  langid = {english}
}

@article{vickreyCongestionTheoryTransportInvestment1969,
  title = {Congestion Theory and Transport Investment},
  author = {Vickrey, William S.},
  year = {1969},
  journal = {The American economic review},
  volume = {59},
  number = {2},
  eprint = {1823678},
  eprinttype = {jstor},
  pages = {251--260},
  publisher = {JSTOR},
  urldate = {2024-11-12}
}

@book{tianCarFollowingDynamicsExperiments2023,
  title = {Car Following {{Dynamics}}: {{Experiments}} and {{Models}}},
  shorttitle = {Car Following {{Dynamics}}},
  author = {Tian, Junfang and Jiang, Rui},
  year = {2023},
  month = dec,
  publisher = {EDP Sciences},
  doi = {10.1051/978-2-7598-3194-4},
  urldate = {2024-11-11},
  isbn = {978-2-7598-3194-4}
}

@article{jiangExperimentalFeaturesCarfollowingBehavior2015,
  title = {On Some Experimental Features of Car-Following Behavior and How to Model Them},
  author = {Jiang, Rui and Hu, Mao-Bin and Zhang, H.M. and Gao, Zi-You and Jia, Bin and Wu, Qing-Song},
  year = {2015},
  month = oct,
  journal = {Transportation Research Part B: Methodological},
  volume = {80},
  pages = {338--354},
  issn = {01912615},
  doi = {10.1016/j.trb.2015.08.003},
  urldate = {2022-05-17},
}

@article{punzoAnalysisComparisonMicroscopicTraffic2005,
  title = {Analysis and {{Comparison}} of {{Microscopic Traffic Flow Models}} with {{Real Traffic Microscopic Data}}},
  author = {Punzo, Vincenzo and Simonelli, Fulvio},
  year = {2005},
  journal = {Transportation Research Record},
  pages = {11},
  doi = {10.1177/0361198105193400106},
  langid = {english}
}

@inproceedings{krajewskiHighDDatasetDroneDataset2018,
  title = {The {{highD Dataset}}: {{A Drone Dataset}} of {{Naturalistic Vehicle Trajectories}} on {{German Highways}} for {{Validation}} of {{Highly Automated Driving Systems}}},
  shorttitle = {The {{highD Dataset}}},
  booktitle = {2018 21st {{International Conference}} on {{Intelligent Transportation Systems}} ({{ITSC}})},
  author = {Krajewski, Robert and Bock, Julian and Kloeker, Laurent and Eckstein, Lutz},
  year = {2018},
  pages = {2118--2125},
  issn = {2153-0017},
  doi = {10.1109/ITSC.2018.8569552},
}

@article{yuanMacroscopicTrafficFlowModeling2021,
  title = {Macroscopic Traffic Flow Modeling with Physics Regularized {{Gaussian}} Process: {{A}} New Insight into Machine Learning Applications in Transportation},
  shorttitle = {Macroscopic Traffic Flow Modeling with Physics Regularized {{Gaussian}} Process},
  author = {Yuan, Yun and Zhang, Zhao and Yang, Xianfeng Terry and Zhe, Shandian},
  year = {2021},
  month = apr,
  journal = {Transportation Research Part B: Methodological},
  volume = {146},
  pages = {88--110},
  issn = {01912615},
  doi = {10.1016/j.trb.2021.02.007},
  urldate = {2022-04-07},
  langid = {english},
}

@article{yuanModelingStochasticMicroscopicTraffic2020,
  title = {Modeling {{Stochastic Microscopic Traffic Behaviors}}: A {{Physics Regularized Gaussian Process Approach}}},
  shorttitle = {Modeling {{Stochastic Microscopic Traffic Behaviors}}},
  author = {Yuan, Yun and Wang, Qinzheng and Yang, Xianfeng Terry},
  year = {2020},
  month = jul,
  journal = {Arxiv:2007.10109 [Cs, Eess, Stat]},
  eprint = {2007.10109},
  primaryclass = {Cs, Eess, Stat},
  urldate = {2022-04-06},
 }

@article{wangComparingHundredsMachineLearning2024,
  title = {Comparing Hundreds of Machine Learning and Discrete Choice Models for Travel Demand Modeling: {{An}} Empirical Benchmark},
  shorttitle = {Comparing Hundreds of Machine Learning and Discrete Choice Models for Travel Demand Modeling},
  author = {Wang, Shenhao and Mo, Baichuan and Zheng, Yunhan and Hess, Stephane and Zhao, Jinhua},
  year = {2024},
  month = dec,
  journal = {Transportation Research Part B: Methodological},
  volume = {190},
  pages = {103061},
  issn = {01912615},
  doi = {10.1016/j.trb.2024.103061},
  urldate = {2024-09-23},
}

@article{hartRobustCarfollowingBasedDeep2024,
  title = {Towards Robust Car-Following Based on Deep Reinforcement Learning},
  author = {Hart, Fabian and Okhrin, Ostap and Treiber, Martin},
  year = {2024},
  month = feb,
  journal = {Transportation Research Part C: Emerging Technologies},
  volume = {159},
  pages = {104486},
  issn = {0968090X},
  doi = {10.1016/j.trc.2024.104486},
  urldate = {2024-04-19},
  langid = {english}
}

@article{kanagaraj_self-driven_2018,
  title = {Self-{{Driven Particle Model}} for {{Mixed Traffic}} and {{Other Disordered Flows}}},
  author = {Kanagaraj, Venkatesan and Treiber, Martin},
  year = {2018},
  month = nov,
  journal = {Physica a-Statistical Mechanics and Its Applications},
  volume = {509},
  pages = {1--11},
  issn = {0378-4371},
  doi = {10.1016/j.physa.2018.05.086},
  langid = {american}
}

@article{kestingAdaptiveCruiseControlDesign2008,
  title = {Adaptive Cruise Control Design for Active Congestion Avoidance},
  author = {Kesting, Arne and Treiber, Martin and Schoenhof, Martin and Helbing, Dirk},
  year = {2008},
  month = dec,
  journal = {Transportation Research Part C-Emerging Technologies},
  volume = {16},
  number = {6},
  pages = {668--683},
  issn = {0968-090X},
  doi = {10.1016/j.trc.2007.12.004},
  langid = {english}
}

@article{kestingHowReactionTimeUpdate2008,
  title = {How {{Reaction Time}}, {{Update Time}}, and {{Adaptation Time Influence}} the {{Stability}} of {{Traffic Flow}}: {{Influence}} of Reaction Time, Update Time, and Adaptation Time},
  shorttitle = {How {{Reaction Time}}, {{Update Time}}, and {{Adaptation Time Influence}} the {{Stability}} of {{Traffic Flow}}},
  author = {Kesting, Arne and Treiber, Martin},
  year = {2008},
  month = jan,
  journal = {Computer-Aided Civil and Infrastructure Engineering},
  volume = {23},
  number = {2},
  pages = {125--137},
  issn = {10939687},
  doi = {10.1111/j.1467-8667.2007.00529.x},
  urldate = {2021-10-07},
  langid = {english},
  lccn = {1}
}

@article{tianCellularAutomatonModelSimulating2016,
  title = {Cellular Automaton Model Simulating Spatiotemporal Patterns, Phase Transitions and Concave Growth Pattern of Oscillations in Traffic Flow},
  author = {Tian, Junfang and Li, Guangyu and Treiber, Martin and Jiang, Rui and Jia, Ning and Ma, Shoufeng},
  year = {2016},
  month = nov,
  journal = {Transportation Research Part B: Methodological},
  volume = {93},
  pages = {560--575},
  issn = {01912615},
  doi = {10.1016/j.trb.2016.08.008},
  urldate = {2023-10-17},
  langid = {english}
}

@article{tianImproved2DIntelligentDriver2016,
  title = {Improved {{2D}} Intelligent Driver Model in the Framework of Three-Phase Traffic Theory Simulating Synchronized Flow and Concave Growth Pattern of Traffic Oscillations},
  author = {Tian, Junfang and Jiang, Rui and Li, Geng and Treiber, Martin and Jia, Bin and Zhu, Chenqiang},
  year = {2016},
  month = aug,
  journal = {Transportation Research Part F: Traffic Psychology and Behaviour},
  volume = {41},
  pages = {55--65},
  issn = {13698478},
  doi = {10.1016/j.trf.2016.06.005},
  urldate = {2022-03-13},
  langid = {english}
}

@article{tianRoleSpeedAdaptationSpacing2019,
  title = {On the Role of Speed Adaptation and Spacing Indifference in Traffic Instability: {{Evidence}} from Car-Following Experiments and Its Stochastic Model},
  shorttitle = {On the Role of Speed Adaptation and Spacing Indifference in Traffic Instability},
  author = {Tian, Junfang and Zhang, H.M. and Treiber, Martin and Jiang, Rui and Gao, Zi-You and Jia, Bin},
  year = {2019},
  month = nov,
  journal = {Transportation Research Part B: Methodological},
  volume = {129},
  pages = {334--350},
  issn = {01912615},
  doi = {10.1016/j.trb.2019.09.014},
  urldate = {2022-03-05},
  langid = {english}
}

@article{treiberComparingNumericalIntegrationSchemes2015,
  title = {Comparing Numerical Integration Schemes for Time-Continuous Car-Following Models},
  author = {Treiber, Martin and Kanagaraj, Venkatesan},
  year = {2015},
  month = feb,
  journal = {Physica a: Statistical Mechanics and Its Applications},
  volume = {419},
  pages = {183--195},
  issn = {03784371},
  doi = {10.1016/j.physa.2014.09.061},
  urldate = {2022-05-20},
  langid = {english}
}

@article{treiberIntelligentDriverModelStochasticity2018,
  title = {The {{Intelligent Driver Model}} with Stochasticity -- {{New}} Insights into Traffic Flow Oscillations},
  author = {Treiber, Martin and Kesting, Arne},
  year = {2018},
  month = nov,
  journal = {Transportation Research Part B: Methodological},
  volume = {117},
  pages = {613--623},
  issn = {01912615},
  doi = {10.1016/j.trb.2017.08.012},
  urldate = {2022-04-11},
  langid = {english}
}

@article{treiberThreephaseTrafficTheoryTwophase2010,
  title = {Three-Phase Traffic Theory and Two-Phase Models with a Fundamental Diagram in the Light of Empirical Stylized Facts},
  author = {Treiber, Martin and Kesting, Arne and Helbing, Dirk},
  year = {2010},
  month = sep,
  journal = {Transportation Research Part B: Methodological},
  volume = {44},
  number = {8-9},
  pages = {983--1000},
  issn = {01912615},
  doi = {10.1016/j.trb.2010.03.004},
  urldate = {2023-09-13},
  langid = {english}
}

@book{treiberTrafficFlowDynamics2013,
  title = {Traffic {{Flow Dynamics}}},
  author = {Treiber, Martin and Kesting, Arne},
  year = {2013},
  publisher = {Springer Berlin Heidelberg},
  address = {Berlin, Heidelberg},
  doi = {10.1007/978-3-642-32460-4},
  urldate = {2021-11-16},
  isbn = {978-3-642-32459-8 978-3-642-32460-4},
  langid = {english}
}

@article{treiberValidationTrafficFlowModels2012,
  title = {Validation of Traffic Flow Models with Respect to the Spatiotemporal Evolution of Congested Traffic Patterns},
  author = {Treiber, Martin and Kesting, Arne},
  year = {2012},
  month = apr,
  journal = {Transportation Research Part C: Emerging Technologies},
  volume = {21},
  number = {1},
  pages = {31--41},
  issn = {0968090X},
  doi = {10.1016/j.trc.2011.09.002},
  urldate = {2021-09-16},
  langid = {english}
}

@article{treiberMicroscopicCalibrationValidationCarFollowing2013,
  title = {Microscopic {{Calibration}} and {{Validation}} of {{Car-Following Models}} -- {{A Systematic Approach}}},
  author = {Treiber, Martin and Kesting, Arne},
  year = {2013},
  month = jun,
  journal = {Procedia - Social and Behavioral Sciences},
  volume = {80},
  pages = {922--939},
  issn = {18770428},
  doi = {10.1016/j.sbspro.2013.05.050},
  urldate = {2021-12-18},
  langid = {english}
}

@article{chandlerTrafficDynamicsStudiesCar1958,
  title = {Traffic {{Dynamics}}: {{Studies}} in {{Car Following}}},
  shorttitle = {Traffic Dynamics},
  author = {Chandler, Robert E. and Herman, Robert and Montroll, Elliott W.},
  year = {1958},
  month = apr,
  journal = {Operations Research},
  volume = {6},
  number = {2},
  pages = {165--184},
  publisher = {INFORMS},
  issn = {0030-364X},
  doi = {10.1287/opre.6.2.165},
  langid = {english},
  lccn = {3}
}

@article{bandoDynamicalModelTrafficCongestion1995,
  title = {Dynamical Model of Traffic Congestion and Numerical Simulation},
  author = {Bando, M. and Hasebe, K. and Nakayama, A. and Shibata, A. and Sugiyama, Y.},
  year = {1995},
  month = feb,
  journal = {Physical Review E},
  volume = {51},
  number = {2},
  pages = {1035--1042},
  issn = {1063-651X, 1095-3787},
  doi = {10.1103/PhysRevE.51.1035},
  urldate = {2024-10-28},
  copyright = {http://link.aps.org/licenses/aps-default-license},
  langid = {english}
}

@article{gazisNonlinearFollowLeaderModelsTraffic1961,
  title = {Nonlinear {{Follow-the-Leader Models}} of {{Traffic Flow}}},
  author = {Gazis, Denos C. and Herman, Robert and Rothery, Richard W.},
  year = {1961},
  month = aug,
  journal = {Operations Research},
  volume = {9},
  number = {4},
  pages = {545--567},
  issn = {0030-364X, 1526-5463},
  doi = {10.1287/opre.9.4.545},
  urldate = {2024-10-28},
  langid = {english}
}

@article{gippsBehaviouralCarfollowingModelComputer1981,
  title = {A Behavioural Car-Following Model for Computer Simulation},
  author = {Gipps, Peter G.},
  year = {1981},
  journal = {Transportation research part B: methodological},
  volume = {15},
  number = {2},
  pages = {105--111},
  publisher = {Elsevier},
  urldate = {2024-10-28}
}

@article{nagelCellularAutomatonModelFreeway1992a,
  title = {A Cellular Automaton Model for Freeway Traffic},
  author = {Nagel, Kai and Schreckenberg, Michael},
  year = {1992},
  journal = {Journal de physique I},
  volume = {2},
  number = {12},
  pages = {2221--2229},
  publisher = {EDP Sciences},
  urldate = {2024-10-28}
}

@article{newellNonlinearEffectsDynamicsCar1961a,
  title = {Nonlinear {{Effects}} in the {{Dynamics}} of {{Car Following}}},
  author = {Newell, G. F.},
  year = {1961},
  month = apr,
  journal = {Operations Research},
  volume = {9},
  number = {2},
  pages = {209--229},
  issn = {0030-364X, 1526-5463},
  doi = {10.1287/opre.9.2.209},
  urldate = {2024-10-28},
  langid = {english}
}

@article{newellSimplifiedCarfollowingTheoryLower2002,
  title = {A Simplified Car-Following Theory: A Lower Order Model},
  shorttitle = {A Simplified Car-Following Theory},
  author = {Newell, Gordon Frank},
  year = {2002},
  journal = {Transportation Research Part B: Methodological},
  volume = {36},
  number = {3},
  pages = {195--205},
  publisher = {Elsevier},
  urldate = {2024-10-28}
}

@article{pipesOperationalAnalysisTrafficDynamics1953,
  title = {An Operational Analysis of Traffic Dynamics},
  author = {Pipes, Louis A.},
  year = {1953},
  journal = {Journal of applied physics},
  volume = {24},
  number = {3},
  pages = {274--281},
  publisher = {American Institute of Physics},
  urldate = {2024-10-28}
}

@article{treiberCongestedTrafficStatesEmpirical2000,
  title = {Congested {{Traffic States}} in {{Empirical Observations}} and {{Microscopic Simulations}}},
  author = {Treiber, Martin and Hennecke, Ansgar and Helbing, Dirk},
  year = {2000},
  month = aug,
  journal = {Physical Review E},
  volume = {62},
  number = {2},
  eprint = {cond-mat/0002177},
  pages = {1805--1824},
  issn = {1063-651X, 1095-3787},
  doi = {10.1103/PhysRevE.62.1805},
  urldate = {2021-10-07},
  archiveprefix = {arXiv},
  langid = {english},
  lccn = {3},
}

@article{alhariqiImpactVehicleArrangementMixed2023a,
  title = {Impact of {{Vehicle Arrangement}} in {{Mixed Autonomy Traffic}} on {{Emissions}}},
  author = {Alhariqi, Abdulrahman and Gu, Ziyuan and Saberi, Meead},
  year = {2023},
  month = dec,
  journal = {Transportation Research Part D-Transport and Environment},
  volume = {125},
  issn = {1361-9209},
  doi = {10.1016/j.trd.2023.103964}
}

@article{baiLongitudinalControlAutomatedVehicles2024a,
  title = {Longitudinal {{Control}} of {{Automated Vehicles}}: {{A Novel Approach}} by {{Integrating Deep Reinforcement Learning}} with {{Intelligent Driver Model}}},
  author = {Bai, Linhan and Zheng, Fangfang and Hou, Kangning and Liu, Xiaobo and Lu, Liang and Liu, Can},
  year = {2024},
  month = aug,
  journal = {IEEE Transactions on Vehicular Technology},
  volume = {73},
  number = {8},
  pages = {11014--11028},
  issn = {0018-9545},
  doi = {10.1109/TVT.2024.3376599},
  langid = {american}
}

@article{baselt_merging_2014,
  title = {Merging {{Lanes-Fairness}} through {{Communication}}},
  author = {Baselt, D. and Knorr, F. and Scheuermann, B. and Schreckenberg, M. and Mauve, M.},
  year = {2014},
  month = apr,
  journal = {Vehicular Communications},
  volume = {1},
  number = {2},
  pages = {97--104},
  issn = {2214-2096},
  doi = {10.1016/j.vehcom.2014.05.005},
  langid = {american}
}

@article{berkhahn_traffic_2022,
  title = {Traffic Dynamics at Intersections Subject to Random Misperception},
  author = {Berkhahn, Volker and Kleiber, Marcel and Langner, Johannes and Timmermann, Chris and Weber, Stefan},
  year = {2022},
  month = may,
  journal = {IEEE Transactions on Intelligent Transportation Systems},
  volume = {23},
  number = {5},
  pages = {4501--4511},
  issn = {1524-9050},
  doi = {10.1109/TITS.2020.3045480}
}

@article{bhattacharyya_hybrid_2022,
  title = {A {{Hybrid Rule-Based}} and {{Data-Driven Approach}} to {{Driver Modeling}} through {{Particle Filtering}}},
  author = {Bhattacharyya, Raunak and Jung, Soyeon and Kruse, Liam A. and Senanayake, Ransalu and Kochenderfer, Mykel J.},
  year = {2022},
  month = aug,
  journal = {IEEE Transactions on Intelligent Transportation Systems},
  volume = {23},
  number = {8},
  pages = {13055--13068},
  issn = {1524-9050},
  doi = {10.1109/TITS.2021.3119415},
  langid = {american}
}

@article{bouadiStochasticFactorsStringStability2022b,
  title = {Stochastic {{Factors}} and {{String Stability}} of {{Traffic Flow}}: {{Analytical Investigation}} and {{Numerical Study Based}} on {{Car-Following Models}}},
  author = {Bouadi, Marouane and Jia, Bin and Jiang, Rui and Li, Xingang and Gao, Zi-You},
  year = {2022},
  month = nov,
  journal = {Transportation Research Part B-Methodological},
  volume = {165},
  pages = {96--122},
  issn = {0191-2615},
  doi = {10.1016/j.trb.2022.09.007}
}

@article{brito_learning_2022,
  title = {Learning {{Interaction-Aware Guidance}} for {{Trajectory Optimization}} in {{Dense Traffic Scenarios}}},
  author = {Brito, Bruno and Agarwal, Achin and {Alonso-Mora}, Javier},
  year = {2022},
  month = oct,
  journal = {IEEE Transactions on Intelligent Transportation Systems},
  volume = {23},
  number = {10},
  pages = {18808--18821},
  issn = {1524-9050},
  doi = {10.1109/TITS.2022.3160936},
  langid = {american}
}

@article{campi_roundabouts_2024,
  title = {Roundabouts: {{Traffic}} Simulations of Connected and Automated Vehicles-a State of the Art},
  author = {Campi, Elena and Mastinu, Gianpiero and Previati, Giorgio and Studer, Luca and Uccello, Lorenzo},
  year = {2024},
  month = may,
  journal = {IEEE Transactions on Intelligent Transportation Systems},
  volume = {25},
  number = {5},
  pages = {3305--3325},
  issn = {1524-9050},
  doi = {10.1109/TITS.2023.3325000}
}

@article{cao_autonomous_2023,
  title = {Autonomous {{Driving Policy Continual Learning}} with {{One-Shot Disengagement Case}}},
  author = {Cao, Zhong and Li, Xiang and Jiang, Kun and Zhou, Weitao and Liu, Xiaoyu and Deng, Nanshan and Yang, Diange},
  year = {2023},
  month = feb,
  journal = {IEEE Transactions on Intelligent Vehicles},
  volume = {8},
  number = {2},
  pages = {1380--1391},
  issn = {2379-8858},
  doi = {10.1109/TIV.2022.3184729},
  langid = {american}
}

@article{cao_gain_2022,
  title = {A {{Gain}} with {{No Pain}}: {{Exploring Intelligent Traffic Signal Control}} for {{Emergency Vehicles}}},
  author = {Cao, Miaomiao and Li, Victor O. K. and Shuai, Qiqi},
  year = {2022},
  month = oct,
  journal = {IEEE Transactions on Intelligent Transportation Systems},
  volume = {23},
  number = {10},
  pages = {17899--17909},
  issn = {1524-9050},
  doi = {10.1109/TITS.2022.3159714},
  langid = {american}
}

@article{caoTrustworthySafetyImprovementAutonomous2022a,
  title = {Trustworthy {{Safety Improvement}} for {{Autonomous Driving Using Reinforcement Learning}}},
  author = {Cao, Zhong and Xu, Shaobing and Jiao, Xinyu and Peng, Huei and Yang, Diange},
  year = {2022},
  month = may,
  journal = {Transportation Research Part C-Emerging Technologies},
  volume = {138},
  issn = {0968-090X},
  doi = {10.1016/j.trc.2022.103656},
  langid = {english}
}

@article{chen_adversarial_2022,
  title = {Adversarial {{Evaluation}} of {{Autonomous Vehicles}} in {{Lane-Change Scenarios}}},
  author = {Chen, Baiming and Chen, Xiang and Wu, Qiong and Li, Liang},
  year = {2022},
  month = aug,
  journal = {IEEE Transactions on Intelligent Transportation Systems},
  volume = {23},
  number = {8},
  pages = {10333--10342},
  issn = {1524-9050},
  doi = {10.1109/TITS.2021.3091477}
}

@article{chen_deep_2023,
  title = {Deep Multi-Agent Reinforcement Learning for Highway on-{{Ramp}} Merging in Mixed Traffic},
  author = {Chen, Dong and Hajidavalloo, Mohammad R. and Li, Zhaojian and Chen, Kaian and Wang, Yongqiang and Jiang, Longsheng and Wang, Yue},
  year = {2023},
  month = nov,
  journal = {IEEE Transactions on Intelligent Transportation Systems},
  volume = {24},
  number = {11},
  pages = {11623--11638},
  issn = {1524-9050},
  doi = {10.1109/TITS.2023.3285442}
}

@article{chen_human-like_2023,
  title = {Human-like Control for Automated Vehicles and Avoiding ``Vehicle Face-off'' in Unprotected Left Turn Scenarios},
  author = {Chen, Jin and Sun, Dihua and Zhao, Min},
  year = {2023},
  month = feb,
  journal = {IEEE Transactions on Intelligent Transportation Systems},
  volume = {24},
  number = {2},
  pages = {1609--1618},
  issn = {1524-9050},
  doi = {10.1109/TITS.2022.3222492}
}

@article{chen_investigating_2020,
  title = {Investigating the {{Long-}} and {{Short-Term Driving Characteristics}} and {{Incorporating Them}} into {{Car-Following Models}}},
  author = {Chen, Xiaoyun and Sun, Jian and Ma, Zian and Sun, Jie and Zheng, Zuduo},
  year = {2020},
  month = aug,
  journal = {Transportation Research Part C-Emerging Technologies},
  volume = {117},
  issn = {0968-090X},
  doi = {10.1016/j.trc.2020.102698},
  langid = {american}
}

@article{chen_safety_2024,
  title = {Safety {{Performance Evaluation}} of {{Freeway Merging Areas}} under {{Autonomous Vehicles Environment Using}} a {{Co-Simulation Platform}}},
  author = {Chen, Peng and Ni, Haoyuan and Wang, Liang and Yu, Guizhen and Sun, Jian},
  year = {2024},
  month = may,
  journal = {Accident Analysis and Prevention},
  volume = {199},
  issn = {0001-4575},
  doi = {10.1016/j.aap.2024.107530},
  langid = {american}
}

@article{chen_sigmoid-based_2024,
  title = {A Sigmoid-Based Car-Following Model to Improve Acceleration Stability in Traffic Oscillation and Following Failure in Free Flow},
  author = {Chen, Xingyu and Zhang, Weihua and Bai, Haijian and Jiang, Rui and Ding, Heng and Wei, Liyang},
  year = {2024},
  month = may,
  journal = {IEEE Transactions on Intelligent Transportation Systems},
  issn = {1524-9050},
  doi = {10.1109/TITS.2024.3393490}
}

@article{claes_decentralized_2011,
  title = {A Decentralized Approach for Anticipatory Vehicle Routing Using Delegate Multiagent Systems},
  author = {Claes, Rutger and Holvoet, Tom and Weyns, Danny},
  year = {2011},
  month = jun,
  journal = {IEEE Transactions on Intelligent Transportation Systems},
  volume = {12},
  number = {2},
  pages = {364--373},
  issn = {1524-9050},
  doi = {10.1109/TITS.2011.2105867}
}

@article{dong_comparative_2022,
  title = {A {{Comparative Study}} of {{Energy-Efficient Driving Strategy}} for {{Connected Internal Combustion Engine}} and {{Electric Vehicles}} at {{Signalized Intersections}}},
  author = {Dong, Haoxuan and Zhuang, Weichao and Chen, Boli and Wang, Yan and Lu, Yanbo and Liu, Ying and Xu, Liwei and Yin, Guodong},
  year = {2022},
  month = mar,
  journal = {Applied Energy},
  volume = {310},
  issn = {0306-2619},
  doi = {10.1016/j.apenergy.2022.118524},
  langid = {american}
}

@article{dongDynamicManagerSelectionAssisted2022a,
  title = {Dynamic {{Manager Selection Assisted Resource Allocation}} in {{URLLC}} with {{Finite Block Length}} for {{5G-V2X Platoons}}},
  author = {Dong, Zhihao and Zhu, Xu and Jiang, Yufei and Zeng, Haiyong and Wei, Zhongxiang and Zheng, Fu-Chun and Leung, Ka-Cheong},
  year = {2022},
  month = nov,
  journal = {IEEE Transactions on Vehicular Technology},
  volume = {71},
  number = {11},
  pages = {11336--11350},
  issn = {0018-9545},
  doi = {10.1109/TVT.2022.3190518}
}

@article{dongFlexibleEcoCruisingStrategyConnected2024a,
  title = {Flexible {{Eco-Cruising Strategy}} for {{Connected}} and {{Automated Vehicles}} with {{Efficient Driving Lane Planning}} and {{Speed Optimization}}},
  author = {Dong, Haoxuan and Wang, Qun and Zhuang, Weichao and Yin, Guodong and Gao, Kun and Li, Zhaojian and Song, Ziyou},
  year = {2024},
  month = mar,
  journal = {IEEE Transactions on Transportation Electrification},
  volume = {10},
  number = {1},
  pages = {1530--1540},
  issn = {2332-7782},
  doi = {10.1109/TTE.2023.3289980}
}

@article{duCoupledVehicleSignalControlMethod2021b,
  title = {A {{Coupled Vehicle-Signal Control Method}} at {{Signalized Intersections}} in {{Mixed Traffic Environment}}},
  author = {Du, Yu and ShangGuan, Wei and Chai, Linguo},
  year = {2021},
  month = mar,
  journal = {IEEE Transactions on Vehicular Technology},
  volume = {70},
  number = {3},
  pages = {2089--2100},
  issn = {0018-9545},
  doi = {10.1109/TVT.2021.3056457}
}

@article{fardCopulaBasedEstimationDistributionAlgorithm2019b,
  title = {A {{Copula-Based Estimation}} of {{Distribution Algorithm}} for {{Calibration}} of {{Microscopic Traffic Models}}},
  author = {Fard, Mehdi Rafati and Mohaymany, Afshin Shariat},
  year = {2019},
  month = jan,
  journal = {Transportation Research Part C-Emerging Technologies},
  volume = {98},
  pages = {449--470},
  issn = {0968-090X},
  doi = {10.1016/j.trc.2018.12.008},
  langid = {english}
}

@article{fernandes_platooning_2012,
  title = {Platooning with {{IVC-enabled}} Autonomous Vehicles: {{Strategies}} to Mitigate Communication Delays, Improve Safety and Traffic Flow},
  author = {Fernandes, Pedro and Nunes, Urbano},
  year = {2012},
  month = mar,
  journal = {IEEE Transactions on Intelligent Transportation Systems},
  volume = {13},
  number = {1},
  pages = {91--106},
  issn = {1524-9050},
  doi = {10.1109/TITS.2011.2179936}
}

@article{ferreira_impact_2012,
  title = {On the Impact of Virtual Traffic Lights on Carbon Emissions Mitigation},
  author = {Ferreira, Michel and {d'Orey}, Pedro M.},
  year = {2012},
  month = mar,
  journal = {IEEE Transactions on Intelligent Transportation Systems},
  volume = {13},
  number = {1},
  pages = {284--295},
  issn = {1524-9050},
  doi = {10.1109/TITS.2011.2169791}
}

@article{ge_intelligent_2022,
  title = {Intelligent {{Simulation Method}} of {{Bridge Traffic Flow Load Combining Machine Vision}} and {{Weigh-in-Motion Monitoring}}},
  author = {Ge, Liangfu and Dan, Danhui and Liu, Zijia and Ruan, Xin},
  year = {2022},
  month = sep,
  journal = {IEEE Transactions on Intelligent Transportation Systems},
  volume = {23},
  number = {9},
  pages = {15313--15328},
  issn = {1524-9050},
  doi = {10.1109/TITS.2022.3140276},
  langid = {american}
}

@article{ge_optimal_2017,
  title = {Optimal Control of Connected Vehicle Systems with Communication Delay and Driver Reaction Time},
  author = {Ge, Jin I. and Orosz, Gabor},
  year = {2017},
  month = aug,
  journal = {IEEE Transactions on Intelligent Transportation Systems},
  volume = {18},
  number = {8},
  pages = {2056--2070},
  issn = {1524-9050},
  doi = {10.1109/TITS.2016.2633164}
}

@article{gengPhysicsInformedTransformerModelVehicle2023b,
  title = {A {{Physics-Informed Transformer Model}} for {{Vehicle Trajectory Prediction}} on {{Highways}}},
  author = {Geng, Maosi and Li, Junyi and Xia, Yingji and Chen, Xiqun (Michael)},
  year = {2023},
  month = sep,
  journal = {Transportation Research Part C-Emerging Technologies},
  volume = {154},
  issn = {0968-090X},
  doi = {10.1016/j.trc.2023.104272},
  langid = {english}
}

@article{ghosh_traffic_2022,
  title = {Traffic Control in a Mixed Autonomy Scenario at Urban Intersections: {{An}} Optimal Control Approach},
  author = {Ghosh, Arnob and Parisini, Thomas},
  year = {2022},
  month = oct,
  journal = {IEEE Transactions on Intelligent Transportation Systems},
  volume = {23},
  number = {10},
  pages = {17325--17341},
  issn = {1524-9050},
  doi = {10.1109/TITS.2022.3166452}
}

@article{gongGameTheoryBasedDecisionMakingIterative2023a,
  title = {Game {{Theory-Based Decision-Making}} and {{Iterative Predictive Lateral Control}} for {{Cooperative Obstacle Avoidance}} of {{Guided Vehicle Platoon}}},
  author = {Gong, Xinle and Liang, Sheng and Wang, Bowen and Zhang, Wei},
  year = {2023},
  month = jun,
  journal = {IEEE Transactions on Vehicular Technology},
  volume = {72},
  number = {6},
  pages = {7051--7066},
  issn = {0018-9545},
  doi = {10.1109/TVT.2023.3237547},
  langid = {american}
}

@article{guIntegratedEcoDrivingAutomationIntelligent2022a,
  title = {Integrated {{Eco-Driving Automation}} of {{Intelligent Vehicles}} in {{Multi-Lane Scenario}} via {{Model-Accelerated Reinforcement Learning}}},
  author = {Gu, Ziqing and Yin, Yuming and Li, Shengbo Eben and Duan, Jingliang and Zhang, Fawang and Zheng, Sifa and Yang, Ruigang},
  year = {2022},
  month = nov,
  journal = {Transportation Research Part C-Emerging Technologies},
  volume = {144},
  issn = {0968-090X},
  doi = {10.1016/j.trc.2022.103863},
  langid = {english}
}

@article{guNetworkTrafficInstabilityAutomated2022b,
  title = {Network {{Traffic Instability}} with {{Automated Driving}} and {{Cooperative Merging}}},
  author = {Gu, Ziyuan and Wang, Zelin and Liu, Zhiyuan and Saberi, Meead},
  year = {2022},
  month = may,
  journal = {Transportation Research Part C-Emerging Technologies},
  volume = {138},
  issn = {0968-090X},
  doi = {10.1016/j.trc.2022.103626},
  langid = {english}
}

@article{guoCoordinationConnectedAutomatedVehicles2023a,
  title = {Coordination for {{Connected}} and {{Automated Vehicles}} at {{Non-Signalized Intersections}}: {{A Value Decomposition-Based Multiagent Deep Reinforcement Learning Approach}}},
  author = {Guo, Zihan and Wu, Yan and Wang, Lifang and Zhang, Junzhi},
  year = {2023},
  month = mar,
  journal = {IEEE Transactions on Vehicular Technology},
  volume = {72},
  number = {3},
  pages = {3025--3034},
  issn = {0018-9545},
  doi = {10.1109/TVT.2022.3219428}
}

@article{heEcoDrivingAdvisoryStrategiesPlatoon2018a,
  title = {Eco-{{Driving Advisory Strategies}} for a {{Platoon}} of {{Mixed Gasoline}} and {{Electric Vehicles}} in a {{Connected Vehicle System}}},
  author = {He, Xiaozheng and Wu, Xinkai},
  year = {2018},
  month = aug,
  journal = {Transportation Research Part D-Transport and Environment},
  volume = {63},
  pages = {907--922},
  issn = {1361-9209},
  doi = {10.1016/j.trd.2018.07.014},
  langid = {american}
}

@article{hou_cooperative_2023,
  title = {Cooperative On-Ramp Merging Control Model for Mixed Traffic on Multi-Lane Freeways},
  author = {Hou, Kangning and Zheng, Fangfang and Liu, Xiaobo and Guo, Ge},
  year = {2023},
  month = oct,
  journal = {IEEE Transactions on Intelligent Transportation Systems},
  volume = {24},
  number = {10},
  pages = {10774--10790},
  issn = {1524-9050},
  doi = {10.1109/TITS.2023.3274586}
}

@article{hu_autonomous_2023,
  title = {Autonomous Vehicle's Impact on Traffic: {{Empirical}} Evidence from Waymo Open Dataset and Implications from Modelling},
  author = {Hu, Xiangwang and Zheng, Zuduo and Chen, Danjue and Sun, Jian},
  year = {2023},
  month = jun,
  journal = {IEEE Transactions on Intelligent Transportation Systems},
  volume = {24},
  number = {6},
  pages = {6711--6724},
  issn = {1524-9050},
  doi = {10.1109/TITS.2023.3258145}
}

@article{huang_characterizing_2023,
  title = {Characterizing the Impact of Autonomous Vehicles on Macroscopic Fundamental Diagrams},
  author = {Huang, Yan and Ye, Yingjun and Sun, Jian and Tian, Ye},
  year = {2023},
  month = jun,
  journal = {IEEE Transactions on Intelligent Transportation Systems},
  volume = {24},
  number = {6},
  pages = {6530--6541},
  issn = {1524-9050},
  doi = {10.1109/TITS.2023.3265647}
}

@article{huang_reservation-based_2023,
  title = {Reservation-Based Cooperative Ecodriving Model for Mixed Autonomous and Manual Vehicles at Intersections},
  author = {Huang, Xin and Lin, Peiqun and Pei, Mingyang and Ran, Bin and Tan, Manchun},
  year = {2023},
  month = sep,
  journal = {IEEE Transactions on Intelligent Transportation Systems},
  volume = {24},
  number = {9},
  pages = {9501--9517},
  issn = {1524-9050},
  doi = {10.1109/TITS.2023.3269803}
}

@article{hwang_autonomous_2022,
  title = {Autonomous {{Vehicle Cut-in Algorithm}} for {{Lane-Merging Scenarios}} via {{Policy-Based Reinforcement Learning Nested}} within {{Finite-State Machine}}},
  author = {Hwang, Seulbin and Lee, Kibeom and Jeon, Hyeongseok and Kum, Dongsuk},
  year = {2022},
  month = oct,
  journal = {IEEE Transactions on Intelligent Transportation Systems},
  volume = {23},
  number = {10},
  pages = {17594--17606},
  issn = {1524-9050},
  doi = {10.1109/TITS.2022.3153848},
  langid = {american}
}

@article{ito_coordination_2019,
  title = {Coordination of Connected Vehicles on Merging Roads Using Pseudo-Perturbation-Based Broadcast Control},
  author = {Ito, Yuji and Kamal, Md Abdus Samad and Yoshimura, Takayoshi and Azuma, Shun-ichi},
  year = {2019},
  month = sep,
  journal = {IEEE Transactions on Intelligent Transportation Systems},
  volume = {20},
  number = {9},
  pages = {3496--3512},
  issn = {1524-9050},
  doi = {10.1109/TITS.2018.2876905}
}

@article{jiaDisturbanceAdaptiveDesignVANETenabledVehicle2014a,
  title = {A {{Disturbance-Adaptive Design}} for {{VANET-enabled Vehicle Platoon}}},
  author = {Jia, Dongyao and Lu, Kejie and Wang, Jianping},
  year = {2014},
  month = feb,
  journal = {IEEE Transactions on Vehicular Technology},
  volume = {63},
  number = {2},
  pages = {527--539},
  issn = {0018-9545},
  doi = {10.1109/TVT.2013.2280721}
}

@article{jiaNetworkConnectivityPlatoonBasedVehicular2014a,
  title = {On the {{Network Connectivity}} of {{Platoon-Based Vehicular Cyber-Physical Systems}}},
  author = {Jia, Dongyao and Lu, Kejie and Wang, Jianping},
  year = {2014},
  month = mar,
  journal = {Transportation Research Part C-Emerging Technologies},
  volume = {40},
  pages = {215--230},
  issn = {0968-090X},
  doi = {10.1016/j.trc.2013.10.006},
  langid = {english}
}

@article{jiangEcoApproachingIsolatedSignalized2017a,
  title = {Eco {{Approaching}} at an {{Isolated Signalized Intersection}} under {{Partially Connected}} and {{Automated Vehicles Environment}}},
  author = {Jiang, Huifu and Hu, Jia and An, Shi and Wang, Meng and Park, Byungkyu Brian},
  year = {2017},
  month = jun,
  journal = {Transportation Research Part C-Emerging Technologies},
  volume = {79},
  pages = {290--307},
  issn = {0968-090X},
  doi = {10.1016/j.trc.2017.04.001},
  langid = {english}
}

@article{jiangTrafficExperimentRevealsNature2014,
  title = {Traffic Experiment Reveals the Nature of Car-Following},
  author = {Jiang, Rui and Hu, Mao-Bin and Zhang, H. M. and Gao, Zi-You and Jia, Bin and Wu, Qing-Song and Wang, Bing and Yang, Ming},
  year = {2014},
  month = apr,
  journal = {Plos One},
  volume = {9},
  number = {4},
  issn = {1932-6203},
  doi = {10.1371/journal.pone.0094351},
  langid = {english}
}

@article{jin_reducing_2014,
  title = {Reducing the Error Accumulation in Car-Following Models Calibrated with Vehicle Trajectory Data},
  author = {Jin, Peter J. and Yang, Da and Ran, Bin},
  year = {2014},
  month = feb,
  journal = {IEEE Transactions on Intelligent Transportation Systems},
  volume = {15},
  number = {1},
  pages = {148--157},
  issn = {1524-9050},
  doi = {10.1109/TITS.2013.2273872}
}

@article{jinEquivalenceContinuumCarFollowingModels2016a,
  title = {On the {{Equivalence}} between {{Continuum}} and {{Car-Following Models}} of {{Traffic Flow}}},
  author = {Jin, Wen-Long},
  year = {2016},
  month = nov,
  journal = {Transportation Research Part B-Methodological},
  volume = {93},
  number = {A},
  pages = {543--559},
  issn = {0191-2615},
  doi = {10.1016/j.trb.2016.08.007}
}

@article{joererVehicularNetworkingPerspectiveEstimating2014a,
  title = {A {{Vehicular Networking Perspective}} on {{Estimating Vehicle Collision Probability}} at {{Intersections}}},
  author = {Joerer, Stefan and Segata, Michele and Bloessl, Bastian and Lo Cigno, Renato and Sommer, Christoph and Dressler, Falko},
  year = {2014},
  month = may,
  journal = {IEEE Transactions on Vehicular Technology},
  volume = {63},
  number = {4},
  pages = {1802--1812},
  issn = {0018-9545},
  doi = {10.1109/TVT.2013.2287343}
}

@article{juPredictiveCruiseControllerElectric2023a,
  title = {Predictive {{Cruise Controller}} for {{Electric Vehicle}} to {{Save Energy}} and {{Extend Battery Lifetime}}},
  author = {Ju, Fei and Murgovski, Nikolce and Zhuang, Weichao and Wang, Qun and Wang, Liangmo},
  year = {2023},
  month = jan,
  journal = {IEEE Transactions on Vehicular Technology},
  volume = {72},
  number = {1},
  pages = {469--482},
  issn = {0018-9545},
  doi = {10.1109/TVT.2022.3208932}
}

@article{kamal_efficient_2016,
  title = {Efficient Driving on Multilane Roads under a Connected Vehicle Environment},
  author = {Kamal, Md Abdus Samad and Taguchi, Shun and Yoshimura, Takayoshi},
  year = {2016},
  month = sep,
  journal = {IEEE Transactions on Intelligent Transportation Systems},
  volume = {17},
  number = {9},
  pages = {2541--2551},
  issn = {1524-9050},
  doi = {10.1109/TITS.2016.2519526}
}

@article{kamalLookAheadDrivingSchemesEfficient2022a,
  title = {Look-{{Ahead Driving Schemes}} for {{Efficient Control}} of {{Automated Vehicles}} on {{Urban Roads}}},
  author = {Kamal, Md Abdus Samad and Hashikura, Kotaro and Hayakawa, Tomohisa and Yamada, Kou and Imura, Jun-ichi},
  year = {2022},
  month = feb,
  journal = {IEEE Transactions on Vehicular Technology},
  volume = {71},
  number = {2},
  pages = {1280--1292},
  issn = {0018-9545},
  doi = {10.1109/TVT.2021.3132936},
  langid = {american}
}

@article{kangTrajectoryBasedEmbeddingRandomCoefficients2023b,
  title = {Trajectory-{{Based Embedding}} for {{Random Coefficients}} of a {{Theory-Based Car-Following Model}}},
  author = {Kang, Yeseul and Kim, Gyeongjun and Jeong, Seungyun and Sohn, Keemin},
  year = {2023},
  month = jul,
  journal = {Transportation Research Part C-Emerging Technologies},
  volume = {152},
  issn = {0968-090X},
  doi = {10.1016/j.trc.2023.104183},
  langid = {english}
}

@article{kerbel_shared_2024,
  title = {Shared {{Learning}} of {{Powertrain Control Policies}} for {{Vehicle Fleets}}},
  author = {Kerbel, Lindsey and Ayalew, Beshah and Ivanco, Andrej},
  year = {2024},
  month = jul,
  journal = {Applied Energy},
  volume = {365},
  issn = {0306-2619},
  doi = {10.1016/j.apenergy.2024.123217},
  langid = {american}
}

@article{khondakerVariableSpeedLimitMicroscopic2015a,
  title = {Variable {{Speed Limit}}: {{A Microscopic Analysis}} in a {{Connected Vehicle Environment}}},
  author = {Khondaker, Bidoura and Kattan, Lina},
  year = {2015},
  month = sep,
  journal = {Transportation Research Part C-Emerging Technologies},
  volume = {58},
  number = {A},
  pages = {146--159},
  issn = {0968-090X},
  doi = {10.1016/j.trc.2015.07.014},
  langid = {english}
}

@article{li_automated_2023,
  title = {Automated Vehicle Identification Based on Car-Following Data with Machine Learning},
  author = {Li, Qianwen and Li, Xiaopeng and Yao, Handong and Liang, Zhaohui and Xie, Weijun},
  year = {2023},
  month = dec,
  journal = {IEEE Transactions on Intelligent Transportation Systems},
  volume = {24},
  number = {12},
  pages = {13893--13902},
  issn = {1524-9050},
  doi = {10.1109/TITS.2023.3304607}
}

@article{li_bounded_2024,
  title = {A Bounded Rationality-Aware Car-Following Strategy for Alleviating Cut-in Events and Traffic Disturbances in Traffic Oscillations},
  author = {Li, Meng and Li, Zhibin and Wang, Bingtong and Wang, Shunchao},
  year = {2024},
  month = jul,
  journal = {IEEE Transactions on Intelligent Transportation Systems},
  issn = {1524-9050},
  doi = {10.1109/TITS.2024.3424447}
}

@article{li_deep_2022,
  title = {Deep {{Reinforcement Learning}} and {{Reward Shaping Based Eco-Driving Control}} for {{Automated HEVs}} among {{Signalized Intersections}}},
  author = {Li, Jie and Wu, Xiaodong and Xu, Min and Liu, Yonggang},
  year = {2022},
  month = jul,
  journal = {Energy},
  volume = {251},
  issn = {0360-5442},
  doi = {10.1016/j.energy.2022.123924},
  langid = {american}
}

@article{li_developing_2022,
  title = {Developing {{Dynamic Speed Limit Strategies}} for {{Mixed Traffic Flow}} to {{Reduce Collision Risks}} at {{Freeway Bottlenecks}}},
  author = {Li, Ye and Pan, Bing and Xing, Lu and Yang, Min and Dai, Jianjun},
  year = {2022},
  month = sep,
  journal = {Accident Analysis and Prevention},
  volume = {175},
  issn = {0001-4575},
  doi = {10.1016/j.aap.2022.106781},
  langid = {american}
}

@article{li_developing_2023,
  title = {Developing a Dynamic Speed Control System for Mixed Traffic Flow to Reduce Collision Risks near Freeway Bottlenecks},
  author = {Li, Ye and Pan, Bin and Chen, Zhibin and Xing, Lu},
  year = {2023},
  month = nov,
  journal = {IEEE Transactions on Intelligent Transportation Systems},
  volume = {24},
  number = {11},
  pages = {12560--12581},
  issn = {1524-9050},
  doi = {10.1109/TITS.2023.3287269}
}

@article{li_effect_2015,
  title = {Effect of {{Pulse-and-Glide Strategy}} on {{Traffic Flow}} for a {{Platoon}} of {{Mixed Automated}} and {{Manually Driven Vehicles}}},
  author = {Li, Shengbo Eben and Deng, Kun and Zheng, Yang and Peng, Huei},
  year = {2015},
  month = nov,
  journal = {Computer-Aided Civil and Infrastructure Engineering},
  volume = {30},
  number = {11},
  pages = {892--905},
  issn = {1093-9687},
  doi = {10.1111/mice.12168},
  langid = {american}
}

@article{li_evaluating_2017,
  title = {Evaluating {{Impacts}} of {{Different Longitudinal Driver Assistance Systems}} on {{Reducing Multi-Vehicle Rear-End Crashes}} during {{Small-Scale Inclement Weather}}},
  author = {Li, Ye and Xing, Lu and Wang, Wei and Wang, Hao and Dong, Changyin and Liu, Shanwen},
  year = {2017},
  month = oct,
  journal = {Accident Analysis and Prevention},
  volume = {107},
  pages = {63--76},
  issn = {0001-4575},
  doi = {10.1016/j.aap.2017.07.014},
  langid = {american}
}

@article{li_evaluation_2017,
  title = {Evaluation of the {{Impacts}} of {{Cooperative Adaptive Cruise Control}} on {{Reducing Rear-End Collision Risks}} on {{Freeways}}},
  author = {Li, Ye and Wang, Hao and Wang, Wei and Xing, Lu and Liu, Shanwen and Wei, Xueyan},
  year = {2017},
  month = jan,
  journal = {Accident Analysis and Prevention},
  volume = {98},
  pages = {87--95},
  issn = {0001-4575},
  doi = {10.1016/j.aap.2016.09.015},
  langid = {american}
}

@article{li_integrated_2017,
  title = {Integrated Cooperative Adaptive Cruise and Variable Speed Limit Controls for Reducing Rear-End Collision Risks near Freeway Bottlenecks Based on Micro-Simulations},
  author = {Li, Ye and Xu, Chengcheng and Xing, Lu and Wang, Wei},
  year = {2017},
  month = nov,
  journal = {IEEE Transactions on Intelligent Transportation Systems},
  volume = {18},
  number = {11},
  pages = {3157--3167},
  issn = {1524-9050},
  doi = {10.1109/TITS.2017.2682193}
}

@article{li_scegene_2022,
  title = {{{SceGene}}: {{Bio-inspired}} Traffic Scenario Generation for Autonomous Driving Testing},
  author = {Li, Ao and Chen, Shitao and Sun, Liting and Zheng, Nanning and Tomizuka, Masayoshi and Zhan, Wei},
  year = {2022},
  month = sep,
  journal = {IEEE Transactions on Intelligent Transportation Systems},
  volume = {23},
  number = {9},
  pages = {14859--14874},
  issn = {1524-9050},
  doi = {10.1109/TITS.2021.3134661}
}

@article{li_unraveling_2024,
  title = {Unraveling the Platoon Dynamics of Connected Automated Vehicles with Information Errors},
  author = {Li, Shihao and Zhou, Bojian and Xu, Min},
  year = {2024},
  month = aug,
  journal = {IEEE Transactions on Intelligent Transportation Systems},
  issn = {1524-9050},
  doi = {10.1109/TITS.2024.3440955}
}

@article{liang_equitable_2020,
  title = {An {{Equitable Traffic Signal Control Scheme}} at {{Isolated Signalized Intersections Using Connected Vehicle Technology}}},
  author = {Liang, Xiao (Joyce) and Guler, S. Ilgin and Gayah, Vikash V.},
  year = {2020},
  month = jan,
  journal = {Transportation Research Part C-Emerging Technologies},
  volume = {110},
  pages = {81--97},
  issn = {0968-090X},
  doi = {10.1016/j.trc.2019.11.005},
  langid = {american}
}

@article{liEcoDrivingSystemElectricVehicles2018a,
  title = {An {{Eco-Driving System}} for {{Electric Vehicles}} with {{Signal Control}} under {{V2X Environment}}},
  author = {Li, Ming and Wu, Xinkai and He, Xiaozheng and Yu, Guizhen and Wang, Yunpeng},
  year = {2018},
  month = aug,
  journal = {Transportation Research Part C-Emerging Technologies},
  volume = {93},
  pages = {335--350},
  issn = {0968-090X},
  doi = {10.1016/j.trc.2018.06.002},
  langid = {english}
}

@article{liGlobalOptimizationAlgorithmTrajectory2016b,
  title = {A {{Global Optimization Algorithm}} for {{Trajectory Data Based Car-Following Model Calibration}}},
  author = {Li, Li and Chen, Xiqun (Micheal) and Zhang, Lei},
  year = {2016},
  month = jul,
  journal = {Transportation Research Part C-Emerging Technologies},
  volume = {68},
  pages = {311--332},
  issn = {0968-090X},
  doi = {10.1016/j.trc.2016.04.011}
}

@article{lindorfer_modeling_2018,
  title = {Modeling the Imperfect Driver: {{Incorporating}} Human Factors in a Microscopic Traffic Model},
  author = {Lindorfer, Manuel and Mecklenbraeuker, Christoph F. and Ostermayer, Gerald},
  year = {2018},
  month = sep,
  journal = {IEEE Transactions on Intelligent Transportation Systems},
  volume = {19},
  number = {9},
  pages = {2856--2870},
  issn = {1524-9050},
  doi = {10.1109/TITS.2017.2765694}
}

@article{liu_connected_2022,
  title = {Connected and {{Automated Vehicle Platoon Maintenance}} under {{Communication Failures}}},
  author = {Liu, Runkun and Ren, Yilong and Yu, Haiyang and Li, Zhiheng and Jiang, Han},
  year = {2022},
  month = jun,
  journal = {Vehicular Communications},
  volume = {35},
  issn = {2214-2096},
  doi = {10.1016/j.vehcom.2022.100467},
  langid = {american}
}

@article{liu_interaction-aware_2023,
  title = {Interaction-Aware Trajectory Prediction and Planning for Autonomous Vehicles in Forced Merge Scenarios},
  author = {Liu, Kaiwen and Li, Nan and Tseng, H. Eric and Kolmanovsky, Ilya and Girard, Anouck},
  year = {2023},
  month = jan,
  journal = {IEEE Transactions on Intelligent Transportation Systems},
  volume = {24},
  number = {1},
  pages = {474--488},
  issn = {1524-9050},
  doi = {10.1109/TITS.2022.3216792}
}

@article{liu_joint_2022,
  title = {Joint {{Communication}} and {{Computation Resource Scheduling}} of a {{UAV-assisted Mobile Edge Computing System}} for {{Platooning Vehicles}}},
  author = {Liu, Yang and Zhou, Jianshan and Tian, Daxin and Sheng, Zhengguo and Duan, Xuting and Qu, Guixian and Leung, Victor C. M.},
  year = {2022},
  month = jul,
  journal = {IEEE Transactions on Intelligent Transportation Systems},
  volume = {23},
  number = {7},
  pages = {8435--8450},
  issn = {1524-9050},
  doi = {10.1109/TITS.2021.3082539},
  langid = {american}
}

@article{liu_potential_2023,
  title = {Potential Game-Based Decision-Making for Autonomous Driving},
  author = {Liu, Mushuang and Kolmanovsky, Ilya and Tseng, H. Eric and Huang, Suzhou and Filev, Dimitar and Girard, Anouck},
  year = {2023},
  month = aug,
  journal = {IEEE Transactions on Intelligent Transportation Systems},
  volume = {24},
  number = {8},
  pages = {8014--8027},
  issn = {1524-9050},
  doi = {10.1109/TITS.2023.3264665}
}

@article{liu_semantic_2023,
  title = {Semantic Traffic Law Adaptive Decision-Making for Self-Driving Vehicles},
  author = {Liu, Jiaxin and Wang, Hong and Cao, Zhong and Yu, Wenhao and Zhao, Chengxiang and Zhao, Ding and Yang, Diange and Li, Jun},
  year = {2023},
  month = dec,
  journal = {IEEE Transactions on Intelligent Transportation Systems},
  volume = {24},
  number = {12},
  pages = {14858--14872},
  issn = {1524-9050},
  doi = {10.1109/TITS.2023.3294579}
}

@article{liuMultistepCooperativeLaneChange2024a,
  title = {A {{Multistep Cooperative Lane Change Strategy}} for {{Connected}} and {{Autonomous Vehicle Platoons Departing}} from {{Dedicated Lanes}}},
  author = {Liu, Chenglin and Liu, Zhiguang and Xu, Zhigang and Li, Xiaopeng},
  year = {2024},
  month = aug,
  journal = {Transportation Research Part C-Emerging Technologies},
  volume = {165},
  issn = {0968-090X},
  doi = {10.1016/j.trc.2024.104720},
  langid = {english}
}

@article{liuQuantileRegressionPhysicsInformedDeepLearning2023c,
  title = {A {{Quantile-Regression Physics-Informed Deep Learning}} for {{Car-Following Model}}},
  author = {Liu, Jiaxin and Jiang, Rui and Zhao, Jiandong and Shen, Wei},
  year = {2023},
  month = sep,
  journal = {Transportation Research Part C-Emerging Technologies},
  volume = {154},
  issn = {0968-090X},
  doi = {10.1016/j.trc.2023.104275},
  langid = {english}
}

@article{liuSpatiotemporalTrajectoryPlanningAutonomous2024a,
  title = {Spatiotemporal {{Trajectory Planning}} for {{Autonomous Vehicle Based}} on {{Reachable Set}} and {{Iterative LQR}}},
  author = {Liu, Yiping and Pei, Xiaofei and Zhou, Honglong and Guo, Xuexun},
  year = {2024},
  month = aug,
  journal = {IEEE Transactions on Vehicular Technology},
  volume = {73},
  number = {8},
  pages = {10932--10947},
  issn = {0018-9545},
  doi = {10.1109/TVT.2024.3371184}
}

@article{longBiScaleCarFollowingModelCalibration2024a,
  title = {Bi-{{Scale Car-Following Model Calibration Based}} on {{Corridor-Level Trajectory}}},
  author = {Long, Keke and Shi, Haotian and Chen, Zhiwei and Liang, Zhaohui and Li, Xiaopeng and {de Souza}, Felipe},
  year = {2024},
  month = jun,
  journal = {Transportation Research Part E-Logistics and Transportation Review},
  volume = {186},
  issn = {1366-5545},
  doi = {10.1016/j.tre.2024.103497}
}

@article{luLearningDriverSpecificBehaviorOvertaking2018a,
  title = {Learning {{Driver-Specific Behavior}} for {{Overtaking}}: {{A Combined Learning Framework}}},
  author = {Lu, Chao and Wang, Huaji and Lv, Chen and Gong, Jianwei and Xi, Junqiang and Cao, Dongpu},
  year = {2018},
  month = aug,
  journal = {IEEE Transactions on Vehicular Technology},
  volume = {67},
  number = {8},
  pages = {6788--6802},
  issn = {0018-9545},
  doi = {10.1109/TVT.2018.2820002}
}

@article{luo_modeling_2024,
  title = {Modeling and {{Analyzing Self-Resistance}} of {{Connected Automated Vehicular Platoons}} under {{Different Cyberattack Injection Modes}}},
  author = {Luo, Dongyu and Wang, Jiangfeng and Wang, Yu and Dong, Jiakuan},
  year = {2024},
  month = apr,
  journal = {Accident Analysis and Prevention},
  volume = {198},
  issn = {0001-4575},
  doi = {10.1016/j.aap.2024.107494},
  langid = {american}
}

@article{luTD3LVSLLaneLevelVariableSpeed2023a,
  title = {{{TD3LVSL}}: {{A Lane-Level Variable Speed Limit Approach Based}} on {{Twin Delayed Deep Deterministic Policy Gradient}} in a {{Connected Automated Vehicle Environment}}},
  author = {Lu, Wenqi and Yi, Ziwei and Gu, Yuanli and Rui, Yikang and Ran, Bin},
  year = {2023},
  month = aug,
  journal = {Transportation Research Part C-Emerging Technologies},
  volume = {153},
  issn = {0968-090X},
  doi = {10.1016/j.trc.2023.104221},
  langid = {english}
}

@article{ma_energetic_2022,
  title = {Energetic {{Impacts Evaluation}} of {{Eco-Driving}} on {{Mixed Traffic}} with {{Driver Behavioral Diversity}}},
  author = {Ma, Yao and Wang, Junmin},
  year = {2022},
  month = apr,
  journal = {IEEE Transactions on Intelligent Transportation Systems},
  volume = {23},
  number = {4},
  pages = {3406--3417},
  issn = {1524-9050},
  doi = {10.1109/TITS.2020.3036326},
  langid = {american}
}

@article{mahajan_improving_2024,
  title = {Improving Traffic Efficiency with Lane Guidance Based on Desired Speeds},
  author = {Mahajan, Niharika and Hegyi, Andreas and Hoogendoorn, Serge P. and {van Arem}, Bart},
  year = {2024},
  month = mar,
  journal = {IEEE Transactions on Intelligent Transportation Systems},
  volume = {25},
  number = {3},
  pages = {2222--2234},
  issn = {1524-9050},
  doi = {10.1109/TITS.2023.3340881}
}

@article{makridis_response_2020,
  title = {Response {{Time}} and {{Time Headway}} of an {{Adaptive Cruise Control}}. {{An Empirical Characterization}} and {{Potential Impacts}} on {{Road Capacity}}},
  author = {Makridis, Michail and Mattas, Konstantinos and Ciuffo, Biagio},
  year = {2020},
  month = apr,
  journal = {IEEE Transactions on Intelligent Transportation Systems},
  volume = {21},
  number = {4},
  pages = {1677--1686},
  issn = {1524-9050},
  doi = {10.1109/TITS.2019.2948646},
  langid = {american}
}

@article{manolis_real_2020,
  title = {Real {{Time Adaptive Cruise Control Strategy}} for {{Motorways}}},
  author = {Manolis, Diamantis and Spiliopoulou, Anastasia and Vandorou, Foteini and Papageorgiou, Markos},
  year = {2020},
  month = jun,
  journal = {Transportation Research Part C-Emerging Technologies},
  volume = {115},
  issn = {0968-090X},
  doi = {10.1016/j.trc.2020.102617},
  langid = {american}
}

@article{maPersonalizedDrivingBehaviorsFuel2022a,
  title = {Personalized {{Driving Behaviors}} and {{Fuel Economy}} over {{Realistic Commute Traffic}}: {{Modeling}}, {{Correlation}}, and {{Prediction}}},
  author = {Ma, Yao and Wang, Junmin},
  year = {2022},
  month = jul,
  journal = {IEEE Transactions on Vehicular Technology},
  volume = {71},
  number = {7},
  pages = {7084--7094},
  issn = {0018-9545},
  doi = {10.1109/TVT.2022.3171165}
}

@article{maProgressiveVirtualRiskBasedVehicle2024a,
  title = {Progressive {{Virtual Risk-Based Vehicle Trajectory Optimization}} in {{Mixed Traffic Flow}}},
  author = {Ma, Haozhan and Qian, Chen and Li, Linheng and Qu, Xu and Ran, Bin},
  year = {2024},
  month = aug,
  journal = {Transportation Research Part C-Emerging Technologies},
  volume = {165},
  issn = {0968-090X},
  doi = {10.1016/j.trc.2024.104701},
  langid = {english}
}

@article{medina_cooperative_2018,
  title = {Cooperative Intersection Control Based on Virtual Platooning},
  author = {Medina, Alejandro Ivan Morales and {van de Wouw}, Nathan and Nijmeijer, Henk},
  year = {2018},
  month = jun,
  journal = {IEEE Transactions on Intelligent Transportation Systems},
  volume = {19},
  number = {6},
  pages = {1727--1740},
  issn = {1524-9050},
  doi = {10.1109/TITS.2017.2735628}
}

@article{milanesModelingCooperativeAutonomousAdaptive2014b,
  title = {Modeling {{Cooperative}} and {{Autonomous Adaptive Cruise Control Dynamic Responses Using Experimental Data}}},
  author = {Milanes, Vicente and Shladover, Steven E.},
  year = {2014},
  month = nov,
  journal = {Transportation Research Part C-Emerging Technologies},
  volume = {48},
  pages = {285--300},
  issn = {0968-090X},
  doi = {10.1016/j.trc.2014.09.001},
  langid = {english}
}

@article{montaninoHomogeneousHeterogeneousTrafficFlows2021b,
  title = {From {{Homogeneous}} to {{Heterogeneous Traffic Flows}}: {\textexclamdown}i{\textquestiondown}{{L}}{\textexclamdown}/{{I}}{\textquestiondown}{\textexclamdown}sub{\textquestiondown}p{\textexclamdown}/{{Sub}}{\textquestiondown} {{String Stability}} under {{Uncertain Model Parameters}}},
  author = {Montanino, Marcello and Monteil, Julien and Punzo, Vincenzo},
  year = {2021},
  month = apr,
  journal = {Transportation Research Part B-Methodological},
  volume = {146},
  pages = {136--154},
  issn = {0191-2615},
  doi = {10.1016/j.trb.2021.01.009},
  langid = {american}
}

@article{montaninoStringStabilityMixedHeterogeneous2021b,
  title = {On {{String Stability}} of a {{Mixed}} and {{Heterogeneous Traffic Flow}}: {{A Unifying Modelling Framework}}},
  author = {Montanino, Marcello and Punzo, Vincenzo},
  year = {2021},
  month = feb,
  journal = {Transportation Research Part B-Methodological},
  volume = {144},
  pages = {133--154},
  issn = {0191-2615},
  doi = {10.1016/j.trb.2020.11.009},
  langid = {american}
}

@article{montaninoTrajectoryDataReconstructionSimulationBased2015b,
  title = {Trajectory {{Data Reconstruction}} and {{Simulation-Based Validation}} against {{Macroscopic Traffic Patterns}}},
  author = {Montanino, Marcello and Punzo, Vincenzo},
  year = {2015},
  month = oct,
  journal = {Transportation Research Part B-Methodological},
  volume = {80},
  pages = {82--106},
  issn = {0191-2615},
  doi = {10.1016/j.trb.2015.06.010}
}

@article{monteil_linear_2014,
  title = {Linear and Weakly Nonlinear Stability Analyses of Cooperative Car-Following Models},
  author = {Monteil, Julien and Billot, Romain and Sau, Jacques and El Faouzi, Nour-Eddin},
  year = {2014},
  month = oct,
  journal = {IEEE Transactions on Intelligent Transportation Systems},
  volume = {15},
  number = {5},
  pages = {2001--2013},
  issn = {1524-9050},
  doi = {10.1109/TITS.2014.2308435}
}

@article{moPhysicsInformedDeepLearningParadigm2021c,
  title = {A {{Physics-Informed Deep Learning Paradigm}} for {{Car-Following Models}}},
  author = {Mo, Zhaobin and Shi, Rongye and Di, Xuan},
  year = {2021},
  month = sep,
  journal = {Transportation Research Part C-Emerging Technologies},
  volume = {130},
  issn = {0968-090X},
  doi = {10.1016/j.trc.2021.103240},
  langid = {english}
}

@article{jiangFullVelocityDifferenceModel2001,
  title = {Full Velocity Difference Model for a Car-Following Theory},
  author = {Jiang, Rui and Wu, Qingsong and Zhu, Zuojin},
  year = {2001},
  month = jun,
  journal = {Physical Review E},
  volume = {64},
  number = {1},
  pages = {017101},
  issn = {1063-651X, 1095-3787},
  doi = {10.1103/PhysRevE.64.017101},
  urldate = {2022-05-11},
  langid = {english}
}

@article{mullakkal-babu_comparative_2022,
  title = {Comparative {{Safety Assessment}} of {{Automated Driving Strategies}} at {{Highway Merges}} in {{Mixed Traffic}}},
  author = {{Mullakkal-Babu}, Freddy Antony and Wang, Meng and {van Arem}, Bart and Happee, Riender},
  year = {2022},
  month = apr,
  journal = {IEEE Transactions on Intelligent Transportation Systems},
  volume = {23},
  number = {4},
  pages = {3626--3639},
  issn = {1524-9050},
  doi = {10.1109/TITS.2020.3038866},
  langid = {american}
}

@article{mullakkal-babu_hybrid_2021,
  title = {A {{Hybrid Submicroscopic-Microscopic Traffic Flow Simulation Framework}}},
  author = {{Mullakkal-Babu}, Freddy Antony and Wang, Meng and {van Arem}, Bart and Shyrokau, Barys and Happee, Riender},
  year = {2021},
  month = jun,
  journal = {IEEE Transactions on Intelligent Transportation Systems},
  volume = {22},
  number = {6},
  pages = {3430--3443},
  issn = {1524-9050},
  doi = {10.1109/TITS.2020.2990376},
  langid = {american}
}

@article{ossenHeterogeneityCarFollowingBehaviorTheory2011b,
  title = {Heterogeneity in {{Car-Following Behavior}}: {{Theory}} and {{Empirics}}},
  author = {Ossen, Saskia and Hoogendoorn, Serge P.},
  year = {2011},
  month = apr,
  journal = {Transportation Research Part C-Emerging Technologies},
  volume = {19},
  number = {2, SI},
  pages = {182--195},
  issn = {0968-090X},
  doi = {10.1016/j.trc.2010.05.006},
  langid = {english}
}

@article{othman_analysis_2022,
  title = {Analysis of the Impact of Variable Speed Limits on Environmental Sustainability and Traffic Performance in Urban Networks},
  author = {Othman, Bassel and De Nunzio, Giovanni and Di Domenico, Domenico and {Canudas-de-Wit}, Carlos},
  year = {2022},
  month = nov,
  journal = {IEEE Transactions on Intelligent Transportation Systems},
  volume = {23},
  number = {11},
  pages = {21766--21776},
  issn = {1524-9050},
  doi = {10.1109/TITS.2022.3192129}
}

@article{punzo_we_2015,
  title = {Do We Really Need to Calibrate All the Parameters? {{Variance-based}} Sensitivity Analysis to Simplify Microscopic Traffic Flow Models},
  author = {Punzo, Vincenzo and Montanino, Marcello and Ciuffo, Biagio},
  year = {2015},
  month = feb,
  journal = {IEEE Transactions on Intelligent Transportation Systems},
  volume = {16},
  number = {1},
  pages = {184--193},
  issn = {1524-9050},
  doi = {10.1109/TITS.2014.2331453}
}

@article{punzoCalibrationCarFollowingDynamicsAutomated2021,
  title = {About {{Calibration}} of {{Car-Following Dynamics}} of {{Automated}} and {{Human-Driven Vehicles}}: {{Methodology}}, {{Guidelines}} and {{Codes}}},
  author = {Punzo, Vincenzo and Zheng, Zuduo and Montanino, Marcello},
  year = {2021},
  month = jul,
  journal = {Transportation Research Part C-Emerging Technologies},
  volume = {128},
  issn = {0968-090X},
  doi = {10.1016/j.trc.2021.103165},
  langid = {english}
}

@article{qin_analytical_2021,
  title = {Analytical {{Framework}} of {{String Stability}} of {{Connected}} and {{Autonomous Platoons}} with {{Electronic Throttle Angle Feedback}}},
  author = {Qin, Yanyan and Wang, Hao},
  year = {2021},
  month = jan,
  journal = {Transportmetrica a-Transport Science},
  volume = {17},
  number = {1, SI},
  pages = {59--80},
  issn = {2324-9935},
  doi = {10.1080/23249935.2018.1518964},
  langid = {american}
}

@article{qinStabilityAnalysisConnectedVehicles2023a,
  title = {Stability {{Analysis}} and {{Connected Vehicles Management}} for {{Mixed Traffic Flow}} with {{Platoons}} of {{Connected Automated Vehicles}}},
  author = {Qin, Yanyan and Luo, Qinzhong and Wang, Hua},
  year = {2023},
  month = dec,
  journal = {Transportation Research Part C-Emerging Technologies},
  volume = {157},
  issn = {0968-090X},
  doi = {10.1016/j.trc.2023.104370}
}

@article{rahman_improving_2015,
  title = {Improving the Efficacy of Car-Following Models with a New Stochastic Parameter Estimation and Calibration Method},
  author = {Rahman, Mizanur and Chowdhury, Mashrur and Khan, Taufiquar and Bhavsar, Parth},
  year = {2015},
  month = oct,
  journal = {IEEE Transactions on Intelligent Transportation Systems},
  volume = {16},
  number = {5},
  pages = {2687--2699},
  issn = {1524-9050},
  doi = {10.1109/TITS.2015.2420542}
}

@article{rahman_longitudinal_2018,
  title = {Longitudinal {{Safety Evaluation}} of {{Connected Vehicles}}' {{Platooning}} on {{Expressways}}},
  author = {Rahman, Md Sharikur and {Abdel-Aty}, Mohamed},
  year = {2018},
  month = aug,
  journal = {Accident Analysis and Prevention},
  volume = {117},
  pages = {381--391},
  issn = {0001-4575},
  doi = {10.1016/j.aap.2017.12.012},
  langid = {american}
}

@article{saifuzzamanIncorporatingHumanFactorsCarFollowingModels2014,
  title = {Incorporating {{Human-Factors}} in {{Car-Following Models}}: {{A Review}} of {{Recent Developments}} and {{Research Needs}}},
  author = {Saifuzzaman, Mohammad and Zheng, Zuduo},
  year = {2014},
  month = nov,
  journal = {Transportation Research Part C-Emerging Technologies},
  volume = {48},
  pages = {379--403},
  issn = {0968-090X},
  doi = {10.1016/j.trc.2014.09.008}
}

@article{schakel_improving_2014,
  title = {Improving Traffic Flow Efficiency by In-Car Advice on Lane, Speed, and Headway},
  author = {Schakel, Wouter J. and {van Arem}, Bart},
  year = {2014},
  month = aug,
  journal = {IEEE Transactions on Intelligent Transportation Systems},
  volume = {15},
  number = {4},
  pages = {1597--1606},
  issn = {1524-9050},
  doi = {10.1109/TITS.2014.2303577}
}

@article{scheel_recurrent_2022,
  title = {Recurrent {{Models}} for {{Lane Change Prediction}} and {{Situation Assessment}}},
  author = {Scheel, Oliver and Nagaraja, Naveen Shankar and Schwarz, Loren and Navab, Nassir and Tombari, Federico},
  year = {2022},
  month = oct,
  journal = {IEEE Transactions on Intelligent Transportation Systems},
  volume = {23},
  number = {10},
  pages = {17284--17300},
  issn = {1524-9050},
  doi = {10.1109/TITS.2022.3163353},
  langid = {american}
}

@article{schrader_global_2024,
  title = {A Global Sensitivity Analysis of Traffic Microsimulation Input Parameters on Performance Metrics},
  author = {Schrader, Maxwell and Bittle, Joshua},
  year = {2024},
  month = mar,
  journal = {IEEE Transactions on Intelligent Transportation Systems},
  issn = {1524-9050},
  doi = {10.1109/TITS.2024.3372334}
}

@article{segataAutomaticEmergencyBrakingRealistic2013a,
  title = {Automatic {{Emergency Braking}}: {{Realistic Analysis}} of {{Car Dynamics}} and {{Network Performance}}},
  author = {Segata, Michele and Lo Cigno, Renato},
  year = {2013},
  month = nov,
  journal = {IEEE Transactions on Vehicular Technology},
  volume = {62},
  number = {9},
  pages = {4150--4161},
  issn = {0018-9545},
  doi = {10.1109/TVT.2013.2277802}
}

@article{shang_impacts_2021,
  title = {Impacts of {{Commercially Available Adaptive Cruise Control Vehicles}} on {{Highway Stability}} and {{Throughput}}},
  author = {Shang, Mingfeng and Stern, Raphael E.},
  year = {2021},
  month = jan,
  journal = {Transportation Research Part C-Emerging Technologies},
  volume = {122},
  issn = {0968-090X},
  doi = {10.1016/j.trc.2020.102897},
  langid = {american}
}

@article{sharath_enhanced_2020,
  title = {Enhanced {{Intelligent Driver Model}} for {{Two-Dimensional Motion Planning}} in {{Mixed Traffic}}},
  author = {Sharath, M. N. and Velaga, Nagendra R.},
  year = {2020},
  month = nov,
  journal = {Transportation Research Part C-Emerging Technologies},
  volume = {120},
  issn = {0968-090X},
  doi = {10.1016/j.trc.2020.102780},
  langid = {american}
}

@article{sharmaMoreAlwaysBetterImpact2019b,
  title = {Is {{More Always Better}}? {{The Impact}} of {{Vehicular Trajectory Completeness}} on {{Car-Following Model Calibration}} and {{Validation}}},
  author = {Sharma, Anshuman and Zheng, Zuduo and Bhaskar, Ashish},
  year = {2019},
  month = feb,
  journal = {Transportation Research Part B-Methodological},
  volume = {120},
  pages = {49--75},
  issn = {0191-2615},
  doi = {10.1016/j.trb.2018.12.016},
  langid = {american}
}

@article{sheng_cooperation-aware_2023,
  title = {A Cooperation-Aware Lane Change Method for Automated Vehicles},
  author = {Sheng, Zihao and Liu, Lin and Xue, Shibei and Zhao, Dezong and Jiang, Min and Li, Dewei},
  year = {2023},
  month = mar,
  journal = {IEEE Transactions on Intelligent Transportation Systems},
  volume = {24},
  number = {3},
  pages = {3236--3251},
  issn = {1524-9050},
  doi = {10.1109/TITS.2022.3225875}
}

@article{shiConnectedAutomatedVehicleCooperative2021b,
  title = {Connected {{Automated Vehicle Cooperative Control}} with a {{Deep Reinforcement Learning Approach}} in a {{Mixed Traffic Environment}}},
  author = {Shi, Haotian and Zhou, Yang and Wu, Keshu and Wang, Xin and Lin, Yangxin and Ran, Bin},
  year = {2021},
  month = dec,
  journal = {Transportation Research Part C-Emerging Technologies},
  volume = {133},
  issn = {0968-090X},
  doi = {10.1016/j.trc.2021.103421}
}

@article{shiDeepReinforcementLearningBased2023b,
  title = {A {{Deep Reinforcement Learning Based Distributed Control Strategy}} for {{Connected Automated Vehicles}} in {{Mixed Traffic Platoon}}},
  author = {Shi, Haotian and Chen, Danjue and Zheng, Nan and Wang, Xin and Zhou, Yang and Ran, Bin},
  year = {2023},
  month = mar,
  journal = {Transportation Research Part C-Emerging Technologies},
  volume = {148},
  issn = {0968-090X},
  doi = {10.1016/j.trc.2023.104019}
}

@article{silgu_combined_2022,
  title = {Combined {{Control}} of {{Freeway Traffic Involving Cooperative Adaptive Cruise Controlled}} and {{Human Driven Vehicles Using Feedback Control}} through {{SUMO}}},
  author = {Silgu, Mehmet Ali and Erdagi, Ismet Goksad and Goksu, Gokhan and Celikoglu, Hilmi Berk},
  year = {2022},
  month = aug,
  journal = {IEEE Transactions on Intelligent Transportation Systems},
  volume = {23},
  number = {8},
  pages = {11011--11025},
  issn = {1524-9050},
  doi = {10.1109/TITS.2021.3098640},
  langid = {american}
}

@article{sohn_waterfilling_2023,
  title = {The ``Waterfilling Algorithm''-an Efficient Approach for Vehicle Velocity Planning with Varying Velocity Limits},
  author = {Sohn, Christian and Andert, Jakob},
  year = {2023},
  month = oct,
  journal = {IEEE Transactions on Intelligent Transportation Systems},
  volume = {24},
  number = {10},
  pages = {10477--10490},
  issn = {1524-9050},
  doi = {10.1109/TITS.2023.3286029}
}

@article{stebbinsCharacterisingGreenLightOptimal2017a,
  title = {Characterising {{Green Light Optimal Speed Advisory Trajectories}} for {{Platoon-Based Optimisation}}},
  author = {Stebbins, Simon and Hickman, Mark and Kim, Jiwon and Vu, Hai L.},
  year = {2017},
  month = sep,
  journal = {Transportation Research Part C-Emerging Technologies},
  volume = {82},
  pages = {43--62},
  issn = {0968-090X},
  doi = {10.1016/j.trc.2017.06.014},
  langid = {english}
}

@article{sun_adaptive_2022,
  title = {Adaptive {{Design}} of {{Experiments}} for {{Safety Evaluation}} of {{Automated Vehicles}}},
  author = {Sun, Jian and Zhou, Huajun and Xi, Haochen and Zhang, He and Tian, Ye},
  year = {2022},
  month = sep,
  journal = {IEEE Transactions on Intelligent Transportation Systems},
  volume = {23},
  number = {9},
  pages = {14497--14508},
  issn = {1524-9050},
  doi = {10.1109/TITS.2021.3130040},
  langid = {american}
}

@article{sun_scenario-based_2022,
  title = {Scenario-{{Based Test Automation}} for {{Highly Automated Vehicles}}: {{A Review}} and {{Paving}} the {{Way}} for {{Systematic Safety Assurance}}},
  author = {Sun, Jian and Zhang, He and Zhou, Huajun and Yu, Rongjie and Tian, Ye},
  year = {2022},
  month = sep,
  journal = {IEEE Transactions on Intelligent Transportation Systems},
  volume = {23},
  number = {9},
  pages = {14088--14103},
  issn = {1524-9050},
  doi = {10.1109/TITS.2021.3136353},
  langid = {american}
}

@article{sunberg_improving_2022,
  title = {Improving Automated Driving through {{POMDP}} Planning with Human Internal States},
  author = {Sunberg, Zachary and Kochenderfer, Mykel J.},
  year = {2022},
  month = nov,
  journal = {IEEE Transactions on Intelligent Transportation Systems},
  volume = {23},
  number = {11},
  pages = {20073--20083},
  issn = {1524-9050},
  doi = {10.1109/TITS.2022.3182687}
}

@article{sunRelationshipCarFollowingString2020b,
  title = {The {{Relationship}} between {{Car Following String Instability}} and {{Traffic Oscillations}} in {{Finite-Sized Platoons}} and {{Its Use}} in {{Easing Congestion}} via {{Connected}} and {{Automated Vehicles}} with {{IDM Based Controller}}},
  author = {Sun, Jie and Zheng, Zuduo and Sun, Jian},
  year = {2020},
  month = dec,
  journal = {Transportation Research Part B-Methodological},
  volume = {142},
  pages = {58--83},
  issn = {0191-2615},
  doi = {10.1016/j.trb.2020.10.004},
  langid = {american}
}

@article{sunStabilityExtensionCarFollowingModel2023b,
  title = {Stability and {{Extension}} of a {{Car-Following Model}} for {{Human-Driven Connected Vehicles}}},
  author = {Sun, Jie and Zheng, Zuduo and Sharma, Anshuman and Sun, Jian},
  year = {2023},
  month = oct,
  journal = {Transportation Research Part C-Emerging Technologies},
  volume = {155},
  issn = {0968-090X},
  doi = {10.1016/j.trc.2023.104317}
}

@article{suo_model-free_2024,
  title = {Model-Free Learning of Corridor Clearance: A near-{{Term}} Deployment Perspective},
  author = {Suo, Dajiang and Jayawardana, Vindula and Wu, Cathy},
  year = {2024},
  month = jun,
  journal = {IEEE Transactions on Intelligent Transportation Systems},
  volume = {25},
  number = {6},
  pages = {4833--4848},
  issn = {1524-9050},
  doi = {10.1109/TITS.2023.3344473}
}

@article{talebpourEffectInformationAvailabilityStability2018a,
  title = {Effect of {{Information Availability}} on {{Stability}} of {{Traffic Flow}}: {{Percolation Theory Approach}}},
  author = {Talebpour, Alireza and Mahmassani, Hani S. and Hamdar, Samer H.},
  year = {2018},
  month = nov,
  journal = {Transportation Research Part B-Methodological},
  volume = {117},
  number = {B, SI},
  pages = {624--638},
  issn = {0191-2615},
  doi = {10.1016/j.trb.2017.09.005}
}

@article{talebpourInfluenceConnectedAutonomousVehicles2016a,
  title = {Influence of {{Connected}} and {{Autonomous Vehicles}} on {{Traffic Flow Stability}} and {{Throughput}}},
  author = {Talebpour, Alireza and Mahmassani, Hani S.},
  year = {2016},
  month = oct,
  journal = {Transportation Research Part C-Emerging Technologies},
  volume = {71},
  pages = {143--163},
  issn = {0968-090X},
  doi = {10.1016/j.trc.2016.07.007},
  langid = {english}
}

@article{tang_grounded_2024,
  title = {Grounded Relational Inference: {{Domain}} Knowledge Driven Explainable Autonomous Driving},
  author = {Tang, Chen and Srishankar, Nishan and Martin, Sujitha and Tomizuka, Masayoshi},
  year = {2024},
  month = jul,
  journal = {IEEE Transactions on Intelligent Transportation Systems},
  issn = {1524-9050},
  doi = {10.1109/TITS.2024.3424667}
}

@article{tangHighwayDecisionMakingMotionPlanning2022a,
  title = {Highway {{Decision-Making}} and {{Motion Planning}} for {{Autonomous Driving}} via {{Soft Actor-Critic}}},
  author = {Tang, Xiaolin and Huang, Bing and Liu, Teng and Lin, Xianke},
  year = {2022},
  month = may,
  journal = {IEEE Transactions on Vehicular Technology},
  volume = {71},
  number = {5},
  pages = {4706--4717},
  issn = {0018-9545},
  doi = {10.1109/TVT.2022.3151651},
  langid = {american}
}

@article{tangNovelHierarchicalCooperativeMerging2022b,
  title = {A {{Novel Hierarchical Cooperative Merging Control Model}} of {{Connected}} and {{Automated Vehicles Featuring Flexible Merging Positions}} in {{System Optimization}}},
  author = {Tang, Zhixian and Zhu, Hong and Zhang, Xin and {Iryo-Asano}, Miho and Nakamura, Hideki},
  year = {2022},
  month = may,
  journal = {Transportation Research Part C-Emerging Technologies},
  volume = {138},
  issn = {0968-090X},
  doi = {10.1016/j.trc.2022.103650},
  langid = {english}
}

@article{tas_making_2018,
  title = {Making Bertha Cooperate-Team {{AnnieWAY}}'s Entry to the 2016 Grand Cooperative Driving Challenge},
  author = {Tas, Omer Sahin and Salscheider, Niels Ole and Poggenhans, Fabian and Wirges, Sascha and Bandera, Claudio and Zofka, Marc Rene and Strauss, Tobias and Zoellner, J. Marius and Stiller, Christoph},
  year = {2018},
  month = apr,
  journal = {IEEE Transactions on Intelligent Transportation Systems},
  volume = {19},
  number = {4},
  pages = {1262--1276},
  issn = {1524-9050},
  doi = {10.1109/TITS.2017.2749974}
}

@article{tianEmpiricalAnalysisSimulationConcave2016,
  title = {Empirical {{Analysis}} and {{Simulation}} of the {{Concave Growth Pattern}} of {{Traffic Oscillations}}},
  author = {Tian, Junfang and Jiang, Rui and Jia, Bin and Gao, Ziyou and Ma, Shoufeng},
  year = {2016},
  month = nov,
  journal = {Transportation Research Part B-Methodological},
  volume = {93},
  number = {A},
  pages = {338--354},
  issn = {0191-2615},
  doi = {10.1016/j.trb.2016.08.001}
}

@article{tianSelfOrganizedRelaySelectionCooperative2017a,
  title = {Self-{{Organized Relay Selection}} for {{Cooperative Transmission}} in {{Vehicular Ad-Hoc Networks}}},
  author = {Tian, Daxin and Zhou, Jianshan and Sheng, Zhengguo and Chen, Min and Ni, Qiang and Leung, Victor C. M.},
  year = {2017},
  month = oct,
  journal = {IEEE Transactions on Vehicular Technology},
  volume = {66},
  number = {10},
  pages = {9534--9549},
  issn = {0018-9545},
  doi = {10.1109/TVT.2017.2715328}
}

@article{toghi_social_2022,
  title = {Social Coordination and Altruism in Autonomous Driving},
  author = {Toghi, Behrad and Valiente, Rodolfo and Sadigh, Dorsa and Pedarsani, Ramtin and Fallah, Yaser P.},
  year = {2022},
  month = dec,
  journal = {IEEE Transactions on Intelligent Transportation Systems},
  volume = {23},
  number = {12},
  pages = {24791--24804},
  issn = {1524-9050},
  doi = {10.1109/TITS.2022.3207872}
}

@article{treiber_delays_2006,
  title = {Delays, {{Inaccuracies}} and {{Anticipation}} in {{Microscopic Traffic Models}}},
  author = {Treiber, M and Kesting, A and Helbing, D},
  year = {2006},
  month = jan,
  journal = {Physica a-Statistical Mechanics and Its Applications},
  volume = {360},
  number = {1},
  pages = {71--88},
  issn = {0378-4371},
  doi = {10.1016/j.physa.2005.05.001},
  langid = {american}
}

@article{valiente_prediction-aware_2024,
  title = {Prediction-Aware and Reinforcement Learning-Based Altruistic Cooperative Driving},
  author = {Valiente, Rodolfo and Razzaghpour, Mahdi and Toghi, Behrad and Shah, Ghayoor and Fallah, Yaser P.},
  year = {2024},
  month = mar,
  journal = {IEEE Transactions on Intelligent Transportation Systems},
  volume = {25},
  number = {3},
  pages = {2450--2465},
  issn = {1524-9050},
  doi = {10.1109/TITS.2023.3323440}
}

@article{wang_analytical_2024,
  title = {Analytical Characterization of Cyberattacks on Adaptive Cruise Control Vehicles},
  author = {Wang, Shian and Shang, Mingfeng and Stern, Raphael},
  year = {2024},
  month = jul,
  journal = {IEEE Transactions on Intelligent Transportation Systems},
  issn = {1524-9050},
  doi = {10.1109/TITS.2024.3428988}
}

@article{wang_cooperative_2020,
  title = {Cooperative {{Eco-Driving}} at {{Signalized Intersections}} in a {{Partially Connected}} and {{Automated Vehicle Environment}}},
  author = {Wang, Ziran and Wu, Guoyuan and Barth, Matthew J.},
  year = {2020},
  month = may,
  journal = {IEEE Transactions on Intelligent Transportation Systems},
  volume = {21},
  number = {5},
  pages = {2029--2038},
  issn = {1524-9050},
  doi = {10.1109/TITS.2019.2911607},
  langid = {american}
}

@article{wang_learning_2023,
  title = {Learning Safe and Human-like High-Level Decisions for Unsignalized Intersections from Naturalistic Human Driving Trajectories},
  author = {Wang, Lingguang and Fernandez, Carlos and Stiller, Christoph},
  year = {2023},
  month = nov,
  journal = {IEEE Transactions on Intelligent Transportation Systems},
  volume = {24},
  number = {11},
  pages = {12477--12490},
  issn = {1524-9050},
  doi = {10.1109/TITS.2023.3286454}
}

@article{wang_modeling_2020,
  title = {Modeling and {{Analyzing Cyberattack Effects}} on {{Connected Automated Vehicular Platoons}}},
  author = {Wang, Pengcheng and Wu, Xinkai and He, Xiaozheng},
  year = {2020},
  month = jun,
  journal = {Transportation Research Part C-Emerging Technologies},
  volume = {115},
  issn = {0968-090X},
  doi = {10.1016/j.trc.2020.102625},
  langid = {american}
}

@article{wang_optimal_2022,
  title = {Optimal {{Control}} of {{Autonomous Vehicles}} for {{Traffic Smoothing}}},
  author = {Wang, Shian and Stern, Raphael and Levin, Michael W.},
  year = {2022},
  month = apr,
  journal = {IEEE Transactions on Intelligent Transportation Systems},
  volume = {23},
  number = {4},
  pages = {3842--3852},
  issn = {1524-9050},
  doi = {10.1109/TITS.2021.3094552},
  langid = {american}
}

@article{wang_stability_2019,
  title = {Stability of {{CACC-manual Heterogeneous Vehicular Flow}} with {{Partial CACC Performance Degrading}}},
  author = {Wang, Hao and Qin, Yanyan and Wang, Wei and Chen, Jun},
  year = {2019},
  month = dec,
  journal = {Transportmetrica B-Transport Dynamics},
  volume = {7},
  number = {1},
  pages = {788--813},
  issn = {2168-0566},
  doi = {10.1080/21680566.2018.1517058},
  langid = {american}
}

@article{wang_trajectory_2022,
  title = {Trajectory {{Jerking Suppression}} for {{Mixed Traffic Flow}} at a {{Signalized Intersection}}: {{A Trajectory Prediction Based Deep Reinforcement Learning Method}}},
  author = {Wang, Shupei and Wang, Ziyang and Jiang, Rui and Yan, Ruidong and Du, Lei},
  year = {2022},
  month = oct,
  journal = {IEEE Transactions on Intelligent Transportation Systems},
  volume = {23},
  number = {10},
  pages = {18989--19000},
  issn = {1524-9050},
  doi = {10.1109/TITS.2022.3152550},
  langid = {american}
}

@article{wangEgoEfficientLaneChangesConnected2022a,
  title = {Ego-{{Efficient Lane Changes}} of {{Connected}} and {{Automated Vehicles}} with {{Impacts}} on {{Traffic Flow}}},
  author = {Wang, Yibing and Wang, Long and Guo, Jingqiu and Papamichail, Ioannis and Papageorgiou, Markos and Wang, Fei-Yue and Bertini, Robert and Hua, Wei and Yang, Qinmin},
  year = {2022},
  month = may,
  journal = {Transportation Research Part C-Emerging Technologies},
  volume = {138},
  issn = {0968-090X},
  doi = {10.1016/j.trc.2021.103478}
}

@article{wangIntegratedSelfConsistentMacroMicroTraffic2024a,
  title = {Integrated {{Self-Consistent Macro-Micro Traffic Flow Modeling}} and {{Calibration Framework Based}} on {{Trajectory Data}}},
  author = {Wang, Zelin and Liu, Zhiyuan and Cheng, Qixiu and Gu, Ziyuan},
  year = {2024},
  month = jan,
  journal = {Transportation Research Part C-Emerging Technologies},
  volume = {158},
  issn = {0968-090X},
  doi = {10.1016/j.trc.2023.104439}
}

@article{wangModelingBoundedRationalityDiscretionary2022a,
  title = {Modeling {{Bounded Rationality}} in {{Discretionary Lane Change}} with the {{Quantal Response Equilibrium}} of {{Game Theory}}},
  author = {Wang, Bingtong and Li, Zhibin and Wang, Shunchao and Li, Meng and Ji, Ang},
  year = {2022},
  month = oct,
  journal = {Transportation Research Part B-Methodological},
  volume = {164},
  pages = {145--161},
  issn = {0191-2615},
  doi = {10.1016/j.trb.2022.08.008},
  langid = {american}
}

@article{wangOptimalFeedbackControlLaw2023a,
  title = {Optimal {{Feedback Control Law}} for {{Automated Vehicles}} in the {{Presence}} of {{Cyberattacks}}: {{A Min-Max Approach}}},
  author = {Wang, Shian and Levin, Michael W. and Stern, Raphael},
  year = {2023},
  month = aug,
  journal = {Transportation Research Part C-Emerging Technologies},
  volume = {153},
  issn = {0968-090X},
  doi = {10.1016/j.trc.2023.104204}
}

@article{wangRollingHorizonControlFramework2014b,
  title = {Rolling {{Horizon Control Framework}} for {{Driver Assistance Systems}}. {{Part II}}: {{Cooperative Sensing}} and {{Cooperative Control}}},
  author = {Wang, Meng and Daamen, Winnie and Hoogendoorn, Serge P. and {van Arem}, Bart},
  year = {2014},
  month = mar,
  journal = {Transportation Research Part C-Emerging Technologies},
  volume = {40},
  pages = {290--311},
  issn = {0968-090X},
  doi = {10.1016/j.trc.2013.11.024}
}

@article{ward_probabilistic_2017,
  title = {Probabilistic {{Model}} for {{Interaction Aware Planning}} in {{Merge Scenarios}}},
  author = {Ward, Erik and Evestedt, Niclas and Axehill, Daniel and Folkesson, John},
  year = {2017},
  month = jun,
  journal = {IEEE Transactions on Intelligent Vehicles},
  volume = {2},
  number = {2},
  pages = {133--146},
  issn = {2379-8858},
  doi = {10.1109/TIV.2017.2730588},
  langid = {american}
}

@article{wegener_automated_2021,
  title = {Automated {{Eco-Driving}} in {{Urban Scenarios Using Deep Reinforcement Learning}}},
  author = {Wegener, Marius and Koch, Lucas and Eisenbarth, Markus and Andert, Jakob},
  year = {2021},
  month = may,
  journal = {Transportation Research Part C-Emerging Technologies},
  volume = {126},
  issn = {0968-090X},
  doi = {10.1016/j.trc.2021.102967},
  langid = {american}
}

@article{wu_hybrid_2023,
  title = {A Hybrid Driving Decision-Making System Integrating Markov Logic Networks and Connectionist {{AI}}},
  author = {Wu, Mengyao and Yu, F. Richard and Liu, Peter Xiaoping and He, Ying},
  year = {2023},
  month = mar,
  journal = {IEEE Transactions on Intelligent Transportation Systems},
  volume = {24},
  number = {3},
  pages = {3514--3527},
  issn = {1524-9050},
  doi = {10.1109/TITS.2022.3227122}
}

@article{wuFlowModularLearningFramework2022a,
  title = {Flow: {{A Modular Learning Framework}} for {{Mixed Autonomy Traffic}}},
  author = {Wu, Cathy and Kreidieh, Abdul Rahman and Parvate, Kanaad and Vinitsky, Eugene and Bayen, Alexandre M.},
  year = {2022},
  month = apr,
  journal = {IEEE Transactions on Robotics},
  volume = {38},
  number = {2},
  pages = {1270--1286},
  issn = {1552-3098},
  doi = {10.1109/TRO.2021.3087314}
}

@article{wuTimeDependentPerformanceAnalysis80211pBased2020a,
  title = {Time-{{Dependent Performance Analysis}} of the 802.11p-{{Based Platooning Communications}} under {{Disturbance}}},
  author = {Wu, Qiong and Ge, Hongmei and Fan, Pingyi and Wang, Jiangzhou and Fan, Qiang and Li, Zhengquan},
  year = {2020},
  month = dec,
  journal = {IEEE Transactions on Vehicular Technology},
  volume = {69},
  number = {12},
  pages = {15760--15773},
  issn = {0018-9545},
  doi = {10.1109/TVT.2020.3034622},
  langid = {american}
}

@article{xiaoDecisionMakingAutonomousVehiclesRandom2024a,
  title = {Decision-{{Making}} for {{Autonomous Vehicles}} in {{Random Task Scenarios}} at {{Unsignalized Intersection Using Deep Reinforcement Learning}}},
  author = {Xiao, Wenxuan and Yang, Yuyou and Mu, Xinyu and Xie, Yi and Tang, Xiaolin and Cao, Dongpu and Liu, Teng},
  year = {2024},
  month = jun,
  journal = {IEEE Transactions on Vehicular Technology},
  volume = {73},
  number = {6},
  pages = {7812--7825},
  issn = {0018-9545},
  doi = {10.1109/TVT.2024.3360445}
}

@article{xiaoUnravellingEffectsCooperativeAdaptive2018a,
  title = {Unravelling {{Effects}} of {{Cooperative Adaptive Cruise Control Deactivation}} on {{Traffic Flow Characteristics}} at {{Merging Bottlenecks}}},
  author = {Xiao, Lin and Wang, Meng and Schakel, Wouter and {van Arem}, Bart},
  year = {2018},
  month = nov,
  journal = {Transportation Research Part C-Emerging Technologies},
  volume = {96},
  pages = {380--397},
  issn = {0968-090X},
  doi = {10.1016/j.trc.2018.10.008}
}

@article{xiong_speed_2022,
  title = {Speed {{Advice}} for {{Connected Vehicles}} at an {{Isolated Signalized Intersection}} in a {{Mixed Traffic Flow Considering Stochasticity}} of {{Human Driven Vehicles}}},
  author = {Xiong, Bang-Kai and Jiang, Rui},
  year = {2022},
  month = aug,
  journal = {IEEE Transactions on Intelligent Transportation Systems},
  volume = {23},
  number = {8},
  pages = {11261--11272},
  issn = {1524-9050},
  doi = {10.1109/TITS.2021.3102430},
  langid = {american}
}

@article{xiongManagingMergingCAVLane2022b,
  title = {Managing {{Merging}} from a {{CAV Lane}} to a {{Human-Driven Vehicle Lane Considering}} the {{Uncertainty}} of {{Human Driving}}},
  author = {Xiong, Bang-Kai and Jiang, Rui and Li, Xiaopeng},
  year = {2022},
  month = sep,
  journal = {Transportation Research Part C-Emerging Technologies},
  volume = {142},
  issn = {0968-090X},
  doi = {10.1016/j.trc.2022.103775}
}

@article{yan_unified_2023,
  title = {Unified {{Automatic Control}} of {{Vehicular Systems}} with {{Reinforcement Learning}}},
  author = {Yan, Zhongxia and Kreidieh, Abdul Rahman and Vinitsky, Eugene and Bayen, Alexandre M. and Wu, Cathy},
  year = {2023},
  month = apr,
  journal = {IEEE Transactions on Automation Science and Engineering},
  volume = {20},
  number = {2},
  pages = {789--804},
  issn = {1545-5955},
  doi = {10.1109/TASE.2022.3168621},
  langid = {american}
}

@article{yang_enhancing_2024,
  title = {Enhancing Robustness of Deep Reinforcement Learning Based Adaptive Traffic Signal Controllers in Mixed Traffic Environments through Data Fusion and Multi-Discrete Actions},
  author = {Yang, Tianjia and Fan, Wei},
  year = {2024},
  month = may,
  journal = {IEEE Transactions on Intelligent Transportation Systems},
  issn = {1524-9050},
  doi = {10.1109/TITS.2024.3399066}
}

@article{yangIsolatedIntersectionControlVarious2016a,
  title = {Isolated {{Intersection Control}} for {{Various Levels}} of {{Vehicle Technology}}: {{Conventional}}, {{Connected}}, and {{Automated Vehicles}}},
  author = {Yang, Kaidi and Guler, S. Ilgin and Menendez, Monica},
  year = {2016},
  month = nov,
  journal = {Transportation Research Part C-Emerging Technologies},
  volume = {72},
  pages = {109--129},
  issn = {0968-090X},
  doi = {10.1016/j.trc.2016.08.009}
}

@article{yao_fuel_2021,
  title = {Fuel {{Consumption}} and {{Transportation Emissions Evaluation}} of {{Mixed Traffic Flow}} with {{Connected Automated Vehicles}} and {{Human-Driven Vehicles}} on {{Expressway}}},
  author = {Yao, Zhihong and Wang, Yi and Liu, Bo and Zhao, Bin and Jiang, Yangsheng},
  year = {2021},
  month = sep,
  journal = {Energy},
  volume = {230},
  issn = {0360-5442},
  doi = {10.1016/j.energy.2021.120766},
  langid = {american}
}

@article{yao_linear_2021,
  title = {Linear {{Stability Analysis}} of {{Heterogeneous Traffic Flow Considering Degradations}} of {{Connected Automated Vehicles}} and {{Reaction Time}}},
  author = {Yao, Zhihong and Xu, Taorang and Jiang, Yangsheng and Hu, Rong},
  year = {2021},
  month = jan,
  journal = {Physica a-Statistical Mechanics and Its Applications},
  volume = {561},
  issn = {0378-4371},
  doi = {10.1016/j.physa.2020.125218},
  langid = {american}
}

@article{yaoAnalysisImpactMaximumPlatoon2023a,
  title = {Analysis of the {{Impact}} of {{Maximum Platoon Size}} of {{CAVs}} on {{Mixed Traffic Flow}}: {{An Analytical}} and {{Simulation Method}}},
  author = {Yao, Zhihong and Wu, Yunxia and Wang, Yi and Zhao, Bin and Jiang, Yangsheng},
  year = {2023},
  month = feb,
  journal = {Transportation Research Part C-Emerging Technologies},
  volume = {147},
  issn = {0968-090X},
  doi = {10.1016/j.trc.2022.103989}
}

@article{yuanPredictiveEnergyManagementStrategy2019,
  title = {Predictive Energy Management Strategy for Connected {{48V}} Hybrid Electric Vehicles},
  author = {Yuan, Jingni and Yang, Lin},
  year = {2019},
  month = nov,
  journal = {Energy},
  volume = {187},
  issn = {0360-5442},
  doi = {10.1016/j.energy.2019.115952},
  langid = {english}
}

@article{yuAssessmentDynamicPlatoonInformation2023a,
  title = {On the {{Assessment}} of the {{Dynamic Platoon}} and {{Information Flow Topology}} on {{Mixed Traffic Flow}} under {{Connected Environment}}},
  author = {Yu, Weijie and Hua, Xuedong and Ngoduy, Dong and Wang, Wei},
  year = {2023},
  month = sep,
  journal = {Transportation Research Part C-Emerging Technologies},
  volume = {154},
  issn = {0968-090X},
  doi = {10.1016/j.trc.2023.104265}
}

@article{zhang_accelerated_2024,
  title = {Accelerated {{Safety Testing}} for {{Highly Automated Vehicles}}: {{Application}} and {{Capability Comparison}} of {{Surrogate Models}}},
  author = {Zhang, He and Sun, Jian and Tian, Ye},
  year = {2024},
  month = jan,
  journal = {IEEE Transactions on Intelligent Vehicles},
  volume = {9},
  number = {1},
  pages = {2409--2418},
  issn = {2379-8858},
  doi = {10.1109/TIV.2023.3319158},
  langid = {american}
}

@article{zhang_bayesian_2024,
  title = {Bayesian Calibration of the Intelligent Driver Model},
  author = {Zhang, Chengyuan and Sun, Lijun},
  year = {2024},
  month = jan,
  journal = {IEEE Transactions on Intelligent Transportation Systems},
  issn = {1524-9050},
  doi = {10.1109/TITS.2024.3354102}
}

@article{zhang_game_2020,
  title = {A {{Game Theoretic Model Predictive Controller}} with {{Aggressiveness Estimation}} for {{Mandatory Lane Change}}},
  author = {Zhang, Qingyu and Langari, Reza and Tseng, H. Eric and Filev, Dimitar and Szwabowski, Steven and Coskun, Serdar},
  year = {2020},
  month = mar,
  journal = {IEEE Transactions on Intelligent Vehicles},
  volume = {5},
  number = {1},
  pages = {75--89},
  issn = {2379-8858},
  doi = {10.1109/TIV.2019.2955367},
  langid = {american}
}

@article{zhang_risk-aware_2023,
  title = {Risk-Aware Decision-Making and Planning Using Prediction-Guided Strategy Tree for the Uncontrolled Intersections},
  author = {Zhang, Ting and Fu, Mengyin and Song, Wenjie},
  year = {2023},
  month = oct,
  journal = {IEEE Transactions on Intelligent Transportation Systems},
  volume = {24},
  number = {10},
  pages = {10791--10803},
  issn = {1524-9050},
  doi = {10.1109/TITS.2023.3280225}
}

@article{zhang_stackelberg_2024,
  title = {Stackelberg Differential Lane Change Game Based on {{MPC}} and Inverse {{MPC}}},
  author = {Zhang, Qingyu and Langari, Reza and Tseng, H. Eric and Mohan, Shankar and Szwabowski, Steven and Filev, Dimitar},
  year = {2024},
  month = apr,
  journal = {IEEE Transactions on Intelligent Transportation Systems},
  issn = {1524-9050},
  doi = {10.1109/TITS.2024.3386790}
}

@article{zhangGenerativeCarFollowingModelConditioned2022c,
  title = {A {{Generative Car-Following Model Conditioned}} on {{Driving Styles}}},
  author = {Zhang, Yifan and Chen, Xinhong and Wang, Jianping and Zheng, Zuduo and Wu, Kui},
  year = {2022},
  month = dec,
  journal = {Transportation Research Part C-Emerging Technologies},
  volume = {145},
  issn = {0968-090X},
  doi = {10.1016/j.trc.2022.103926}
}

@article{zhangNoMoreRoadBullying2023a,
  title = {No {{More Road Bullying}}: {{An Integrated Behavioral}} and {{Motion Planner}} with {{Proactive Right-of-Way Acquisition Capability}}},
  author = {Zhang, Zihan and Yan, Xuerun and Wang, Hong and Ding, Chuan and Xiong, Lu and Hu, Jia},
  year = {2023},
  month = nov,
  journal = {Transportation Research Part C-Emerging Technologies},
  volume = {156},
  issn = {0968-090X},
  doi = {10.1016/j.trc.2023.104363}
}

@article{zhangPlatoonCenteredControlEcoDrivingSignalized2023b,
  title = {Platoon-{{Centered Control}} for {{Eco-Driving}} at {{Signalized Intersection Built}} upon {{Hybrid MPC System}}, {{Online Learning}} and {{Distributed Optimization Part I}}: {{Modeling}} and {{Solution Algorithm Design}}},
  author = {Zhang, Hanyu and Du, Lili},
  year = {2023},
  month = jun,
  journal = {Transportation Research Part B-Methodological},
  volume = {172},
  pages = {174--198},
  issn = {0191-2615},
  doi = {10.1016/j.trb.2023.02.006},
  langid = {english}
}

@article{zhengMultiObjectiveCalibrationFrameworkCapturing2023b,
  title = {A {{Multi-Objective Calibration Framework}} for {{Capturing}} the {{Behavioral Patterns}} of {{Autonomously-Driven Vehicles}}},
  author = {Zheng, Shi-Teng and Makridis, Michail A. and Kouvelas, Anastasios and Jiang, Rui and Jia, Bin},
  year = {2023},
  month = jul,
  journal = {Transportation Research Part C-Emerging Technologies},
  volume = {152},
  issn = {0968-090X},
  doi = {10.1016/j.trc.2023.104151},
  langid = {english}
}

@article{zhongCrossEntropyMethodProbabilisticSensitivity2016b,
  title = {A {{Cross-Entropy Method}} and {{Probabilistic Sensitivity Analysis Framework}} for {{Calibrating Microscopic Traffic Models}}},
  author = {Zhong, R. X. and Fu, K. Y. and Sumalee, A. and Ngoduy, D. and Lam, W. H. K.},
  year = {2016},
  month = feb,
  journal = {Transportation Research Part C-Emerging Technologies},
  volume = {63},
  pages = {147--169},
  issn = {0968-090X},
  doi = {10.1016/j.trc.2015.12.006}
}

@article{zhou_analytical_2021,
  title = {Analytical {{Analysis}} of the {{Effect}} of {{Maximum Platoon Size}} of {{Connected}} and {{Automated Vehicles}}},
  author = {Zhou, Jiazu and Zhu, Feng},
  year = {2021},
  month = jan,
  journal = {Transportation Research Part C-Emerging Technologies},
  volume = {122},
  issn = {0968-090X},
  doi = {10.1016/j.trc.2020.102882},
  langid = {american}
}

@article{zhou_cooperative_2023,
  title = {Cooperative {{Control}} of a {{Platoon}} of {{Connected Autonomous Vehicles}} and {{Unconnected Human-Driven Vehicles}}},
  author = {Zhou, Anye and Peeta, Srinivas and Wang, Jian},
  year = {2023},
  month = dec,
  journal = {Computer-Aided Civil and Infrastructure Engineering},
  volume = {38},
  number = {18},
  pages = {2513--2536},
  issn = {1093-9687},
  doi = {10.1111/mice.12995},
  langid = {american}
}

@article{zhou_development_2020,
  title = {Development of an {{Efficient Driving Strategy}} for {{Connected}} and {{Automated Vehicles}} at {{Signalized Intersections}}: {{A Reinforcement Learning Approach}}},
  author = {Zhou, Mofan and Yu, Yang and Qu, Xiaobo},
  year = {2020},
  month = jan,
  journal = {IEEE Transactions on Intelligent Transportation Systems},
  volume = {21},
  number = {1},
  pages = {433--443},
  issn = {1524-9050},
  doi = {10.1109/TITS.2019.2942014},
  langid = {american}
}

@article{zhou_hybrid_2024,
  title = {A {{Hybrid Virtual-Real Traffic Simulation Approach}} to {{Reproducing}} the {{Spatiotemporal Distribution}} of {{Bridge Loads}}},
  author = {Zhou, Junyong and Wu, Wenrong and Caprani, Colin C. and Tan, Zeyin and Wei, Bin and Zhang, Junping},
  year = {2024},
  month = jun,
  journal = {Computer-Aided Civil and Infrastructure Engineering},
  volume = {39},
  number = {11},
  pages = {1699--1723},
  issn = {1093-9687},
  doi = {10.1111/mice.13154},
  langid = {american}
}

@article{zhou_impact_2017,
  title = {On the Impact of Cooperative Autonomous Vehicles in Improving Freeway Merging: A Modified Intelligent Driver Model-Based Approach},
  author = {Zhou, Mofan and Qu, Xiaobo and Jin, Sheng},
  year = {2017},
  month = jun,
  journal = {IEEE Transactions on Intelligent Transportation Systems},
  volume = {18},
  number = {6},
  pages = {1422--1428},
  issn = {1524-9050},
  doi = {10.1109/TITS.2016.2606492}
}

@article{zhou_impact_2021,
  title = {Impact of {{CAV Platoon Management}} on {{Traffic Flow Considering Degradation}} of {{Control Mode}}},
  author = {Zhou, Linjie and Ruan, Tiancheng and Ma, Ke and Dong, Changyin and Wang, Hao},
  year = {2021},
  month = nov,
  journal = {Physica a-Statistical Mechanics and Its Applications},
  volume = {581},
  issn = {0378-4371},
  doi = {10.1016/j.physa.2021.126193},
  langid = {american}
}

@article{zhou_modeling_2020,
  title = {Modeling the {{Fundamental Diagram}} of {{Mixed Human-Driven}} and {{Connected Automated Vehicles}}},
  author = {Zhou, Jiazu and Zhu, Feng},
  year = {2020},
  month = jun,
  journal = {Transportation Research Part C-Emerging Technologies},
  volume = {115},
  issn = {0968-090X},
  doi = {10.1016/j.trc.2020.102614},
  langid = {american}
}

@article{zhou_state-constrained_2019,
  title = {A {{State-Constrained Optimal Control Based Trajectory Planning Strategy}} for {{Cooperative Freeway Mainline Facilitating}} and on-{{Ramp Merging Maneuvers}} under {{Congested Traffic}}},
  author = {Zhou, Yue and Chung, Edward and Bhaskar, Ashish and Cholette, Michael E.},
  year = {2019},
  month = dec,
  journal = {Transportation Research Part C-Emerging Technologies},
  volume = {109},
  pages = {321--342},
  issn = {0968-090X},
  doi = {10.1016/j.trc.2019.10.017},
  langid = {american}
}

@article{zhouCooperativeSignalFreeIntersectionControl2022b,
  title = {Cooperative {{Signal-Free Intersection Control Using Virtual Platooning}} and {{Traffic Flow Regulation}}},
  author = {Zhou, Anye and Peeta, Srinivas and Yang, Menglin and Wang, Jian},
  year = {2022},
  month = may,
  journal = {Transportation Research Part C-Emerging Technologies},
  volume = {138},
  issn = {0968-090X},
  doi = {10.1016/j.trc.2022.103610}
}

@article{zhouExperimentalFeaturesEmissionsFuel2023b,
  title = {Experimental {{Features}} of {{Emissions}} and {{Fuel Consumption}} in a {{Car-Following Platoon}}},
  author = {Zhou, Shirui and Tian, Junfang and Ge, Ying-En and Yu, Shaowei and Jiang, Rui},
  year = {2023},
  month = aug,
  journal = {Transportation Research Part D-Transport and Environment},
  volume = {121},
  issn = {1361-9209},
  doi = {10.1016/j.trd.2023.103823}
}

@article{zhuModelingCarFollowingBehaviorUrban2018b,
  title = {Modeling {{Car-Following Behavior}} on {{Urban Expressways}} in {{Shanghai}}: {{A Naturalistic Driving Study}}},
  author = {Zhu, Meixin and Wang, Xuesong and Tarko, Andrew and Fang, Shou'en},
  year = {2018},
  month = aug,
  journal = {Transportation Research Part C-Emerging Technologies},
  volume = {93},
  pages = {425--445},
  issn = {0968-090X},
  doi = {10.1016/j.trc.2018.06.009}
}

\bio{}
\endbio
\printcredits
\end{document}